\documentclass[graybox]{svmult}

\usepackage{mathptmx}       % selects Times Roman as basic font
\usepackage{helvet}         % selects Helvetica as sans-serif font
\usepackage{courier}        % selects Courier as typewriter font
\usepackage{type1cm}        % activate if the above 3 fonts are

\usepackage{makeidx}         
\usepackage{graphicx}      
            
\usepackage{multicol}       
\usepackage[bottom]{footmisc}
\usepackage{multirow}

\usepackage{color}
\usepackage[numbers]{natbib}
\usepackage[breaklinks, colorlinks, citecolor=blue]{hyperref}
\newcommand{\be}{\,\begin{equation}}
\newcommand{\ee}{\,\end{equation}}

\def\fermilat{{\it Fermi}-LAT }

\makeindex      

\begin{document}

\title*{SELECTED TOPICS \\ IN COSMIC RAY PHYSICS}
\author{Roberto Aloisio, Pasquale Blasi, Ivan De Mitri, Sergio Petrera}
\institute{R. Aloisio\at Gran Sasso Science Institute, viale F. Crispi 7, L'Aquila, Italy \at Laboratori Nazionali Gran Sasso INFN, Assergi (L'Aquila), Italy\\
\email{roberto.aloisio@gssi.infn.it}
\and P. Blasi \at INAF/Osservatorio Astrofisico di Arcetri, largo E. Fermi 5, Firenze, Italy \at Gran Sasso Science Institute, viale F. Crispi 7, L'Aquila, Italy\\ 
\email{blasi@arcetri.astro.it}
\and I. De Mitri \at Dipartimento di Matematica e Fisica "E. De Giorgi", Universit\`a del Salento, Lecce, Italy \at INFN, Sezione di Lecce, Italy\\
\email{ivan.demitri@le.infn.it}
\and S. Petrera \at Dipartimento di Scienze Fisiche e Chimiche Universit\`a dell'Aquila, L'Aquila, Italy \at Gran Sasso Science Institute, viale F. Crispi 7, L'Aquila, Italy \at Laboratori Nazionali Gran Sasso INFN, Assergi (L'Aquila), Italy\\
\email{sergio.petrera@aquila.infn.it}
}

\maketitle

\abstract{The search for the origin of cosmic rays is as active as ever, mainly driven by new insights provided by recent pieces of observation. Much effort is being channelled in putting the so called supernova paradigm for the origin of galactic cosmic rays on firmer grounds, while at the highest energies we are trying to understand the observed cosmic ray spectra and mass composition and relating them to potential sources of extragalactic cosmic rays. Interestingly, a topic that has acquired a dignity of its own is the investigation of the transition region between the galactic and extragalactic components, once associated with the ankle and now increasingly thought to be taking place at somewhat lower energies. Here we summarize recent developments in the observation and understanding of galactic and extragalactic cosmic rays and we discuss the implications of such findings for the modelling of the transition between the two.}

%\tableofcontent

\section{Introduction}
\label{sec:intro}

There are different levels of understanding of the origin of Cosmic Rays (CRs), but a reasonable starting point is to establish some separation between the CRs that can potentially be accelerated inside the Galaxy and the ones that are thought to be produced outside the Milky Way. This separation is somewhat arbitrary and hides our fundamental ignorance of the actual conditions required to define CRs as extragalactic. Nevertheless, the physical problems associated with galactic CRs appear to be qualitatively different from the ones involved in ultra high energy CRs, hence we will adopt this separation in this review as well, paying special attention to the underlying assumptions and possibly their failure. While there is a substantial consensus that galactic CRs are somehow related to one or more types of supernova (SN) explosions and that acceleration is mainly due to diffusive transport in the proximity of strong shocks formed as a consequence of these explosions, less consensus exists on whether all or a subset of SNe can actually reach the knee energy. At a few PeV, there is some evidence that chemical composition changes, thereby leading to the formation of the knee in the all-particle spectrum \cite{Apel:2008cd}, although the details of how this takes place are not well understood: some observations suggest that the knee is made by light elements \cite{Apel:2008cd}, while others \cite{Bartoli:2015vca} find that light elements disappear at lower energies and the knee gets dominated by elements with intermediate mass. This type of problem is to be considered essentially of experimental nature at this time. 

The transport of CRs inside the accelerators and throughout the Galaxy is described by models based on the same physical ingredients: spatial diffusion induced by resonant scattering of charged particles off plasma waves. Such waves are likely to be, at least partially, generated by the same particles during transport, due to instabilities induced by local streaming. This apparently simple picture is in fact deceiving, in that it hides the essentially non-linear nature of the transport phenomenon: the large scale behaviour of CRs is determined by the superposition of microphysical particle-wave interactions. In this sense, the transport of CRs has become an instance of the so-called inner space - outer space conundrum, well known in the field of cosmology. In much the same way that the laws of particle physics shape the evolution of the universe, the laws of plasma physics on small scales shape the behaviour of CRs on large scales. 

Several instances of self-regulation have been found in such systems, ranging from particle acceleration at supernova shocks to propagation in the Galaxy in a background of self-generated turbulence. The complexity of these situations is often overwhelming and one resolves to adopt effective approaches that, while retaining the main underlying physical aspects, may still allow us to describe nature in a satisfactory way. 

The ever increasing quality of observations reveals aspects of Nature that force us to improve the quality of the effective models that we adopt to describe it. 
This trend is a fair description of the history of CRs in the last few decades:  the simple energetic argument that led to propose supernova remnants (SNRs) as sources of the bulk of galactic CRs and the diffusive paradigm for the transport of CRs in the Galaxy explain, by themselves, the main aspects of the origin of CRs. On the other hand, this simple picture fails to describe many other pieces of observation that have come about with time. 

The standard SNR paradigm predicts that the spectrum of CRs accelerated at strong shocks is very close to $\sim E^{-2}$ \cite{Bell:1978zc} but in the few cases in which gamma ray emission can be unequivocally attributed to hadronic interactions, the inferred CR spectrum appears to be steeper than $E^{-2}$. Moreover, since the spectrum observed at the Earth is $\sim E^{-2.7}$, the SNR paradigm would naively suggest that CR transport be described by a diffusion coefficient $D(E)\propto E^{0.7}$, which however is in contradiction with the measured large scale anisotropy \cite{Ptuskin:2006xz,Blasi:2011fm}. In addition, such a diffusion coefficient leads to an energy dependence of the B/C ratio, proportional to the grammage traversed by CRs, that is not consistent, at high energy, with measurements from the PAMELA experiment \cite{Adriani:2014xoa} and AMS-02 collaboration \cite{2016PhRvL.117w1102A}. 

If used to estimate the maximum energy $E_{max}$ of CRs accelerated at SNR shocks, the same diffusion coefficient would lead to expect that $E_{max}\leq 1 GeV$, quite at odds with observations. This fact alone is a signature that the process of particle acceleration at SNR shocks works in a much more complex manner than the basic paradigm would suggest. The recent detection of narrow rims of X-ray emission from virtually all young SNRs \cite{Vink:2011ei} provided indirect confirmation that the magnetic field in the shock proximity is amplified by a factor $\sim 10-100$. Although the nature of the amplification process is not clear as yet, streaming instability excited by CRs themselves provides the correct order of magnitude to explain the observed rims as a result of synchrotron emission from very high energy electrons. At the same time, the inferred magnetic field would make the acceleration of CRs up to $100-1000$ TeV at SNR shocks plausible \cite{Voelk:2008bf,Volk:2004vi,Bell:2013kq,Cardillo:2015zda}. Interestingly, in order to explain CR energetics on galactic scales, SNRs are required to accelerate CRs with a $\sim 10\%$ efficiency, which is also required for magnetic field amplification, which in turn leads to high values of $E_{max}$: particle acceleration at a SNR shock is a typical example of a self-regulated non-linear system, in which well known plasma physics laws combine to provide a complex outcome. 

It is likely, though less clear, that a similar chain of processes also works for CR transport through the Galaxy. At present, propagation of CRs is described as diffusive with a diffusion coefficient that is tailored to fit observations. Advection with a wind is treated as an option in most propagation codes (e.g. GALPROP, DRAGON, PICARD and Usine \cite{1998ApJ...509..212S,2008JCAP...10..018E,2014APh....55...37K,2001ApJ...555..585M}). 

The spectrum of different elements in CRs has been recently measured at the Earth location by PAMELA \cite{Adriani:2011cu} and AMS-02 \cite{Aguilar:2015ooa,Aguilar:2015ctt} and found to be characterised by small spectral breaks at a few hundred GV rigidity, quite at odds with the standard view of power law injection and diffusion. This phenomenon might be the manifestation of several effects: for instance it might result from the stochastic overlap of discrete sources around the Sun \cite{Thoudam:2013sia}, from reacceleration in weak SN shocks \cite{Thoudam:2014sta}, or it might result from a spatially inhomogeneous diffusion coefficient \cite{Tomassetti:2012ga}. Finally, it might show that non-linear production of waves and pre-existing waves are both responsible for CR diffusion, each one of them being important at different energies \cite{Blasi:2012yr,Aloisio:2013tda}.

Traditionally, the ratios $e^{+}/(e^{-}+e^{+})$ and $\bar p/p$ have also been used to infer the propagation properties of galactic CRs. However this is possible only if both positrons and antiprotons are solely generated as secondary products of CR interactions in the Galaxy, and in this case one expects both these ratios to be monotonically decreasing functions of energy above $\sim 10$ GeV. 
The PAMELA experiment measured the positron ratio and found that it grows with energy \cite{Adriani:2008zr} at least up to $\sim 100$ GeV. This result was later confirmed and extended to higher energies by AMS-02 \cite{Aguilar:2013qda}. The absence of an increasing trend with energy in the $\bar p/p$ ratio \cite{Adriani:2008zq} leads to the conclusion that the positron excess is likely due to new sources of astrophysical positrons that do not produce antiprotons. The same concept also imposes strong constraints on possible Dark Matter (DM) related models of the positrons excess (see \cite{Serpico:2011wg} for a comprehensive review). The measurement of the spectra of electrons and positrons separately \cite{Adriani:2013uda,Aguilar:2014mma} allowed us to conclude that the positron excess stems out of an extra contribution to the positron flux rather than a deficit of the electron flux, namely sources of positrons (but not antiprotons) are required to explain data. It has been speculated that such sources might be old SNRs \cite{Blasi:2009hv,Blasi:2009bd} or pulsar wind nebulae \cite{Hooper:2008kg,2012CEJPh..10....1P}. 

On the other hand, the recent extension of the measurement of the antiproton flux and the $\bar p/p$ ratio to higher energies by AMS-02 \cite{2016PhRvL.117i1103A} has stimulated an exciting discussion on a radically new view of the anomalies in secondary to primary ratios: it has been pointed out \cite{Lipari:2016vqk,2016PhRvL.117i1103A} that the energy spectra of positrons and antiprotons have very similar slopes and such slope is, in turn, very close to that of the proton spectrum at high energies. A similar consideration was put forward earlier in Refs. \cite{2010MNRAS.405.1458K,2013PhRvL.111u1101B} based on the positron and proton spectra alone. This simple consideration is used by the authors to suggest that both positrons and antiprotons are purely secondary products of CR interactions. Clearly these scenarios are not problem free: for instance, an alternative explanation of the B/C ratio should be sought \cite{Lipari:2016vqk} since no apparent anomaly has been measured in this quantity.

A general consequence of the SNR paradigm outlined above is that the flux of galactic CRs should end with an iron dominated composition at energies $\sim 26$ times larger than the knee in the proton spectrum. If such knee is indeed at PeV energies, as KASCADE data suggest, then galactic CRs should end below $\sim 10^{17}$ eV, well below the ankle. 

The measurements carried out by the Pierre Auger Observatory \cite{Aab:2014kda} have shown that the mass composition of CRs, from prevalently light at $\sim 10^{18}$ eV, becomes increasingly heavier towards higher energies. Several independent calculations \cite{Allard:2011aa,Fang:2012rx,Aloisio:2013hya} showed that the observed spectrum and composition can be well explained only if sources of ultra-high energy CRs (UHECRs) provide very hard spectra and a maximum rigidity $\sim 5\times 10^{18}$ V. One should appreciate the change of paradigm that these recent observations forced us towards: ten years ago, the general consensus was that UHECRs are protons and that sources should accelerate them to $> 10^{20}$ eV, something that would not be consistent with current Auger data. On the other hand, the Telescope Array (TA), operational in the northern hemisphere, collected data that suggest a somewhat different scenario \cite{Abbasi:2014sfa}, where the mass composition is compatible with being light for energies above $10^{18}$ eV, with no apparent transition to a heavier mass composition. A joint working group made of members of both collaborations has recently concluded that the results of the two experiments are not in conflict once systematic and statistical uncertainties have been taken into account. This conclusion, though encouraging on one hand, casts serious doubts on the possibility of reliably measuring the mass composition at the highest energies, unless some new piece of information becomes available. It should be noted that the spectra measured by the two experiments, though being in general agreement, differ beyond the systematic error at the highest energies: the TA spectrum shows a suppression that is consistent with the Greisen-Zatsepin-Kuzmin cutoff, while the shape of the spectrum measured by Auger appears to be in better agreement with propagation of nuclei. 

On the other hand, the results of the two experiments in terms of spectra and mass composition show good agreement around $10^{18}$ eV, where CRs are found to be light. The fact that between $10^{17}$ and $10^{18}$ eV the mass composition changes from heavy to light is suggestive of a possible transition from galactic to extragalactic CRs in the same region, well below the ankle. 

This paper is structured as follows: in \S \ref{sec:observ} we discuss the status of observations. The transport of galactic CRs is discussed in \S \ref{sec:transport}, while the status of investigation on CR acceleration is summarized in \S \ref{sec:accelera}. The transport of ultra high energy cosmic rays is discussed in \S \ref{sec:prop} while some considerations about the sources are reported in \S \ref{sec:source}. 
The possibility to infer useful information on exotic physics (such as top-down models and violations of Lorentz invariance) are discussed in \S \ref{sec:td}. 
In \S \ref{sec:trans} we summarize different models of the transition from galactic to extragalactic cosmic rays. We conclude in \S \ref{sec:conclu}.

\section{Cosmic rays observations}
\label{sec:observ}

In this section, a short review of experimental results on some selected topics on CR physics will be given.
For each topic, a discussion on new and future projects has also been added in order to focus on the key issues 
that, from the experimental side, could bring to more and better information for the understanding of the relevant
physics phenomena. After a section dedicated to the observation of electrons/positrons and antiprotons, the measurements
on protons and nuclei will be discussed starting from ballon and space borne experiments up to the highest energies, 
currently covered with giant ground arrays.

\subsection{Observations of electrons, positrons and antiprotons}
\label{sec:ele-posi-antip}

Even if the electron/positron component, i.e. ($e^- + e^+$), accounts for approximatively 1\% only of the total CR flux, it is deeply studied in order to infer important information on propagation processes. In the stardard scenario, secondary electrons and positrons are (equally) produced via interactions of primary CRs with the InterStellar Medium (ISM), therefore the observed overabundance of electrons on positrons is a clear indication that most of the electrons have a primary origin.
Because of the low mass, this component suffers significant energy losses during propagation in the Galaxy. At high energies, such losses produce a steeper energy spectrum compared to that of protons and actually place upper limits on the age and distance (at about 1$0^5\,$yr and 1\,kpc, respectively) of the astrophysical sources of TeV electrons. Since the number of such nearby objects is limited, the electron energy spectrum above 1\,TeV is then expected to exhibit spectral features, and a sizeable anisotropy in the arrival directions is also foreseen at very high energies \cite{Kobayashi:2003kp}.

Measurements of the CR electron/positron fluxes have been pursued since many years by balloon-borne and space-based experiments. Because of the low intensity of the signals and the large proton-induced background, the main requirements for these instruments are a large exposure time and a sufficient e/p separation capability. Calorimeters can be used to measure the inclusive so called all-electron, i.e. ($e^- + e^+$), spectrum, while separating electrons form positrons obviously requires the determination of the sign of the charge through a magnetic spectrometer, that puts anyway severe limits to the highest possible detectable energy, this being limited by the Maximum Detectable Rigidity (MDR).
Important progress was made in the field in the last years, due to the use of magnetic spectrometers in space.  The positron fraction was shown to grow with energy by the PAMELA experiment \cite{Adriani:2008zr} at least up to $\sim 100$ GeV, this result being confirmed with precision measurements by AMS-02 \cite{Aguilar:2013qda}, that also extended the covered range up to about 500\,GeV.

These findings were also confirmed, even though with larger systematic uncertainties by the Fermi-LAT experiment \cite{Ackerman:2012ph}, which is not equipped with a magnetic spectrometer but used the Earth magnetic field as a charge sign separation tool.

\begin{figure}[t]
\begin{center}
\includegraphics[scale=.483]{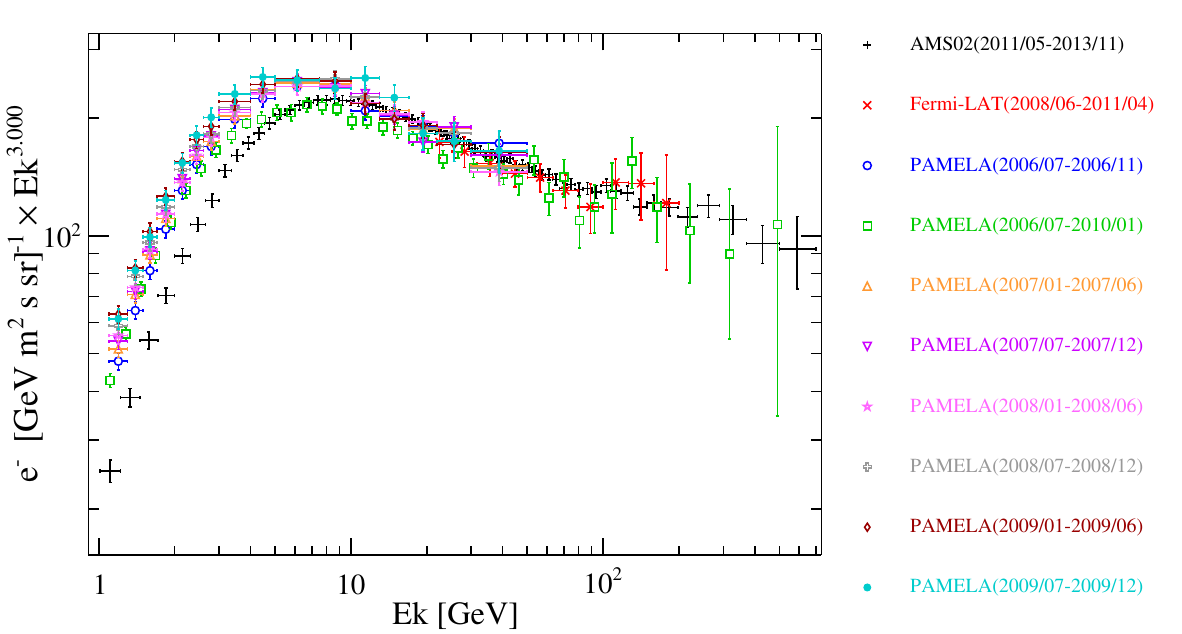} 
\includegraphics[scale=.483]{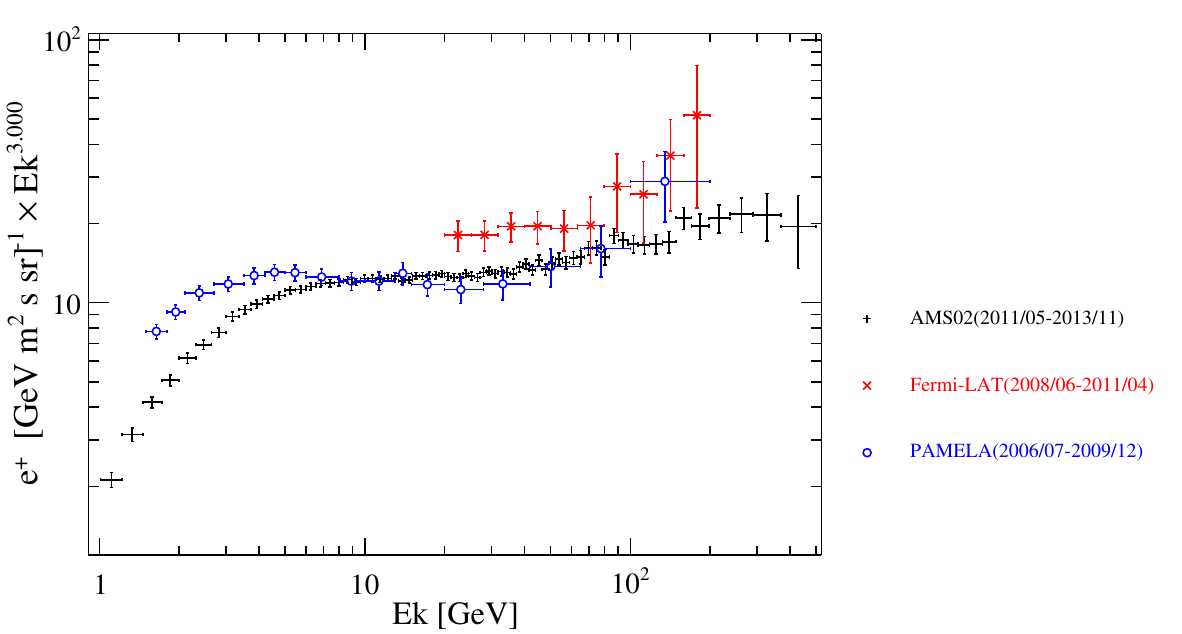} 
\includegraphics[scale=.483]{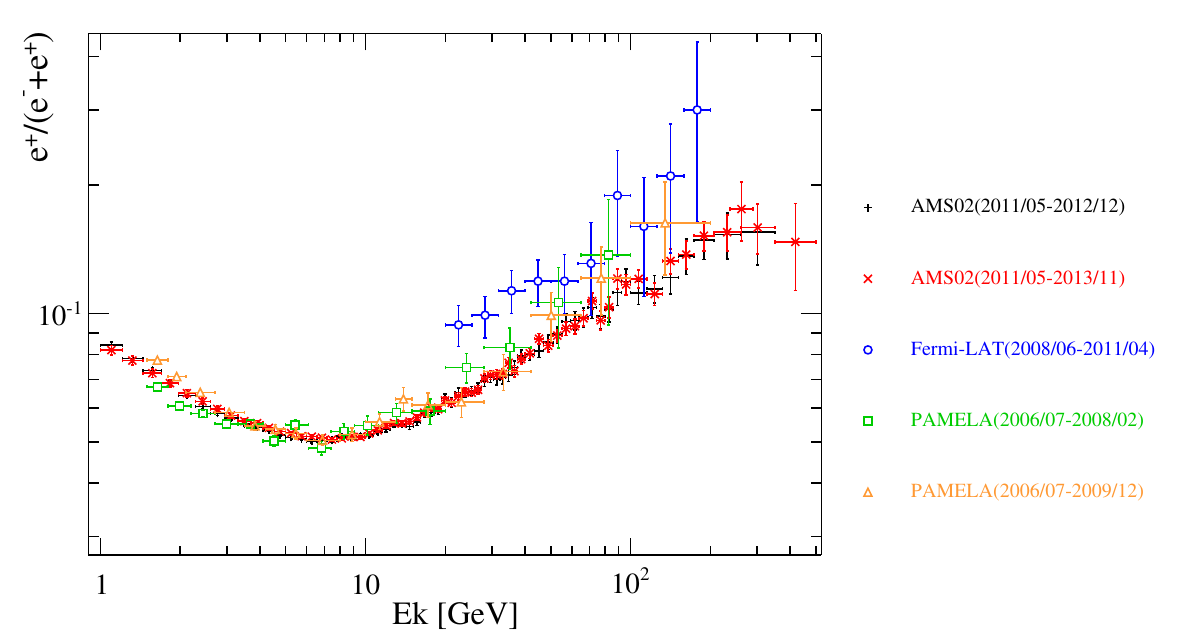} 
\end{center}
\caption{Fluxes of cosmic electrons and positrons (upper and central panel, respectively) as measured by AMS-02, PAMELA and Fermi-LAT experiments. 
The positron fraction is shown in the lower panel. The dates in the experiment labels refer to the analyzed data sample \cite{crdb}.}
\label{fig:epm}
\end{figure}

Experimental results show evidence for an excess of the positron fraction with respect to the standard production mechanism (i.e. primary CR interaction with the ISM), in the form of an increase with energy above approximatively 10\,GeV (see Fig.\ref{fig:epm}).
The latest precision measurements of the AMS-02 experiment \cite{Aguilar:2014mma} ascribe the positron fraction excess to a hardening of the positron flux, 
showing a spectral index above 50 GeV compatible with that of primary protons. 
Moreover, as Fermi-LAT recently showed \cite{Abdollahi:2017kyf}, no anisotropy signal has been detected in the inclusive electron spectrum with current sensitivities.

% new reference
% \bibitem{Abdollahi:2017kyf}
% S. Abdollahi et al., Phys. Rev. Lett. 118 (2017)  091103 
% https://doi.org/10.1103/PhysRevLett.118.091103
% Abdollahi:2017kyf

%Besides the interpretation in terms of dark matter (DM) contribution \cite{Giesen:2015ufa}, explanations involving (nearby) astrophysical sources are also viable. 
Understanding the origin of this excess of positrons in the cosmic radiation requires measurements up to the highest possible energies, where both spectral features 
and/or anisotropies might be detected. 
In this context, the multi-TeV, largely unexplored, region is very interesting because of the high potential for studying local sources. Indirect measurements made by imaging atmospheric \v{C}erenkov  telescopes suggest, even though with large uncertainties, an exponential cutoff at about 2\,TeV 
\cite{Aharonian:2009ah,Staszak:2015kza}. 
The analysis of seven years Fermi-LAT data \cite{Abdollahi:2017nat} recently extended the spectral measurements up to 2\,TeV. Fermi-LAT data alone exclude an exponential cutoff below 1.8\,TeV at 95\% C.L., while a combined fit of Fermi-LAT and HESS data 
would lead to a cutoff at energies larger than 2.1\,TeV.
The exploration of the high energy part of the spectrum with high precision direct measurements is then mandatory.

\begin{figure}[t]
\begin{center}
\includegraphics[scale=.65]{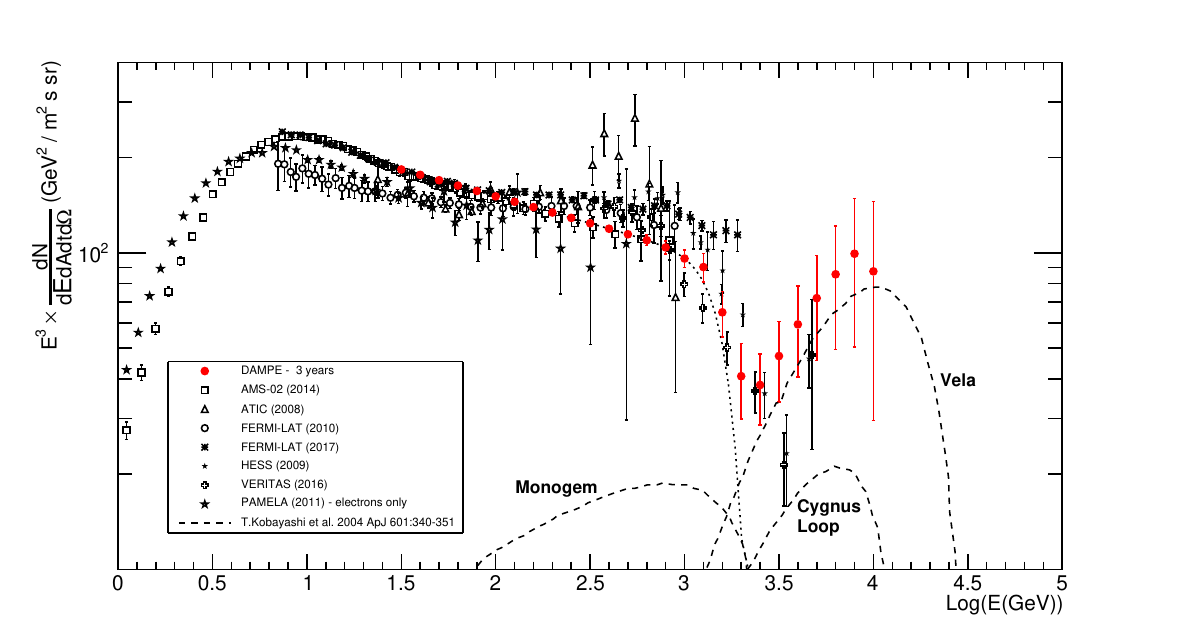} 
\end{center}
\caption{Inclusive all-electron energy spectrum as measured by several experiments. The possible contribution of the DAMPE mission after 3 years of operation
is also shown (see text for details). Plot taken from \cite{DeMitri:2017idm}.}
\label{fig:DAMPE-elec}
\end{figure}

New technologies might extend current MDR values up to few TeV for future missions, while deep homogeneous calorimeters in space, with large geometric factors, will reach even higher energies, but obviously without matter/antimatter separation. 
The recently launched CALET and DAMPE detectors might return interesting results on the high energy all-electron component.

CALET (CALorimetric Electron Telescope) is a space mission led by the Japanese Space Agency (JAXA) with the participation of the Italian Space Agency (ASI) and NASA. The payload was launched on August 19th, 2015 and  installed on the Japanese Experiment Module Exposure Facility (JEM-EF) of the International Space Station (ISS) on August 24th. The mission is foreseen to last for two years, with a possible first extension to 5 years \cite{Marrocchesi:2012nk}. 
The main scientific goal is to search for possible nearby sources of high energy electrons or signatures of DM, by measuring accurately the all electron spectrum from 1 GeV up to several TeV. It will also measure the energy spectra and elemental composition of CR nuclei from H to Fe up to hundreds of TeV (see below). The instrument consists of two layers of segmented plastic scintillators (for particle charge determination), a thin tungsten-scintillating fiber imaging calorimeter providing accurate particle tracking and identification by multiple dE/dx sampling, and a thick PWO crystal calorimeter to measure the energy of CRs with excellent resolution and electron/hadron separation up to the multi-TeV scale. The total thickness is equivalent to 30 radiation lenghts and 1.3 proton interaction lengths with a geometric factor of about 0.1 m$^2$sr. An extensive campaign of beam tests for calibration was carried out at GSI and CERN \cite{Asaoka:2017qfm}.

% new reference
% \bibitem{Asaoka:2017qfm}
% Y. Asaoka et al., Astroparticle Physics 91 (2017) 1-10
% http://doi.org/10.1016/j.astropartphys.2017.03.002
% Asaoka:2017qfm

The DAMPE (DArk Matter Particle Explorer) satellite was launched on December 17th, 2015 and is in smooth data taking since few days after. It was designed in order to properly work for at least three years and, thanks to its large geometric factor (about $0.3\,$m$^2$sr for protons and nuclei and even larger for electrons), it already integrated one of the largest exposures for galactic CR studies in space. The detector, built and operated by a collaboration of Chinese, Italian and Swiss institutions, is made by 12 layers of Si-W tracker followed by a 32\,X$_0$ BGO calorimeter. A plastic scintillator detector on top and a neutron detector on bottom, for ion charge and shower neutron content measurements respectively, actually complete the setup. 
As also resulted from a large set of beam test measurements with a full scale detector prototype at CERN, the BGO calorimeter actually provides an energy resolution for electrons at the level of 5\% at 1\,GeV and better than 2\% above 10\,GeV. The information from the various subdetectors (e.g. ion charge measurement, precision tracking, shower topology) allows an efficient identification of the electron signal over the large (mainly proton induced) background. As a result, the all-electron spectrum will be measured with excellent resolution form few GeV up to few TeV \cite{DeMitri:2017idm}.
%
% new reference
% \bibitem{DeMitri:2017idm}
% I. De Mitri et al.,  EPJ Web of Conferences 136 , 02010 (2017) 
% DOI: 10.1051/epjconf/201713602010
% DeMitri:2017idm
%
%
The DAMPE contribution to the measurement of the all-electron energy spectrum, after 3 years of operation, is shown in Fig.\ref{fig:DAMPE-elec}. The DAMPE spectrum was simulated by assuming a power law with a spectral index as given by AMS-02 \cite{Aguilar:2014fea} data above 30\,GeV, a cutoff at about 1.5\,TeV as suggested by HESS and VERITAS in \cite{Aharonian:2009ah,Staszak:2015kza}, and then a possible contribution of three nearby sources as parametrized in \cite{Kobayashi:2003kp}. In the figure, the result is compared with existing measurements and with the model given in \cite{Kobayashi:2003kp}.

As can be seen, this will allow a direct and precise detection of a possible cutoff at about 1-2\,TeV. Moreover further structures/excesses due to nearby sources will be clearly identified below few TeV, together with possible indirect evidence for a DM-induced excess.

In the case of the HERD mission (see below) a larger acceptance and an even deeper calorimeter would provide a unique tool to investigate all the spectral features also above the TeV region. In particular, the contribution of nearby sources could be clearly identified and studied. In the case of sizeable contribution of nearby sources, a large anisotropy is expected at high energy, which could be easily detected by HERD, giving important clues to the understanding of diffusion processes in the Galaxy.

%Given the large amounts of events, in few years of data taking also the detection of an anisotropy of the hadronic component might be possible,
%mainly at large/medium angular scales, this being the first such observation from space.
%Other important CR studies could be made: as an example the boron-to-carbon ratio would be studied up to about 30\,TeV/n. 
%Finally, the large geometric factor and the excellent tracking capabilities would allow studying anisotropies in the arrival directions.

\begin{figure}[t]
%\sidecaption[b]
\begin{center}
\includegraphics[scale=.6]{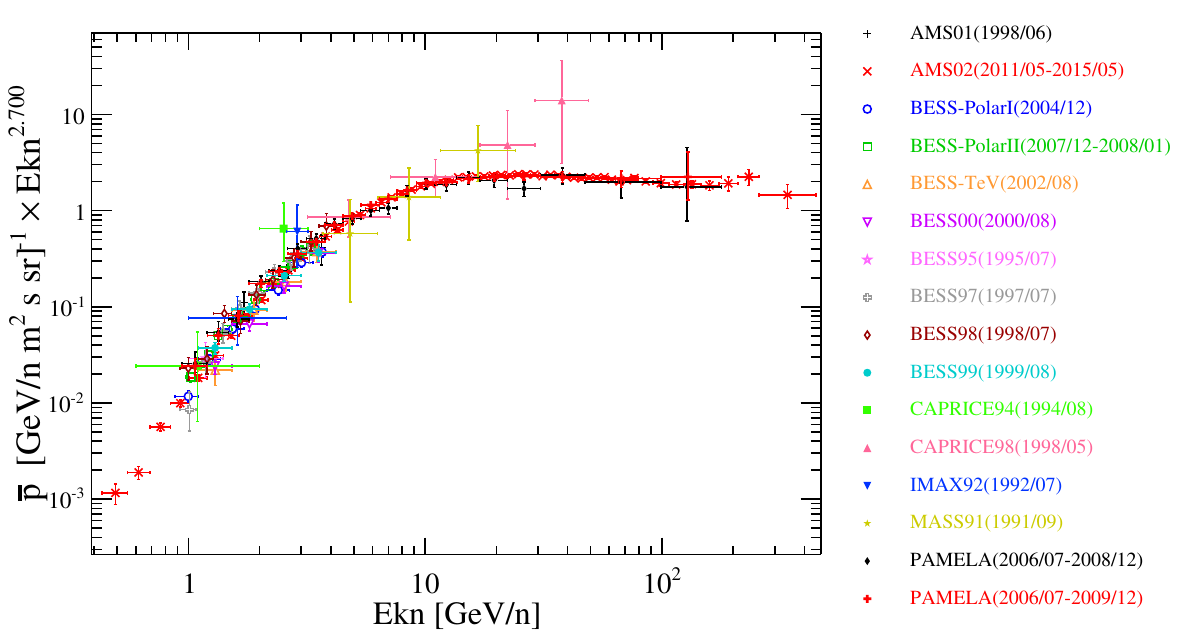} 
\end{center}
\caption{Fluxes of anti-protons as measured by several experiments. 
The dates in the experiment labels refer to the analyzed data sample \cite{crdb}.}
\label{fig:antiprotons}
\end{figure}

The CR anti-proton component can only be identified by using magnetic spectrometers together with sufficient MDR and particle identification capability.
Many balloon and space born experiments contributed to this field with important progress due to the PAMELA \cite{Adriani:2008zq}, AMS-02 \cite{2016PhRvL.117i1103A}, and (at low energies) BESS \cite{Sakai:2015enb} experiments (see Fig. \ref{fig:antiprotons}).
The measurements, currently carried out up to few hundreds GeV, are in fair agreement with secondary production due to primary CR interactions with the interstellar medium. Interestingly enough, the same spectral index as the one for protons is suggested by data, for both positrons and anti-protons (see Sec.\ref{eq:positrons} for a discussion). New important inputs on this topic might be provided by the search for antinuclei (e.g. anti-deuterons) in the CR flux by both current, e.g. AMS-02, and future experiments, such as GAPS \cite{Aramaki:2015laa}.
% new reference
% \bibitem{Sakai:2015enb}
% K. Abe et al. Phys. Rev. Lett. 108, 051102 (2012)
% https://doi.org/10.1103/PhysRevLett.108.051102
%  Sakai:2015enb  caombiato paper

\subsection{Observations of protons and nuclei up to hundreds PeV}
\label{sec:hard}

\subsubsection{From low energies up to 100 TeV: flux hardenings and secondary-to-primary abundances}
\begin{figure}[t]
%\sidecaption[b]
\begin{center}
\includegraphics[scale=0.22]{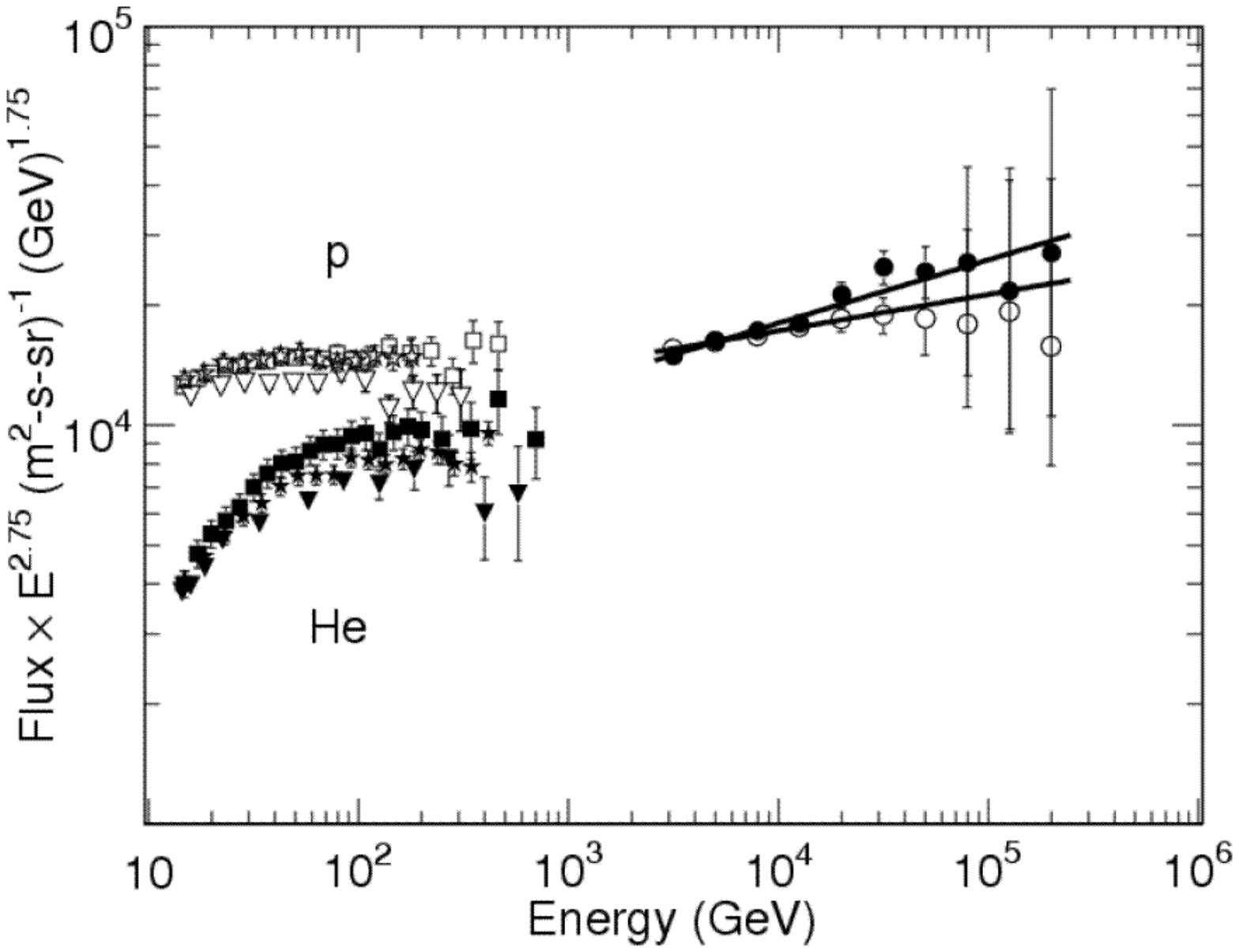}
\includegraphics[scale=0.22]{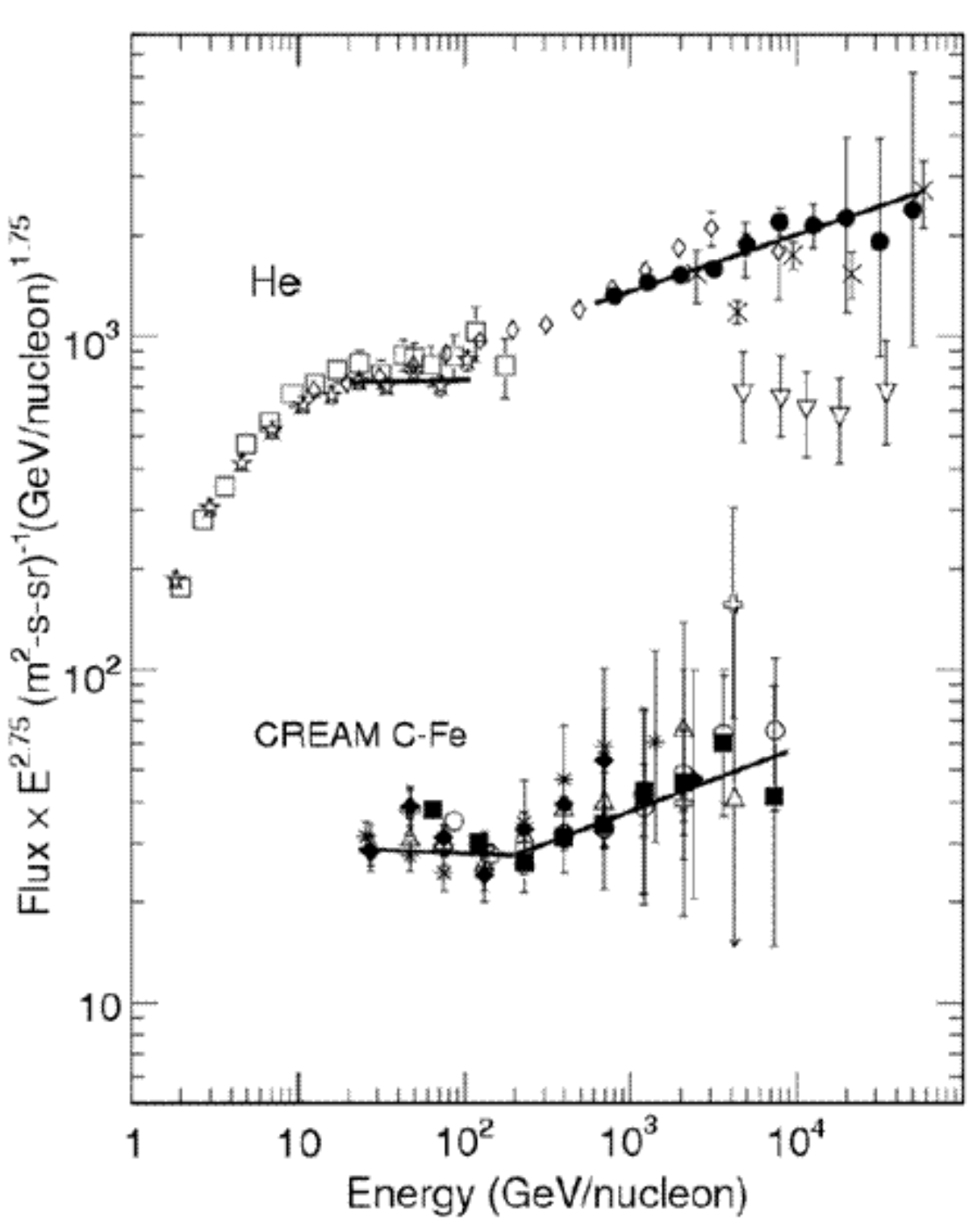}
\end{center}
\caption{Measurements of proton, helium and nuclei fluxes as for year 2010: first evidence for discrepant hardenings by the CREAM experiment (see text). Plots taken from \cite{Ahn:2010gv}.}
\label{fig:discrepant}
\end{figure}

Recent direct measurements of primary protons and nuclei shed new light on acceleration and propagation mechanisms. The paradigm of a unique power law energy spectrum below the knee, down to the region where solar modulation effects become sizeable, might have been invalidated. In 2010 the CREAM (Cosmic Ray Energetics And Mass) experiment showed evidence for a hardening in the spectra of protons and nuclei with different ("discrepant'') spectral index changes. This is summarised in Fig. \ref{fig:discrepant} where CREAM data, also fitted by (broken) power laws, are shown together with other measurements \cite{Ahn:2010gv}. Even with large error bars (mainly at high energy and/or heavy primaries), a change of spectral index is suggested at about 200 GeV/n. Both the energy ranges and the flux uncertainties prevented anyway a clear claim for a break in the proton and helium spectra.

\begin{figure}
%\sidecaption[b]
\begin{center}
\includegraphics[scale=0.3]{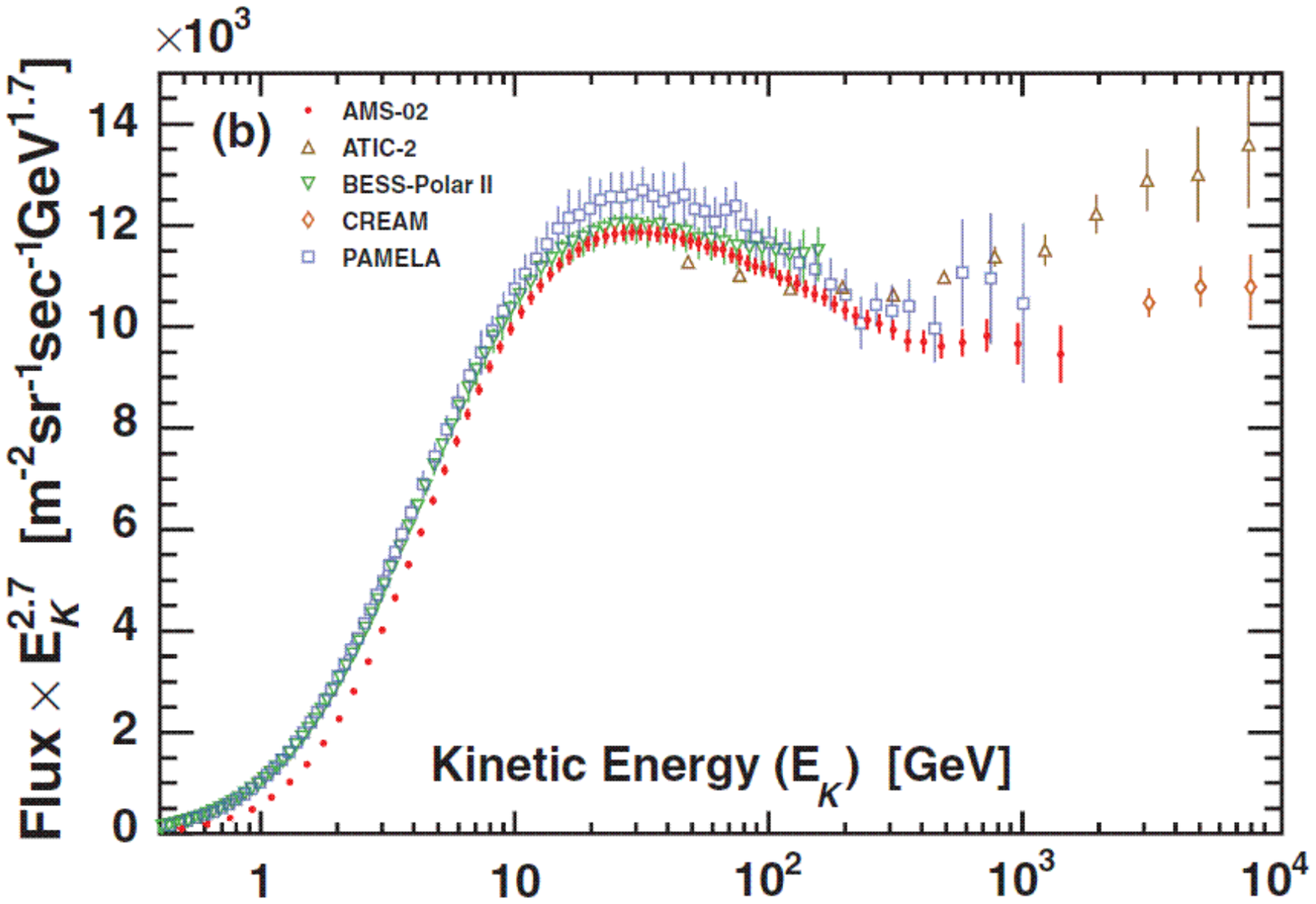} 
\includegraphics[scale=0.3]{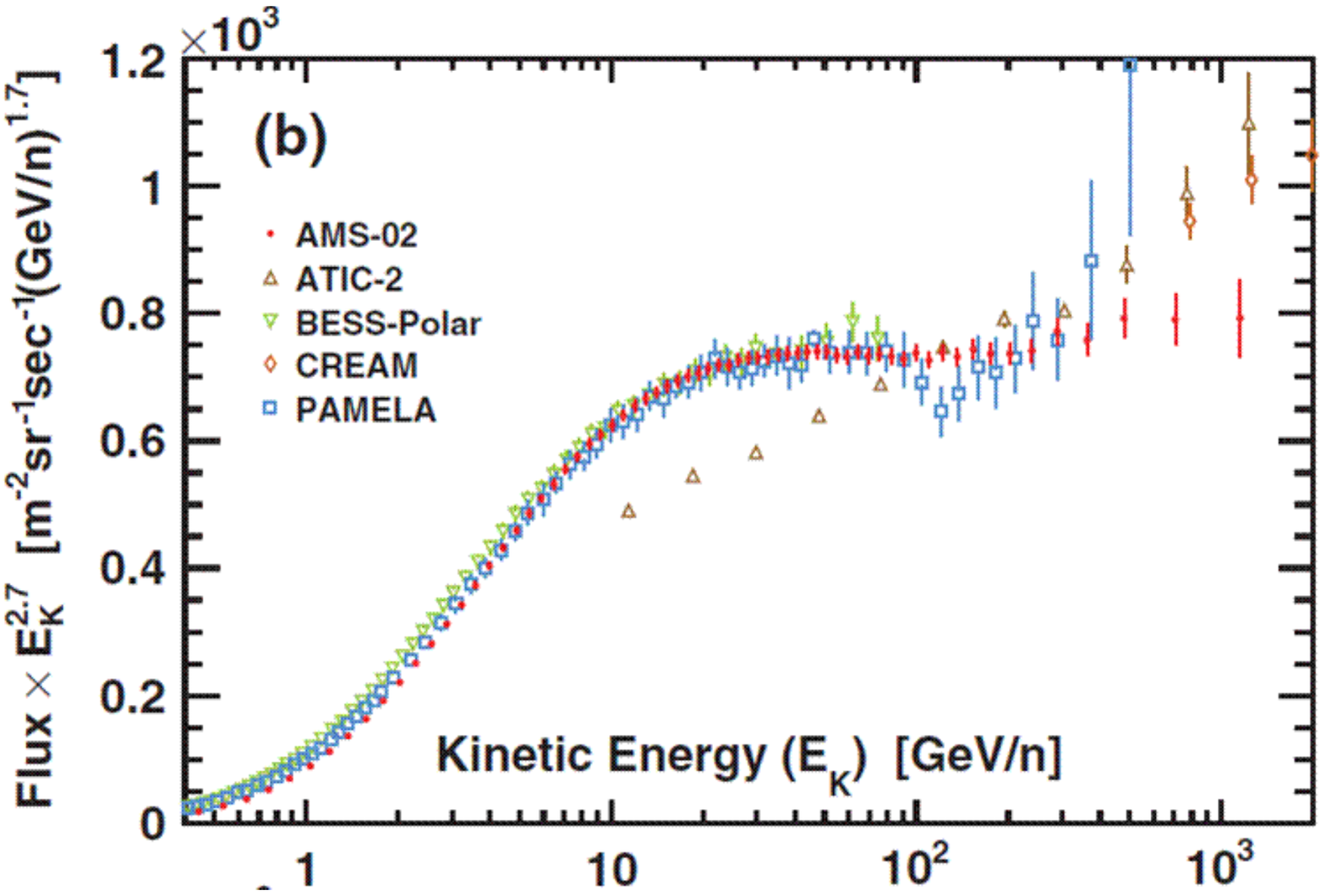}
\end{center}
\caption{Recent measurements on proton and helium fluxes (upper and lower panel respectively). Plots taken from \cite{Aguilar:2015ooa,Aguilar:2015ctt}.}
\label{fig:p-and-he}
\end{figure}

This became possible with the analysis of PAMELA results \cite{Adriani:2011cu}, later confirmed by AMS-02 \cite{Aguilar:2015ooa,Aguilar:2015ctt}.  As can be seen in Fig. \ref{fig:p-and-he} a clear change of spectral index is shown by data, even though different experiments return slightly different slopes at energies above the breaks. 

Recent results of the analysis of CREAM-III \cite{Yoon:2017qjx} and NUCLEON \cite{Atkin:2017cwa} data did confirm the scenario up to about 100\,TeV but with large uncertainties. More data are then needed at high energy in order to measure, with a single experiment, both the region across the breaks and the high energy one, with sufficiently small uncertainties.
Current missions like CALET and DAMPE have the size and the needed resolution in order to check the break region and uniquely determine the spectral behaviour up to more than 100 TeV.

% new reference 
% \bibitem{Yoon:2017qjx}
% Y.S. Yoon et al., The Astrophysical Journal, 839 (2017) 5
% http://stacks.iop.org/0004-637X/839/i=1/a=5
% Yoon:2017qjx

% new reference
% \bibitem{Atkin:2017cwa}
% E. Atkin et al., Astroparticle Physics 90 (2017) 69
% http://dx.doi.org/10.1016/j.astropartphys.2017.02.006
% Atkin:2017cwa

Even though primarily optimized for the study of electrons and gamma rays (see above), the DAMPE detector provides good tracking and calorimetric performances also in the case of protons and nuclei, together with the possibility of ion identification through charge measurements in the top scintillator layer, in the tracker, and in the calorimeter itself. This allows precise measurements of proton and nuclei energy spectra from tens of GeV up to about 100\,TeV, the high energy limit being essentially determined by the overall geometric factor and the calorimeter's dynamic range.

In particular, the energy region above about 50 GeV will be explored with higher precision compared to previous experiments \cite{Seo:2012pw}. Spectral indexes for individual species could then be well measured and evidence for the observed hardenings could be checked and better quantified. This would be very important for a comparison with state-of-the-art models of galactic CR acceleration/propagation mechanisms, and to assess the contribution of nearby sources. Moreover measurements of important quantities like the boron-to-carbon ratio will be improved and extended to higher energies. Similar contributions are expected from the CALET mission, even if the smaller geometric factor (by about a factor three) would result in larger uncertainties.
Further extensions in energy, towards the all-particle knee, are expected for the ISS-CREAM and HERD projects (see below).

\begin{figure}[t]
\begin{center}
\includegraphics[scale=.6]{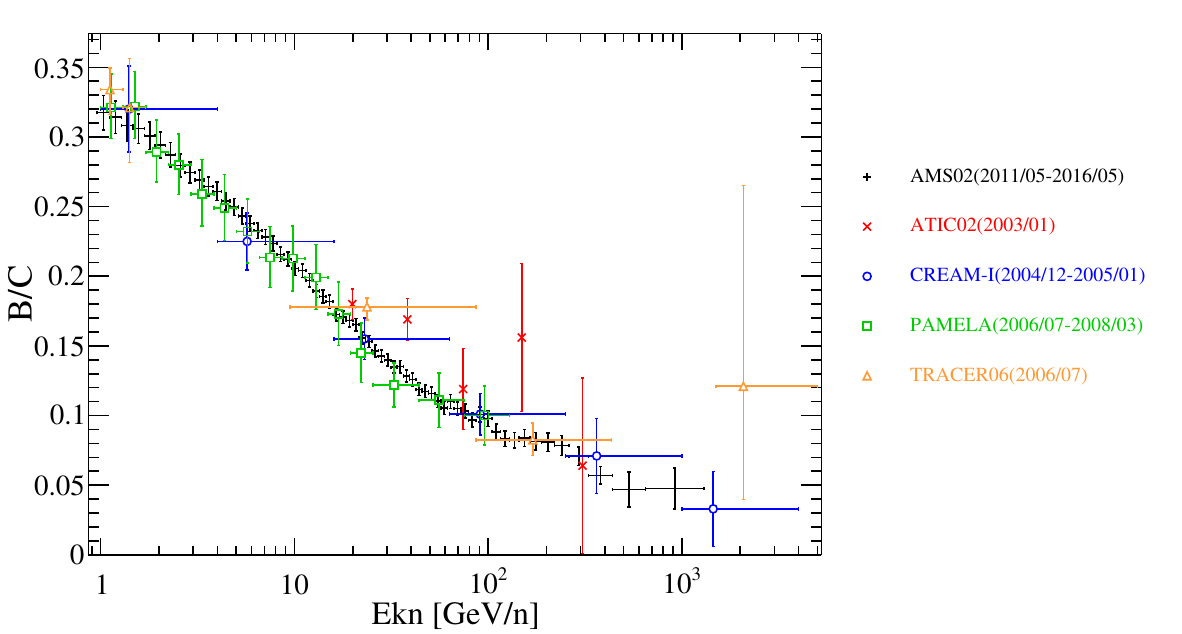} 
\end{center}
\caption{Measurements B/C ratio made by several experiments.
The dates in the experiment labels refer to the analyzed data sample \cite{crdb}.}
\label{fig:BoverC}
\end{figure}

In the low energy range, important information can be provided by the study of the production rate of secondary CRs.
Recent results from PAMELA and BESS-Polar \cite{Picot:2015cda} on the isotopic abundance ratios $^2$H/$^1$H and $^3$He/$^4$He
in the range 0.1-2 GeV/n provide essential information to better understand the history of cosmic-ray propagation in the Galaxy. 
On the other hand, as discussed in Sec.\ref{sec:transport}, elemental secondary-to-primary ratios 
(such as Boron/Carbon or subFe/Fe) can be employed, to infer information on the nature and size of the cosmic-ray confinement region and on the propagation properties of CRs in the Galaxy.
Current measurements of the B/C ratio are shown in Fig. \ref{fig:BoverC}. While measurements performed by balloon-born experiments suffer
from small statistics and large systematic errors (due to short exposure times and to the effects of the residual overburden atmosphere, respectively),
data from space spectrometers like PAMELA and AMS-02, can provide an accurate spectral measurement up to about 1\,TeV/n \cite{Adriani:2014xoa,Oliva:2015cda}.
Also in this case, from the experimental point of view, the challenge is then to extend the energy range to the multi TeV/n region by using 
large geometric factor instruments in space.

\subsubsection{Approaching the knee(s) with direct measurements}

Since the first experimental evidence in 1958 \cite{Kulikov:1958aa}, the energy region around the knee in the all-particle cosmic ray spectrum, at about 3$\,$PeV, has been investigated by many experiments with different approaches \cite{Bluemer:2009zf}.

Several theoretical explanations have been proposed exploiting different hypotheses on source properties/populations, acceleration/propagation mechanisms and particle physics issues at high energies \cite{Hoerandel:2004gv}. After the first results at the Large Hadron Collider (LHC), the ``particle physics'' origin of the knee seems to be disfavoured, confirming that it is a genuine property of the CR spectrum itself \cite{d'Enterria:2011kw}. It remains still unsolved whether the (dominant) origin of the knee is due to the reaching of the maximum energy achievable at the sources or to diffusion processes in the Galaxy. In both cases a rigidity dependent sequence of knees in individual elemental spectra is the most likely scenario \cite{Hoerandel:2004gv}. 

For the analysis of the CR flux, direct measurements carried on space or stratospheric balloons actually provide the best performance in terms of both energy resolution and charge identification. However, due to their limited acceptance and the steeply falling fluxes, they could hardly reach, up until now, energies of hundreds of TeV and then did not yet provide clear information on the steepening of the spectrum of various elements nor on the knee of each species or of the all-particle spectrum itself \cite{Seo:2012pw,Boezio:2012rr}.

As shown in the previous section, current data suggest a hardening of the spectra above about 0.2\,TeV/nucleon and spectral indexes $\gamma$ (above that energy) of about -2.6 for all considered elements but for protons, that show a softer spectrum with $\gamma \simeq -2.7$ \cite{Ahn:2010gv,Yoon:2011aa}. Moreover the chemical composition is shown to evolve towards heavier nuclei, with helium becoming more abundant than hydrogen at energies of about 10-20$\,$TeV \cite{Seo:2012pw}.
It is then mandatory to explore the sub PeV region with high precision direct measurements in order to study the energy spectra of each nuclear species, to measure the various spectral indexes, to detect any possible hardening and to establish mass composition below the knee of the all-particle spectrum. The measurement of the spectrum of individual elements and an understanding of the nature of the knee in the all-particle spectrum would represent a result of unprecedented importance in CR physics, and would provide a crucial insight into unveiling the transition between galactic and extragalactic CRs.

The ISS-CREAM detector has been built by transforming the CREAM payload, flown in several flights over Antarctica, for accommodation on the ISS for a 3 year mission \cite{Seo:2015aa}. The exposure will be increased by about one order of magnitude allowing to extend the measurements on nuclei up to hundreds of TeV. The detector includes: four layers of Si pixels to measure the particle charge; a carbon target to induce the inelastic interaction of the incoming nuclei; a sampling calorimeter made of 20 layers of alternating tungsten plates and scintillating fibres, providing energy measurement, particle tracking and trigger; top and bottom plastic scintillator counters and a boronated scintillator detector for e/p separation. The Si charge detector and the calorimeter were already used in CREAM, while the last two detectors have been newly developed for the space mission, in order to add sensitivity also to CR electrons.

The HERD (High Energy Radiation Detector) experiment \cite{Xu:2016ibc} is being proposed by an international collaboration as a space mission for the study of the high energy cosmic radiation with a detector characterized by an unprecedented geometric factor, to be installed onboard the CSS (Chinese Space Station) around 2023. Current detector design includes a cubic calorimeter (about 55X$_0$ and 3$\Lambda$ in depth) made by (3cm$\times$3cm$\times$3cm ) LYSO crystals, readout individually, 
a high precision Si-W tracker covering 5 out of the 6 calorimeter faces and an array of plastic scintillator for the charge measurements. 
This innovative setup allows a jump in the geometric factor, with respect to previous experiments, of more than one order of magnitude for an estimated value of the geometric factor $\sim 3\,{ m^2\,sr}$. Together with the unprecedented depth of the calorimeter and the high resolution tracker, this will allow the extension of high precision measurements on proton and nuclei spectra up to PeV energies. Moreover a clear identification of each nuclear species will be possible through the charge measurements made by the plastic scintillators, the Si-W tracker and by the calorimeter itself. 
Energy resolution for the electromagnetic and hadronic showers will be at the 1\% and 30\% level, respectively.

% \bibitem{Xu:2016ibc}
% M. Xu et al., Nuclear and Particle Physics Proceedings, vol. 279-281 , (2016) 161-165
% https://doi.org/10.1016/j.nuclphysbps.2016.10.023
% Xu:2016ibc

\begin{figure}[t]
%\sidecaption[b]    
\begin{center}
\includegraphics[scale=0.65]{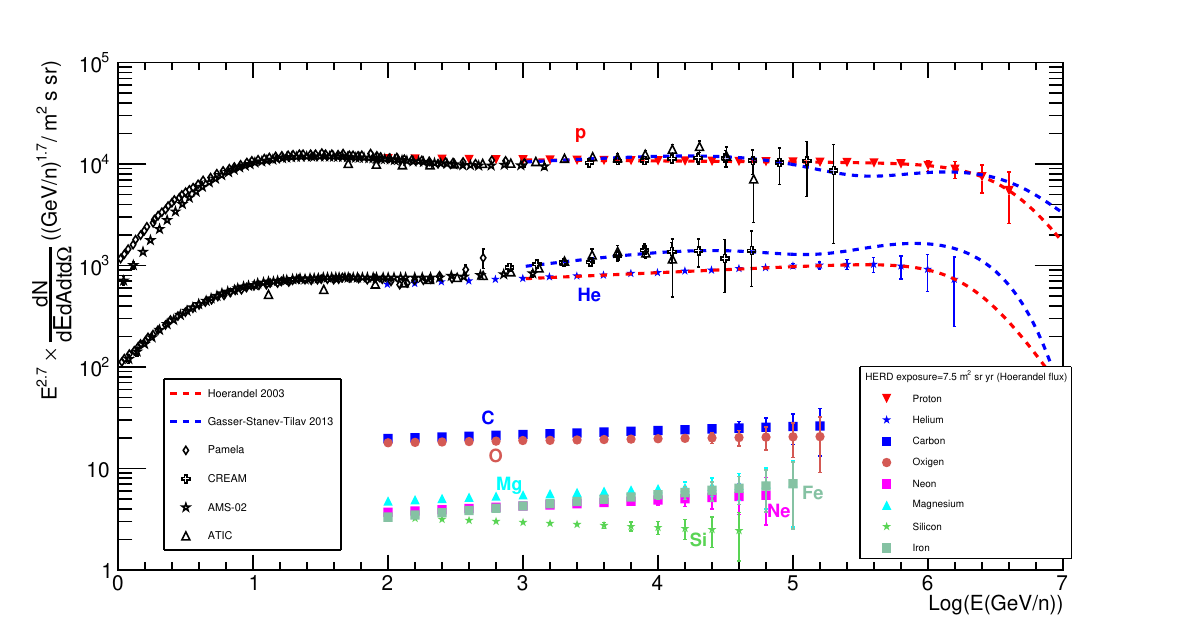} 
\end{center}
\caption{Contributions of the HERD mission after 3 years of operation: individual proton and nuclei energy spectra. (see text for details).}
\label{fig:HERD-pro-nuc}
\end{figure}

Simulated HERD results for the measurement of protons and nuclei, after 3 years of operation, obtained by assuming flux parametrisation as given in \cite{Hoerandel:2002yg}, are shown in Fig. \ref{fig:HERD-pro-nuc}. HERD points (in color) are compared with measurements from PAMELA, CREAM, AMS-02 and ATIC experiments (shown in black) and theoretical models \cite{Hoerandel:2002yg,Gaisser:2013bla}.

As can be seen, features like single element spectral indexes and spectral hardenings/steepenings could be carefully studied from hundreds GeV/n up to hundreds TeV/n. In particular the proton and helium component could be measured up to PeV energies, thus providing a test of the origin of the knee in the spectrum of the light CR component, possibly associated with the maximum energy reached by CR at the source (see next section). 
%This result would be very important for understanding galactic CR acceleration/propagation processes.

\subsubsection{Ground-based CR observations up to hundreds PeV}

Indirect measurements fully explore the energy region above 0.5-1\,PeV (or even few TeV, if located at high altitude) through the detection of Extensive Air Showers (EAS) in the atmosphere. Data show a general agreement, within the systematic uncertainties, on the all-particle spectrum, also suggesting evidence of a second knee at about $100\,$PeV \cite{Apel:2011mi,Budnev:2013noa,Prosin:2014dxa,Aartsen:2013wda}. However systematic uncertainties related to the experimental procedure itself and intrinsic in the assumptions adopted for the hadronic interaction models do not allow an easy and straightforward estimate of the mass composition nor of the single species (or mass group) energy spectra \cite{Bluemer:2009zf,Kampert:2012mx}.

\begin{figure}[t]
%\sidecaption[b]
\begin{center}
\includegraphics[scale=.4]{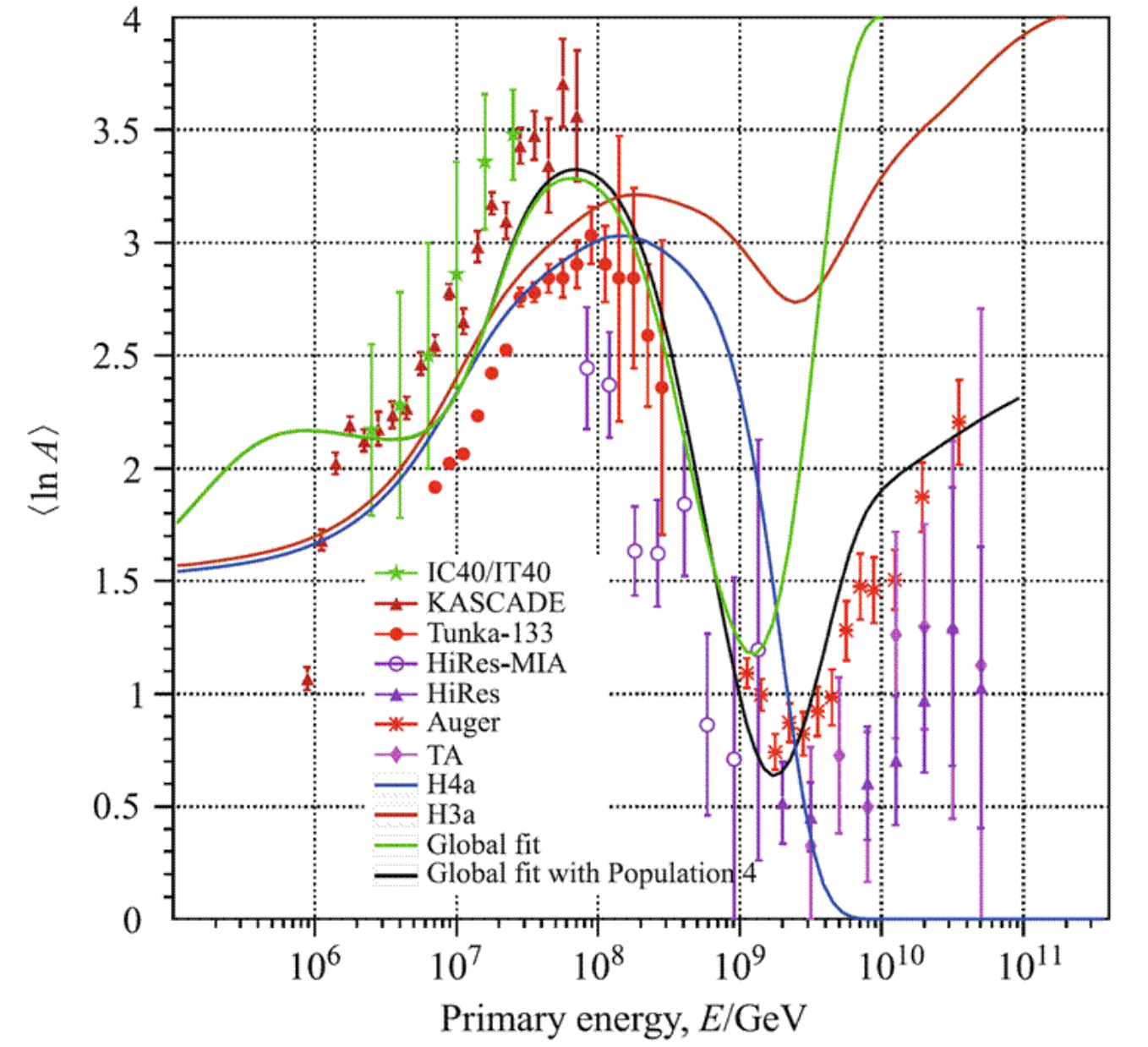} 
\end{center}
\caption{ Average value of the estimated CR logarithmic mass, $ <lnA> $, from several ground-based experiments. Superimposed are the lines corresponding to mixed composition resulting from muti-population models as given in \cite{Gaisser:2013bla}. Plot taken from \cite{Gaisser:2013bla}.}
\label{fig:logA}
\end{figure}

One or more EAS observables (e.g. the lateral distribution of particles at the ground, the longitudinal development in the atmosphere, the muon content of the shower, etc.) are measured in order to  estimate, by adopting a given assumption on the primary interaction and the shower development in the atmosphere, the CR composition. Results are often given in terms of the energy dependence of the mean logarithmic mass, defined as $ \langle ln A \rangle = \sum_{i} \eta_i ln A_i$, where $\eta_i$ is the fraction of nuclei of mass $A_i$ in the CR beam. A compilation of $ \langle ln A \rangle$ measurements can be found in \cite{Bluemer:2009zf} and \cite{Kampert:2012mx}, with a comprehensive discussion on the results and their uncertainties. As can be seen in Fig. \ref{fig:logA}, data show large uncertainties, mainly coming from the systematics associated to the adopted interaction models. Moreover a somewhat different trend with energy might be identified by dividing experiments in two large classes: the ones measuring charged particles at the ground and those detecting the \v{C}erenkov or fluorescence light emissions in the atmosphere \cite{Hoerandel:2002yg}. Even with these uncertainties, data collected across the knee region show an evolution towards heavier mass groups as expected from acceleration/propagation models. A tendency towards lighter elements is then observed starting at energies compatible with the position of the second knee, while a new trend towards the medium mass group can be recognised above the ankle region (see next section).

The energy spectra of individual elements (or mass groups) are even more difficult to be measured. 
%and large uncertainties are introduced by EAS fluctuations, the adopted hadronic interaction models, the data unfolding procedure, etc.
%In this case the results do not agree each other as far as the position of the {\it knees} for the individual species are concerned.
%
Results from the KASCADE experiment, even with sizeable systematic uncertainties on the individual fluxes mainly coming from the dependence on the hadronic interaction model, imply an average composition at the knee that is dominated by light elements, and the knee itself is interpreted as the steepening of the p and He spectra \cite{Apel:2008cd}. The KASCADE-Grande experiment returned results consistent, at higher energies, with this scenario \cite{Apel:2013dga,Apel:2014aa}, and ascribed the second knee to the steepening of the heavy component \cite{Apel:2011mi}.

Several different experimental results suggested a somewhat heavier composition at energies around the knee. For instance a hybrid measurement was carried out by the EAS/TOP and MACRO experiments (by detecting, in coincidence, EAS \v{C}erenkov  light at 2000 m a.s.l. and underground muons below about 3000 m of water equivalent depth, respectively). The result implied a decreasing proton contribution to the primary flux at energies well below the observed knee in 
the primary spectrum \cite{Aglietta:2004ws}.

The same indication was previously obtained through the analysis of the underground muon component alone in the MACRO experiment \cite{Ambrosio:1996eu}. In addition, the results of the Tibet AS$\gamma$ experiment, located at 4300 m a.s.l, do favour a heavier composition because the proton component is no longer dominant at the knee \cite{Amenomori:2005nx}. In particular, the fraction of the light component (i.e. protons and helium nuclei) is shown to be of 50\% at about 500\,TeV and decreasing with energy \cite{Amenomori:2005nx}, this also being consistent with later measurements of the upgraded Tibet array showing a steepening of the proton spectrum above few hundred TeV \cite{Amenomori:2011zza}.

This is also in agreement with results from the CASA-MIA experiment, showing a decreasing proton content at about 600\,TeV \cite{Glasmacher:1999xn} and with a series of measurements on Mount Chacaltaya (about 5200\,m a.s.l.) giving a steady increase of the average mass number of primary CRs with energy above $10^{14.5}\,$eV \cite{Ogio:2004sc,Tokuno:2008zzb}.

Furthermore, indications for a substantial fraction of nuclei heavier than helium at 1$\,$PeV have also been obtained in measurements with delayed hadrons \cite{Freudenreich:1990uy}. Finally, the compilation of measurements of the energy spectrum of the so-called CNO group (i.e. Carbon-Nitrogen-Oxygen) show a knee at energies not larger than about 7\,PeV (see for instance \cite{Bluemer:2009zf}). In a scenario with a rigidity dependent knee position, this is not consistent with a position of the proton knee at about 3\,PeV.

\begin{figure}[t]
%\sidecaption[b]
\includegraphics[scale=0.6]{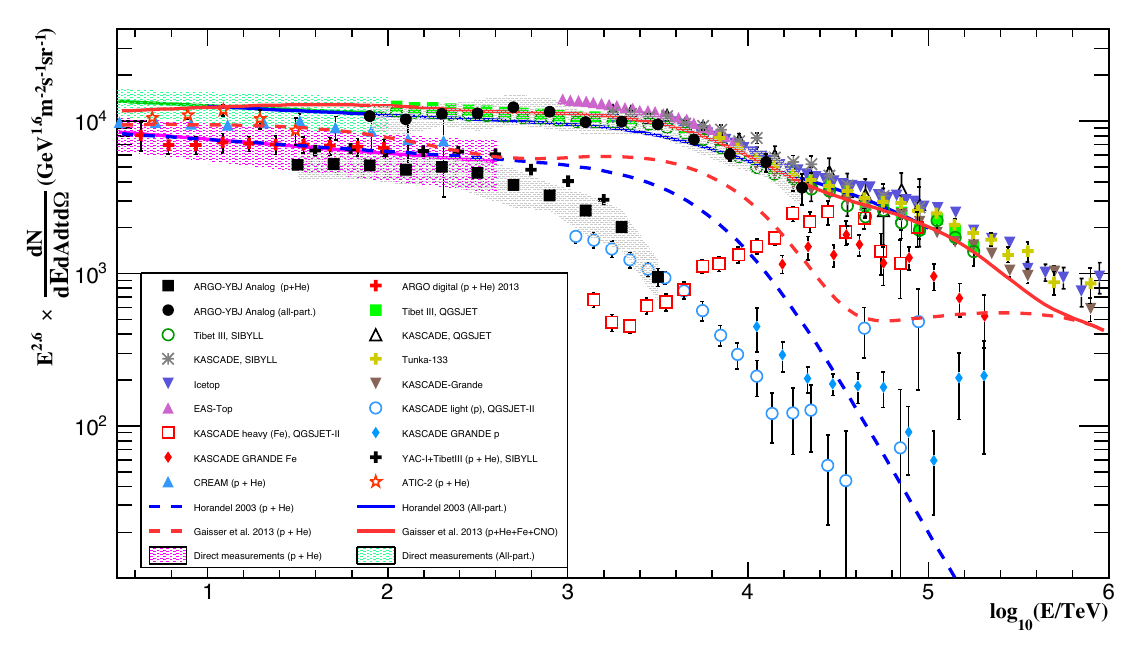} 
\caption{Indirect measurements of the all particle CR energy spectrum below $10^{18}\,$eV.  Also shown are the combination of high energy direct measurements, and the energy spectrum for the light (i.e. proton + helium) component. Plot taken from \cite{DeMitri:2015cbd}.}
\label{fig:allp-light}
\end{figure}

Similar conclusions have been reached by the combined analysis of data coming from the ARGO-YBJ experiment and a wide field of view \v{C}erenkov  telescope (a prototype of the future LHAASO experiment \cite{Bartoli:2015vca}): the measured energy spectrum of the proton + helium component shows a break at $(700 \pm 230 \pm 70 ) \,$TeV \cite{Bartoli:2015vca}. Preliminary results from two independent analyses of the ARGO-YBJ data alone do confirm this picture, 
within the quoted uncertainties \cite{DeMitri:2015cbd,Mari:2015jda}.

An overall picture of indirect measurements of the all particle spectrum, below $10^{18}\,$eV, is shown in Fig. \ref{fig:allp-light}. In the same plot, the measurements of the so called light-component (i.e. proton+helium) is also given, showing a clear bending at energies below the knee of the all-particle spectrum. For comparison, the combination of indirect measurements is shown at lower energies, while the results for the light and heavy component as identified by KASCADE and KASCADE-Grande are shown for higher energies (\cite{Apel:2008cd,Apel:2013dga} and references therein).
Besides a knee-like behaviour in the heavy elements at about 10$^{17}$\, eV (consistent with the second knee), KASCADE-Grande data also suggest an ankle-like structure in the light elements at the same energies \cite{Apel:2013dga}.
The uncertainties on both energy spectra and mass composition are reduced at higher energies, due to the possibility to detect fluorescence light emission along the whole shower development in the atmosphere (see next session).

\begin{figure}[t]
\begin{center}
\includegraphics[scale=.5]{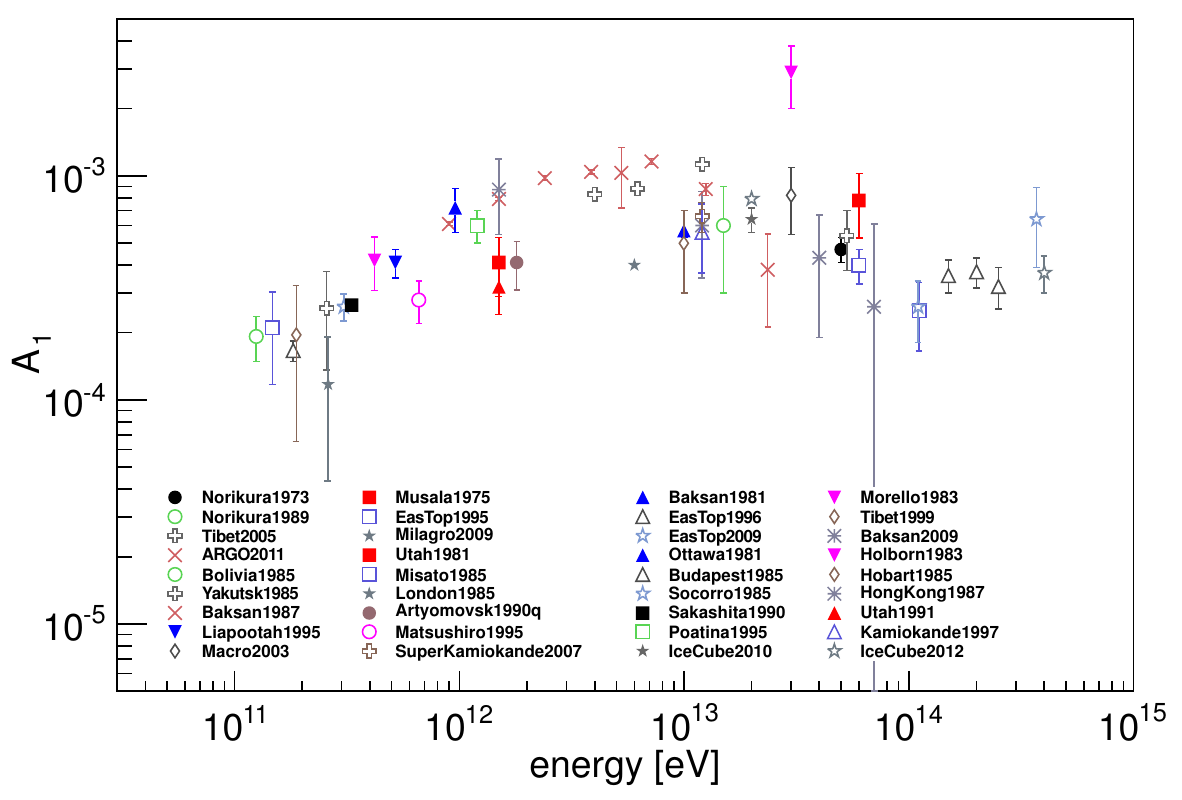} 
\includegraphics[scale=.5]{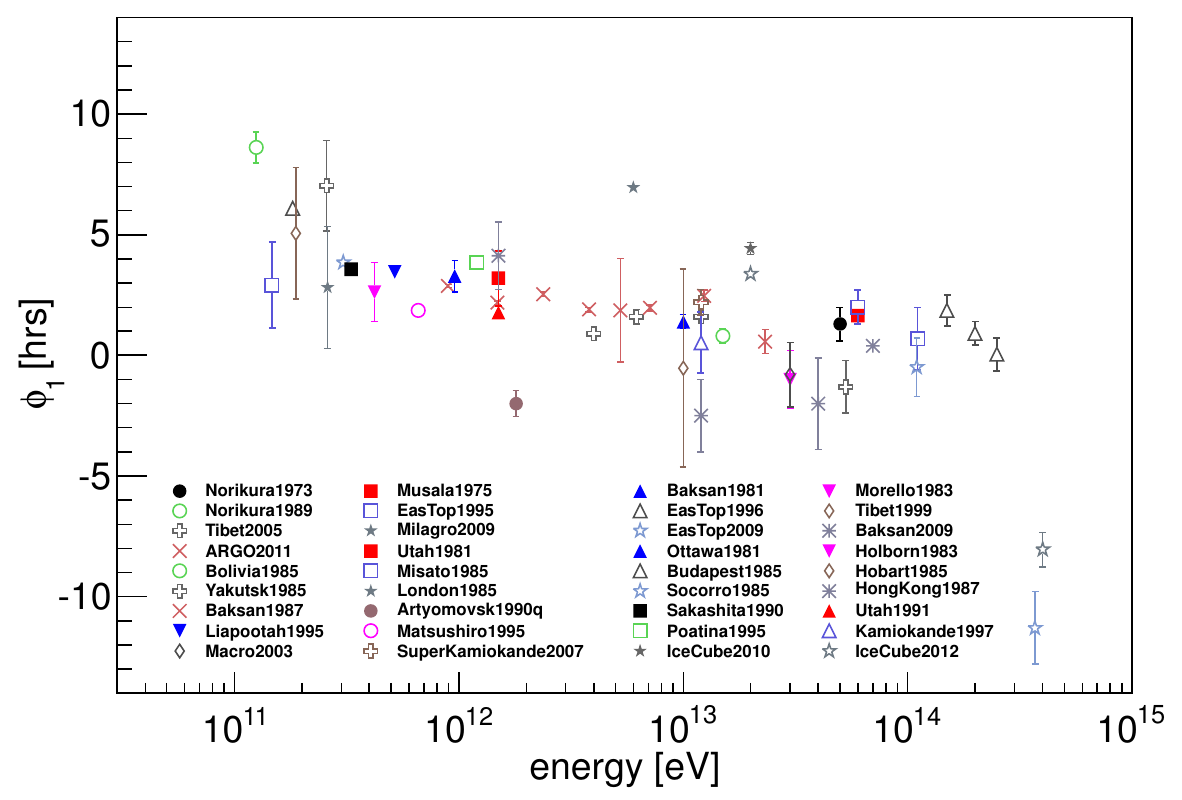} 
\end{center}
\caption{Amplitude and phase (upper and lower panel respectively) of the sidereal CR flux 
daily variation (first harmonic) as measured by several (under)ground experiments. 
Plot taken from \cite{DiSciascio:2014jwa}.}
\label{fig:LSA}
\end{figure}

\subsubsection{Flux anisotropies}
A complementary approch to the study of CR sources and propagation, with respect to the analysis of energy spectra and composition, is provided by the measurement of
anisotropy signals. This also possibly leads to some information on the galactic magnetic field, which is mainly responsible for the highly isotropic CR flux.
Even though the first evidences for anisotropies (resulting from the CR intensity variations with sidereal time) dates back to Hess and Steinmaurer in 1932
\cite{Wollan:1939wol}, in recent years the huge event statistics collected by several experiments with good pointing accuracy allowed a detailed analysis of two dimensional arrival direction distribution maps (right ascension and declination) and their evolution with time.
As a consequence, anisotropy signals at the level of $10^{-4}-10^{-3}$ were found at different angular scales in both hemispheres 
(see for instance \cite{DiSciascio:2014jwa} and refs. therein).

% new reference
% \bibitem{Wollan:1939wol}
% E.O. Wollan:1939wol, Rev. Mod. Phys. 11, 160 (1939)
% https://doi.org/10.1103/RevModPhys.11.160
% Wollan:1939wol:1939wol

% new reference
% \bibitem{DiSciascio:2014jwa}
% G. Di Sciascio and R.Iuppa, in "Homage to the discovery cosmic rays", chapter 9, pag.221-257, Nova Science Publishers, Inc., New York 2013
% arXiv:1407.2144 (2013)
% DiSciascio:2014jwa

A so-called Large Scale Anisotropy (LSA) has been measured by several experiments 
(e.g. Tibet-AS$\gamma$ \cite{Amenomori:2012uda}, Milagro \cite{Abdo:2008aw}, ARGO-YBJ \cite{Bartoli:2015ysa}, IceCube \cite{Aartsen:2016ivj}) 
showing an approximate dipole-like feature with an excess region between $40^{\circ}-90^{\circ}$ in right ascension (around the heliospheric tail) 
and a deficit between $150^{\circ}-240^{\circ}$ (in the direction of the galactic north pole), referred to as {\it tail-in} and {\it loss cone} regions respectively.

These observations are likely to reflect the combination of several effects, namely the relative motion of the solar system with respect to the frame in which CRs are isotropic (Compton-Getting effect \cite{Compton:1935wde}), the orientation of the local magnetic field \cite{2015PhRvL.114b1101M} and the overall gradient in the CR local density (see for instance \cite{2006APh....25..183E,2006AdSpR..37.1909P,Blasi:2011fm}).

As can be seen in Fig.\ref{fig:LSA}, the amplitude of the observed signal is of the order of $10^{-4}-10^{-3}$ with a wide maximum in the multi-TeV region 
and a stable phase.  An increase in the amplitude and a dramatic change of phase (pointing to the opposite direction) are then suggested by data 
above 100\,TeV up to 5\,PeV (see \cite{DiSciascio:2014jwa} and ref. therein).

% new reference 
% \bibitem{Amenomori:2012uda}
% M. Amenomori et al., Astroparticle Physics 36 (2012) 237–241
% http://doi.org/10.1016/j.astropartphys.2012.06.005
% Amenomori:2012uda

% new reference 
% \bibitem{Abdo:2008aw}
% A.A. Abdo et al., The Astrophysical Journal, 698 (2009) 2121–2130 
% doi:10.1088/0004-637X/698/2/2121
% Abdo:2008aw

% new reference 
% \bibitem{Bartoli:2015ysa}
% B. Bartoli et al., The Astrophysical Journal, 809 (2015) 90
% doi:10.1088/0004-637X/809/1/90
% Bartoli:2015ysa

% new reference 
% \bibitem{Aartsen:2016ivj}
% M.G. Aartsen et al., The Astrophysical Journal 826 (2016) 220
% http://stacks.iop.org/0004-637X/826/i=2/a=220
% Aartsen:2016ivj

% new reference
% \bibitem{Compton:1935wde}
% A.H. Compton and I.A. Getting I.A., Phys. Rev., 47, 817 (1935)
% doi:10.1103/PhysRev.47.817, 1935
% Compton:1935wde

Recently an additional anisotropy signal has been found in the few TeV energy region, with excesses at angular scales of about 10$^{\circ}$ (the so-called Medium/Small Scale Anisotropy , MSA ) by Tibet-AS$\gamma$ \cite{Amenomori:2006bx}, Milagro \cite{Abdo:2008kr}, ARGO-YBJ \cite{ARGO-YBJ:2013gya}, IceCube \cite{Aartsen:2016ivj} and HAWC \cite{Abeysekara:2014sna}. Such signals have a quite large statistical significance (up to 15 standard deviations) and a nice matching between observations from both hemispheres. Moreover, there are hints for a harder energy spectrum in the excess regions with respect to the isotropic CR background. For both LSA and MSA signals, most observations suggest time stability over several years time scales, which would exclude a correlation with the solar activity.

% new reference 
% \bibitem{Amenomori:2006bx}
% M. Amenomori et al., Science  314 (2006) 439
% DOI: 10.1126/science.1131702
% Amenomori:2006bx

% new reference 
% \bibitem{ARGO-YBJ:2013gya}
% B. Bartoli et al., PHYSICAL REVIEW D 88, 082001 (2013)
% DOI: 10.1103/PhysRevD.88.082001
% ARGO-YBJ:2013gya

% new reference
% \bibitem{Abdo:2008kr}
% A. A. Abdo et al., Phys. Rev. Lett. 101 (2008) 221101
% https://doi.org/10.1103/PhysRevLett.101.221101
% Abdo:2008kr

% new reference
% \bibitem{Abeysekara:2014sna}
% A. U. Abeysekara et al, The Astrophysical Journal 796 (2014) 108
% doi: 10.1088/0004-637X/796/2/108
% Abeysekara:2014sna

The discovery of anisotropy on small angular scales was rather surprising in that the basic expectation of the theory of CR diffusive transport is that only a dipole anisotropy should be expected. On the other hand, it has been noted by several authors that small scale anisotropies may develop because of the local configuration of the magnetic field, within, say, a few pc from Earth. For instance, in Ref. \cite{2012PhRvL.109g1101G} the author describes the propagation of CRs arriving in the neighborhood of the solar system in several realizations of the local magnetic field and small scale anisotropies are in fact found, mainly as a result of the fact that fluctuations in the deflections are not averaged to zero. In other words, the transport in not fully in the diffusive limit. An elegant derivation of the same result was found by \cite{2014PhRvL.112b1101A}, in which the author shows that these small-scale fluctuations naturally arise as a consequence of the LiouvilleÕs theorem.

A better knowledge of CR physics up to the ankle will need measurements of energy spectra and anisotropy maps of individual species (or at least mass groups) with better resolution and larger statistics. This will also depend on future experiments trying to use new observables (e.g. radio emission \cite{Aab:2016eeq}) and/or to combine several techniques to be used at the same time (e.g. the LHAASO project \cite{Vernetto:2016gro}).

\subsection{Observations of ultra-high energy cosmic rays}
\label{sec:UHEdet}

Above 10$^{18}$ eV (UHE), the two largest and most precise detectors to date are the Pierre Auger Observatory in Argentina (Mendoza) and the Telescope Array in the USA (Utah). Both detectors exploit the hybrid concept, combining an array of surface detectors to sample extensive air showers when they reach the ground and telescopes, overlooking the surface array, to collect the fluorescence light of the excited atmospheric nitrogen. The advent of the hybrid approach has been a major breakthrough in the detection of UHECRs since the method allows to have the same energy scale in the surface detectors and the fluorescence telescopes. In fact the absence of an energy scale common to both detection methods had led to the puzzle about the existence of the flux suppression around $5\times 10^{19}$ eV, which was observed by HiRes~\cite{Abbasi:2007sv}  but not present in AGASA data~\cite{Takeda:2002at}, whose energy calibration was based on Monte Carlo simulations. The first hybrid measurements were done in HiRes/MIA~\cite{AbuZayyad:1999xa} with a detector array of limited size; the Auger project, for the first time, adopted the hybrid approach~\cite{Abreu:2010aa} as the basis of the detector design to definitely attack the suppression puzzle.

\begin{table}
\label{table:detfeatures}
%\begin{tabular}{lrrr}
%\begin{tabular}{lccc}
\centering
\begin{tabular}{|c|cc|c|c|}
\hline
& & & Auger & TA \\ [0.5ex]
\hline \hline
   & \multicolumn{2}{c}{Average latitude} \vline & $35.3^\circ$~S& $39.4^\circ$~N \\
   & \multicolumn{2}{c}{Average altitude} \vline& 1,400~m& 1,400~m \\
   & \multicolumn{2}{c}{Surface area} \vline & 3,000~km$^2$ & 700~km$^2$ \\
SD & \multicolumn{2}{c}{Lattice} \vline & 1.5~km hexagon & 1.2~km square \\[0.5ex]
\cline{2-5}
    &               & Type         & water-\v{C}erenkov  & Plastic scintillator \\
    & Detector & Size          & 10 m$^2 \times 1.2$~m  & ($2 \times$) 3~m$^2 \times 1.2$~cm \\
    &                & Sampling & 25 ns & 20 ns \\ [0.5ex]
\hline
     & \multicolumn{2}{c}{Sites} \vline & 4 & 3 \\ [0.5ex]
\cline{2-5}
    & \multirow{4}{7em}{~~Telescopes} & Number & 24 & 36 \\
FD & & Size & 13~m$^2$ & 6.8~m$^2/3$~m$^2$ \\
     & & Field of view & $28.5^\circ \times 30^\circ $
                                        & $16^\circ \times 14^\circ / 18^\circ \times 15^\circ $  \\
     & & Pixels & 440 & 256 \\ [0.5ex]
\hline
\end{tabular}
\caption{Comparison of characteristics of the Pierre Auger Observatory
 and the Telescope Array. The low energy extensions for each observatory,
HEAT and TALE, are not included.}
\end{table}

\subsubsection{Pierre Auger Observatory and Telescope Array}

The two detectors are very similar, but the different sizes and operation times make them differ sizably in the collected data sets and exposures. They are both located at similar average elevation, about 1,400 m a.s.l., and roughly similar longitudes, Auger in the southern hemisphere and TA in the northern. A detailed description of the experiments can be found in~\cite{ThePierreAuger:2015rma,AbuZayyad:2012kk}; the main features of the basic detectors are summarized in
Table \ref{table:detfeatures} \cite{Zas:2015fpa}.

The most remarkable difference lies in their surface detectors (SD) which are based on different detection methods. The particle detectors in the Auger SD are cylindrical tanks of 10 m$^2$ surface and 1.2 m height, filled with purified water, with three photomultiplier tubes (PMT) to detect the \v{C}erenkov  light of particles in the shower front. In the TA they consist of two 3 m$^2$ slabs of plastic scintillator on top of each other which give light pulses also read by PMTs. The water tanks are relatively much more sensitive to shower muons which usually traverse the tank from wall to wall while the counts in the TA detectors are dominated by electrons and positrons in the shower front. Furthermore, because of their height, the Auger detectors are well suited to detect highly inclined showers, thereby increasing the exposure and the sky coverage. Inclined showers are also used for neutrino searches and to establish the muon content of the showers.
In Auger, an array of radio antennas (AERA) complements the data with the detection of the shower radiation in the hundred MHz region.

The fluorescence telescopes are located on the boundary of the two observatories to overlook the whole atmospheric volume just above the surface arrays and are based on similar detector components. The Pierre Auger Observatory contains a smaller area of 23.5 km$^2$ with stations separated by 750 m ({\it infilled} array) which can be combined with three additional telescopes pointing at higher elevations (HEAT) for lower energy measurements. Similarly TA has two sub-arrays of 46 and 35 stations separated by about 600 m and 400 m over a surface of 20 km$^2$, together with ten telescopes covering from 31$^\circ$ to 59$^\circ$ of elevation (TALE). 

\subsection{Event classes and energy calibration}
A hybrid experiment collects shower events of different classes. The separation into classes is a natural consequence of the different on-time (generally called {\it duty cycle}) of the two detector components: the surface array is able to collect showers at any time, whereas the fluorescence detectors can operate only during clear moonless nights ($\approx$ 15\% duty cycle).  Taking into account geometry and quality cuts applied at the event reconstruction level, the common data-set is only few percent. Therefore only a small part of the SD showers are actually reconstructed by the FD. Nonetheless this sub-sample (the {\it hybrid} data-set) is very valuable, including events having both the footprint of the shower at ground and the longitudinal profile measured. The advantage of this approach is twofold: 
\begin{itemize}
\item the energy estimator used in the surface detector can be compared on an event-by-event basis with the shower energy reconstructed by the FD. The latter measurement is based on the total amount of light emitted along the shower, which is in turn proportional to the energy deposited by the shower particles. Apart from the missing energy carried by neutrinos and energetic muons, the FD perfoms a calorimetric measurement of the shower energy. The SD estimator is given by the particle density at 1000 m (800 m) from the shower core in Auger (TA), corrected for the shower attenuation in the atmosphere depending on the zenith angle. The correlation between the FD energy and the SD energy estimator provides the energy calibration that is used for the whole SD data set~\cite{Zas:2015fpa}\cite{Verzi:rap}.
\item The hybrid events are higher quality showers, because the availability of the longitudinal profiles allows to access the most prominent information about the primary mass (the maximum of the shower depth). These events have also a superior definition of the shower geometry, even if the SD data are coming from a single surface detector~\cite{Abreu:2010aa}. Therefore the hybrid data-set, though being reduced in size, constitues a selection of well reconstructed events and a reference for all methods, based on SD data, aiming to obtain mass discriminating parameters.
\end{itemize}

The SD energy calibration through the hybrid data-set is a technique adopted by both collaborations. The Pierre Auger collaboration calculates the correlation between the energy estimators (for the three classes: standard, inclined and infill)  and the FD energy~\cite{Abraham:2008ru}.  TA has found that the SD energy derived from the energy estimator via Monte Carlo simulation has to be multiplied by a constant factor of 1.27 to ensure a good matching to the measured FD energy~\cite{AbuZayyad:2012ru}. The latter result is a remarkable evidence of the inadequacy of energy calibrations based on Monte Carlo methods as done in the past.

\begin{figure}[t]
%\sidecaption[b]
\begin{center}
\includegraphics[scale=0.3]{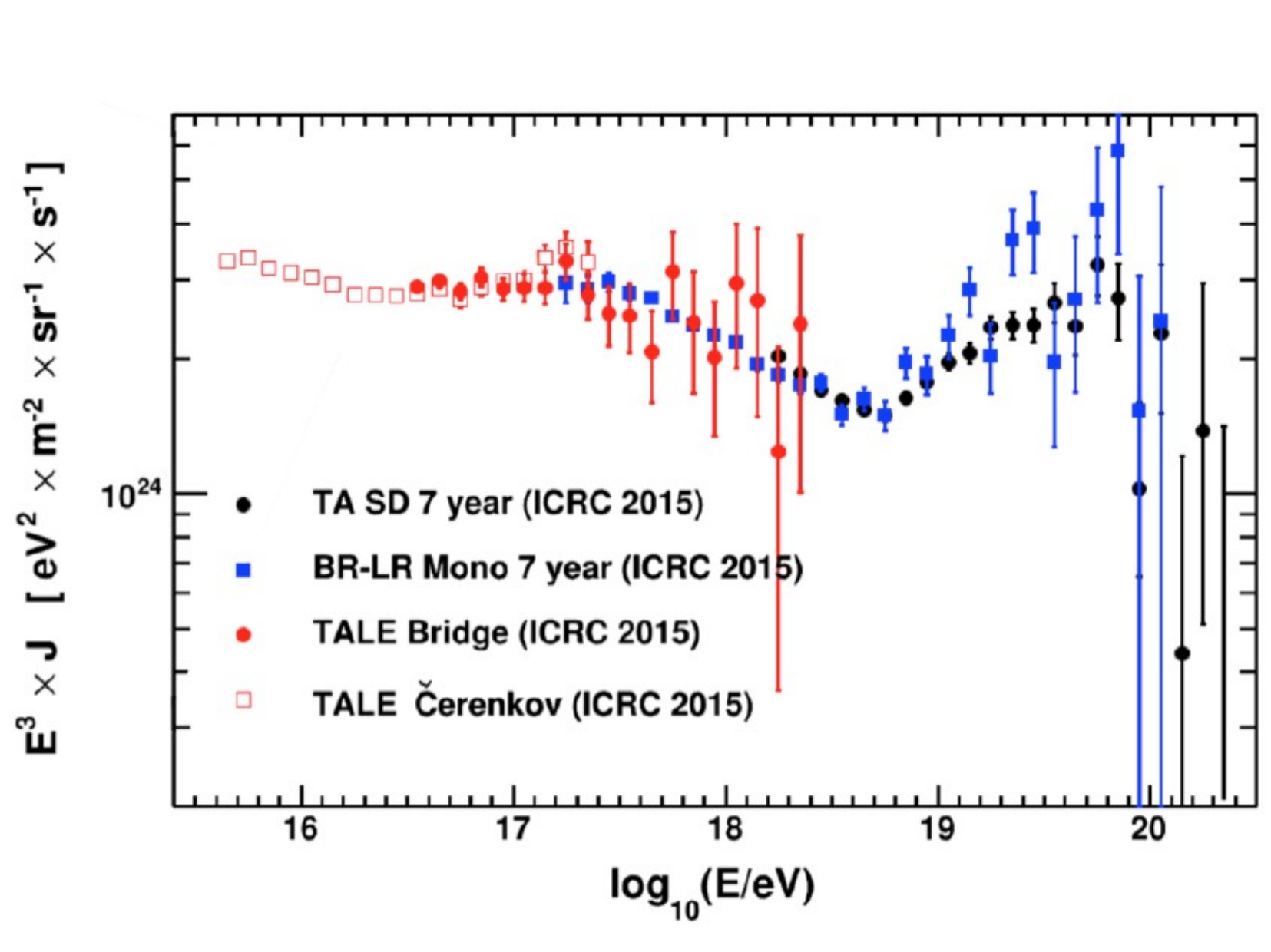}
\includegraphics[scale=0.42]{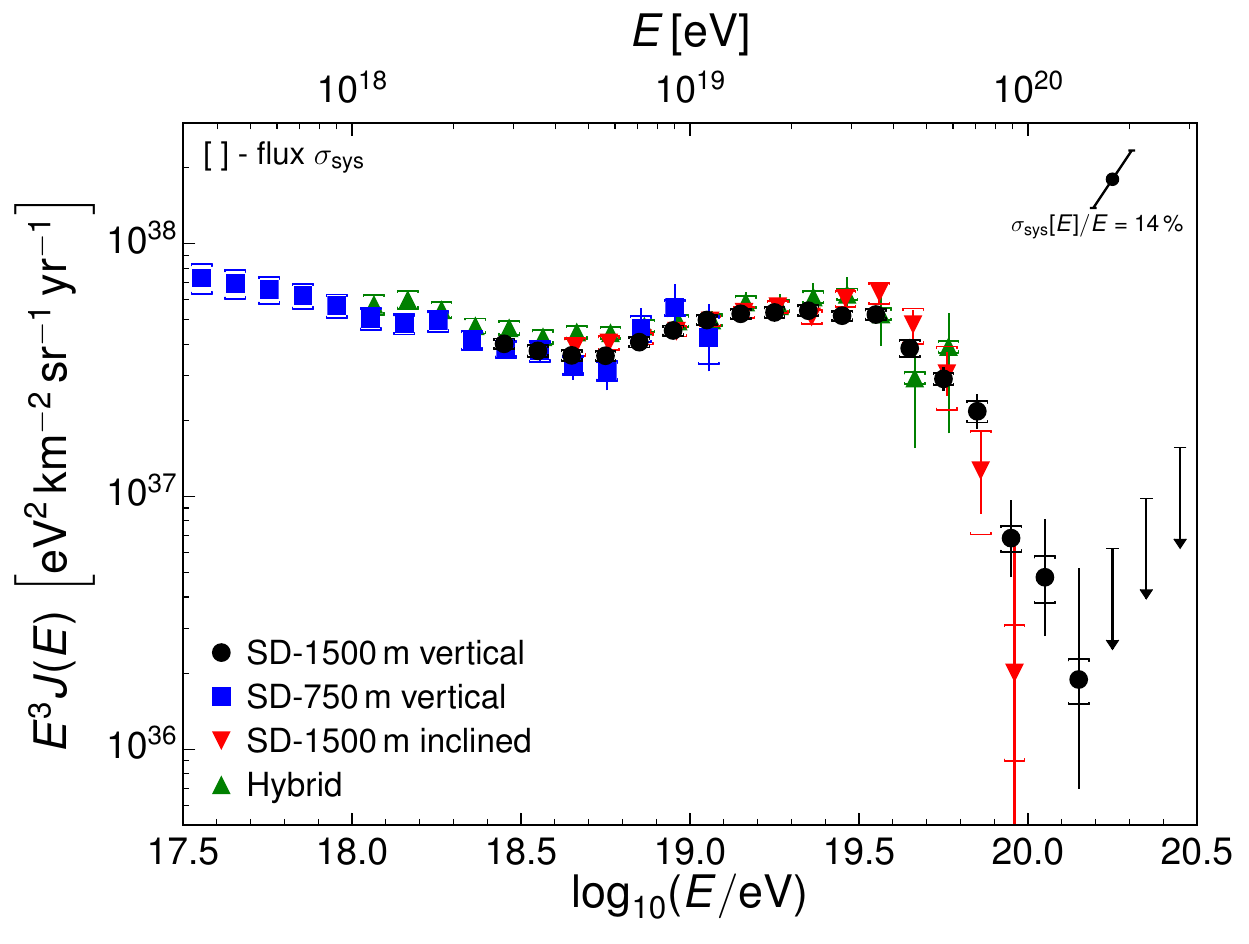}
\end{center}
\caption{Energy spectra presented at ICRC 2015 by the Telescope Array (upper panel) and Auger (lower panel) collaborations. The data from the different sub-detectors are 
shown separately.}
\label{fig:UHEspectra}
\end{figure}

It has finally to be noted that the relation between the longitudinal profile density and the measured light is provided by the combination of the fluorescence yield and the measured light transmission. The former has been established experimentally, the latter is obtained from the atmospheric monitoring data system operated at each site. Unfortunately the collaborations use different parameterizations of the fluorescence yield. Including all that the quoted systematic uncertainty in the energy scale is 14\% for Auger and 20\% for TA.

\subsubsection{The energy spectrum}
The energy spectra measured at the two observatories are shown in Fig.~\ref{fig:UHEspectra}. A more comprehensive review of spectrum data, including other experiments, e.g. IceCube and Yakutzk, can be found in~\cite{Verzi:rap}. Yet, especially for energies above 10$^{18}$ eV, the bulk of the data comes from Auger and TA. The two panels show the spectra as originating from different detector components for TA (left) and Auger (right). 

The most prominent features appear similar in the common energy interval with a break (the {\it ankle}) at around 10$^{18.7}$ eV and a flux suppression, quite evident (at several standard deviations for both experiment) in both cases, but exhibiting somewhat different shapes. It has to be noted that for both experiments the data above the ankle are dominated by the respective ground arrays.

\begin{figure}[t]
%\sidecaption[b] 
\begin{center}
\includegraphics[scale=0.4]{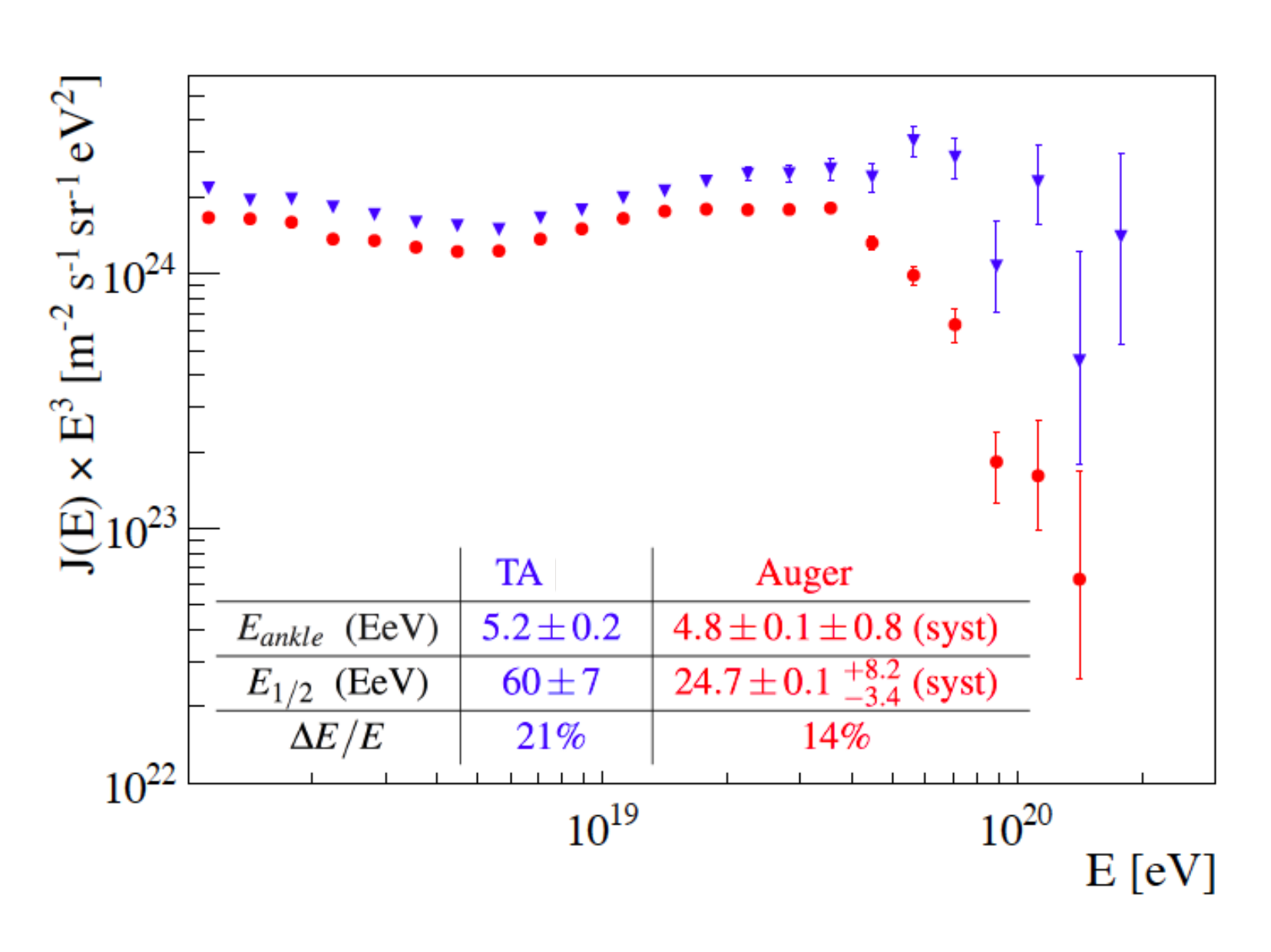}
\end{center}
\caption{Comparison between TA (blue) and Auger (red) combined energy spectra.}
\label{fig:CompSpectra}
\end{figure}

Both the collaborations exploit procedures to combine the different spectrum components into a unique spectrum. For a better comparison, the combined energy spectra are superimposed in Fig.~\ref{fig:CompSpectra}, which provides also the values of the main spectral features~\cite{Verzi:rap}. The corresponding exposures are about 6,300 km$^2$~sr~yr for TA and 50,000 km$^2$~sr~yr for Auger. Comparing the values of the ankle energy ($E_{ankle}$) and of the cut-off ($E_{1/2}$) (the energy at which the integral flux drops to half of what is expected in the absence of a cut-off) one finds that the ankle energies are consistent within the systematic uncertainties in the energy scale, but the discrepancy between the cut-off energies is not explained by systematics.

The different behavior in the cut-off region is apparently beyond the expectations of the current knowledge of systematics. A possible contribution to this difference in terms of declination dependence of the flux has been investigated by Auger~\cite{Aab:2015bza}. No significant variation in the flux measured with the SD in four declination bands has been found that could account for the discrepancy between spectra measured from different hemispheres.  The differences found between the measurements in two separate declination bands are instead compatible with the variations expected from a dipolar modulation (see below) of 
the flux~\cite{ThePierreAuger:2014nja}.

\subsubsection{Mass composition}
Composition is addressed using the depth of the position of the maximum in the number of shower particles, $X_\mathrm{max}$, which is measured by the FD. In a simplistic picture, the sensitivity of $X_\mathrm{max}$ to mass composition relies on the fact that showers from heavier (lighter) nuclei develop higher (deeper) in the atmosphere and their profiles fluctuate less (more). 

\begin{figure}[t]
%\sidecaption[b]
\begin{center}
\includegraphics[scale=1.2]{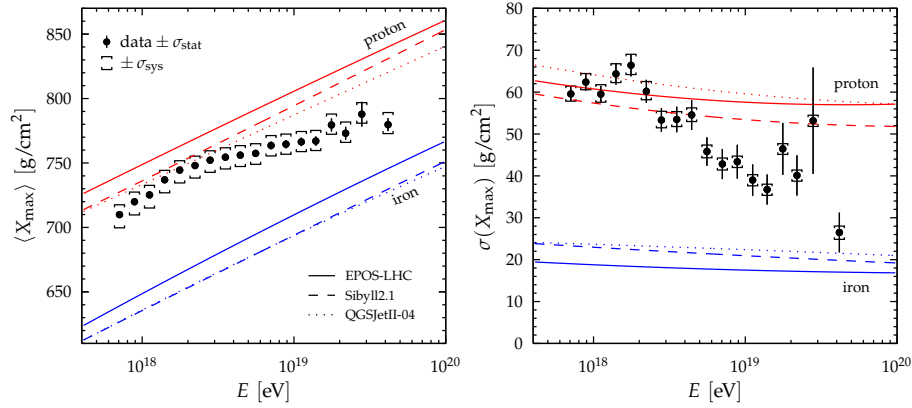}
\end{center}
\caption{The mean (left) and the standard deviation (right) of the $X_\mathrm{max}$
distributions measured by Auger, as a function of energy compared to air-shower simulations for protons and iron primaries.}
\label{fig:AugerXmax}
\end{figure}

The measurements by Auger are the most robust for both the data selection and the quality of the $X_\mathrm{max}$ distributions that are obtained. For the limited field of view of the telescopes, depending on the zenith angle and impact point of the shower, a fluorescence detector views a different range of $X_\mathrm{max}$. The Auger analysis adopts event selection and quality cuts that allow to get rid of this bias and thus obtain unbiased $X_\mathrm{max}$ distributions. Correcting for detector resolution and acceptance, the first two moments of the distributions (mean and standard deviations) can be directly compared to air shower simulations. The Auger collaboration has published $X_\mathrm{max}$ measurements for hybrid showers having energies above 10$^{17.8}$ eV~\cite{Aab:2014kda} and recently reported preliminary results extending these measurements down to 10$^{17}$ eV~\cite{Aab:2015bza}.  Fig. \ref{fig:AugerXmax} shows the published data. 

\begin{figure}[t]
%\sidecaption[b]
\begin{center}
 \includegraphics[scale=0.22]{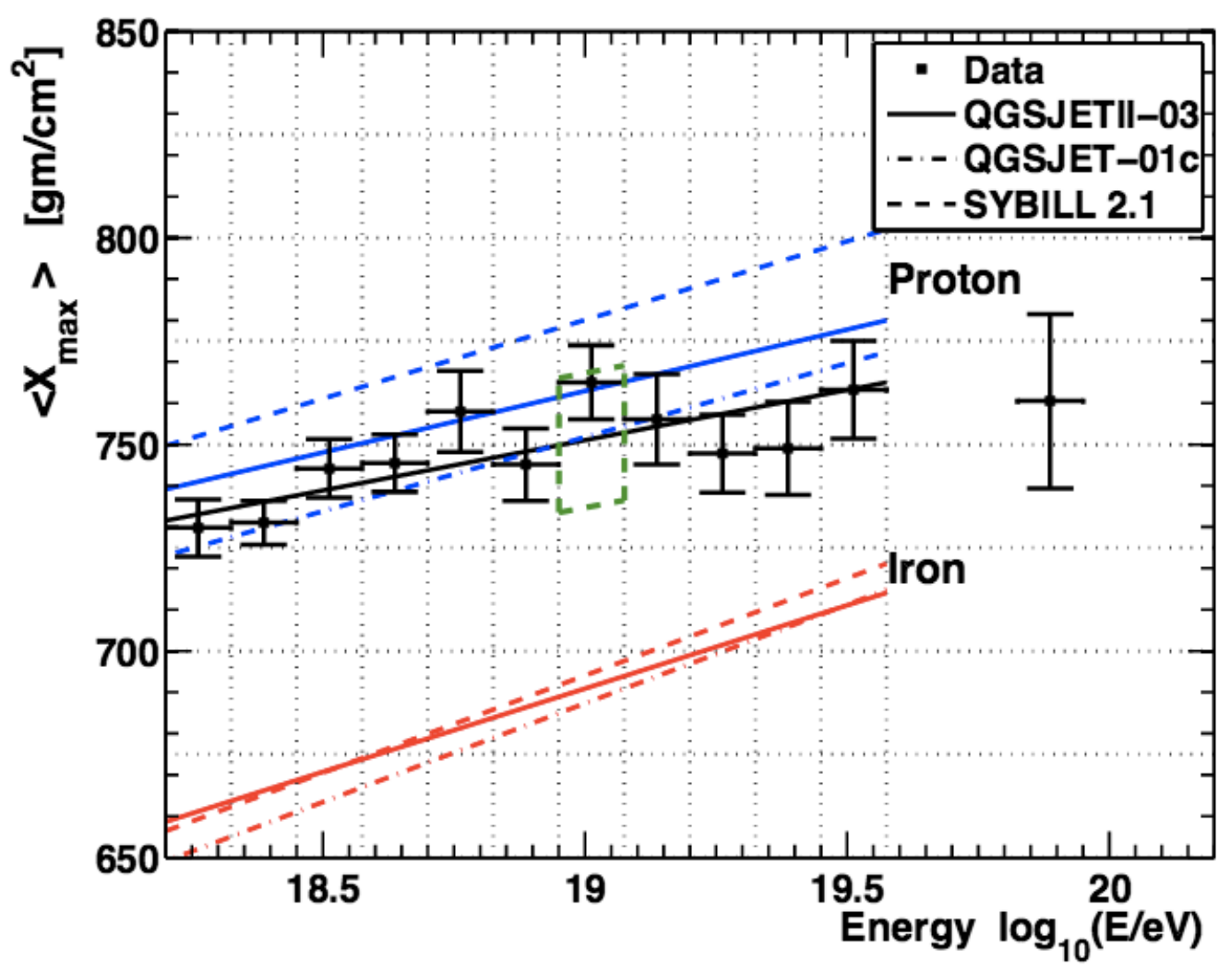}
  \includegraphics[scale=0.22]{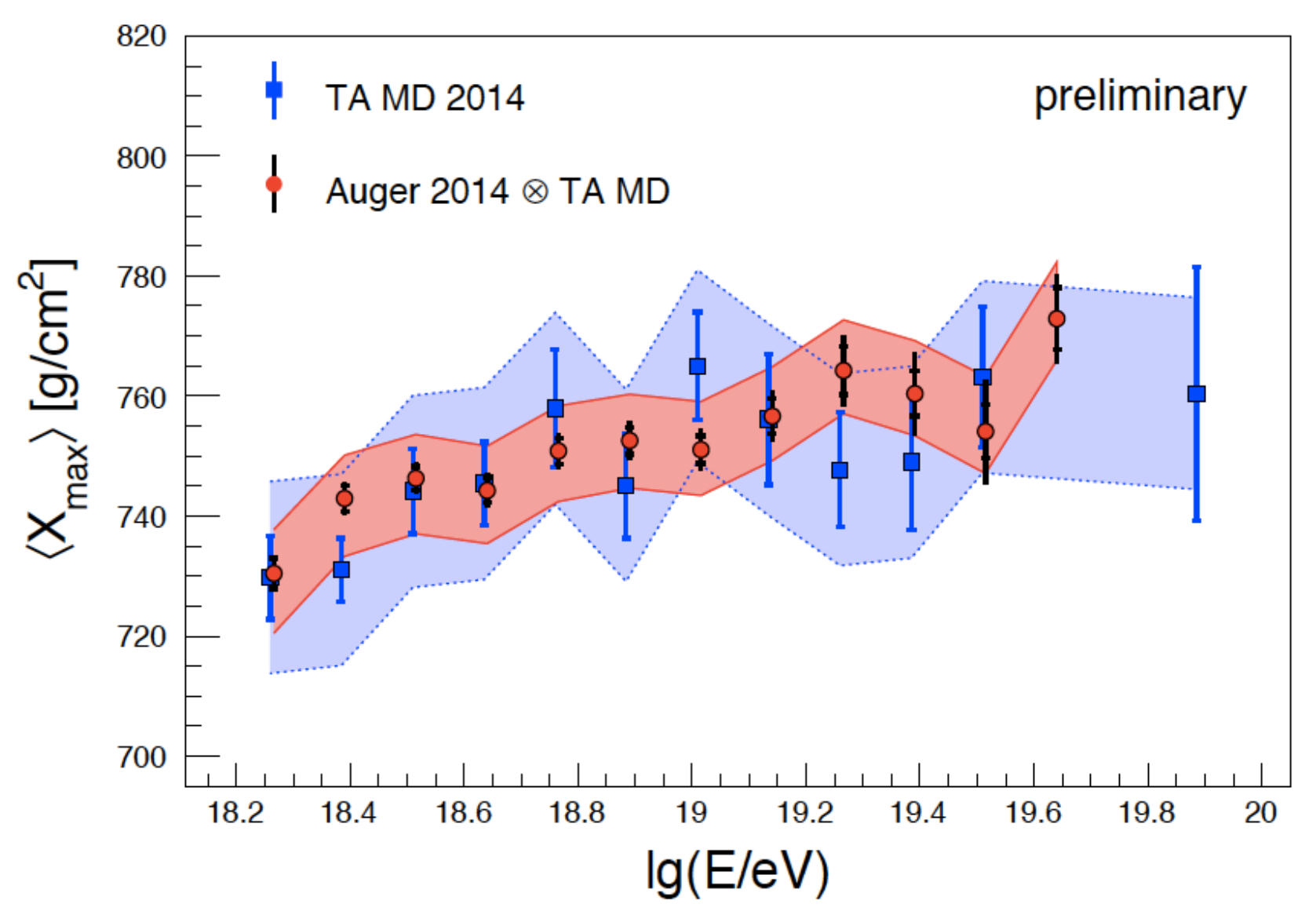}
\end{center}
\caption{[Left Panel] The TA Middle Drum hybrid composition result using geometry and pattern recognition cuts. The solid black line is a fit to the data. Coloured lines are fits to MC, for the used hadronic models. The green hashed box indicates the  total systematic error on $\langle X_\mathrm{max} \rangle$. [Right Panel] Comparison of $\langle X_\mathrm{max} \rangle$ as measured with the MD of TA (blue squares) and the $\langle X_\mathrm{max} \rangle$ of the Auger data folded with the MD acceptance.  The colored bands show the systematic uncertainties of the  $X_\mathrm{max}$ scales of each experiment.}
\label{fig:AugerTAcomp}
\end{figure}

Telescope Array has reported data for different data selections: monocular, stereo and hybrid data. The measurements of the mean depth for hybrid events observed from the Middle Drum fluorescence detector~\cite{Abbasi:2014sfa} are shown in Fig. \ref{fig:AugerTAcomp} (left panel).

It has to be noted that, contrary to Auger, the TA  $X_\mathrm{max}$ distributions are  folded with the detector effects such as the selection efficiency and acceptance, and therefore biased by experimental effects. This approach is not only due to an analysis choice, but is primarily determined by the limited size of
the data that prevents from applying cuts so selective as the ones adopted by Auger. The interpretation of TA data is then performed using Monte Carlo predictions folded with the same detector efficiency and resolution.

The different approaches of the two collaborations imply that the $\langle X_\mathrm{max} \rangle$ values obtained cannot be directly compared to one another. Also the MC predictions for the measurements are based on different hadronic interaction models. A joint working group has been setup for comparing the two approaches. The primary abundances which best describe the Auger data have been simulated and analyzed by the TA collaboration using the same procedure as applied to their data. The result, which is a simulated version of the Auger data as it would be observed by  TA,  is shown in Fig. \ref{fig:AugerTAcomp} (right panel). The average difference between the two data sets was found to be (2.9 $\pm$ 2.7 (stat.) $\pm$ 18 (syst.)) g/cm$^2$.

The Auger $X_\mathrm{max}$ data (moments and distributions) enable a step further in the interpretation of mass composition. In fact the mean log mass can be derived from the measurement of $\langle X_\mathrm{max} \rangle$, on the basis of the superposition  model or a simple parametric extension of the same model~\cite{Abreu:2013env}. Yet from the mean $X_\mathrm{max}$ alone it is not possible to retrieve information about the relative weights of primaries contributing to the actual value of $\langle \ln A \rangle$. Using their data the Auger collaboration has obtained the evolution with energy of the first two moments of $\ln A$~\cite{Abreu:2013env} and of the fractions of four mass groups (H, He, N and Fe) from the fit of the $X_\mathrm{max}$ distributions~\cite{Aab:2014aea}.

\subsubsection{Anisotropy}
The search for anisotropies in the arrival directions of cosmic rays aims at spotting their sources or a global inhomogeneity in the source distribution on scales comparable with the loss length at given energy. At the highest energies, finding small angular scale anisotropies would represent the access gate to the beginning of charged particle astronomy. This possibility is however tightly related to the composition of the CRs at such high energies, since only for protons the deflection angles induced by the (poorly known) magnetic fields are expected to be small. Both the Auger and TA collaborations have intense programs to search for anisotropies. These include several tools like auto-correlation, correlation with source catalogs, search for flux excesses ({\it hotspots}) in the visible sky and correlation with other experiments. 

At present, none of the tests show statistically significant evidence of anisotropy \cite{PierreAuger:2014yba} \cite{Abbasi:2014lda}. Yet remarkable flux excesses are observed at intermediate scales in the North (South) hemisphere by TA (Auger). Using cosmic ray events with energy $E > 57$ EeV, TA have observed a cluster of events, centered at R.A. = 146.7$^\circ$, Dec.= 43.2$^\circ$~\cite{Abbasi:2014lda}, of about 20$^\circ$ radius and with a calculated probability of appearing by chance in an isotropic cosmic-ray sky of 3.7$\times 10^{-4}~(3.4 \sigma)$. In Auger the strongest departures from isotropy (post-trial probability $\sim 1.4\%$) are obtained for cosmic rays with $E > 58$ EeV around the direction of Cen A (15$^\circ$ radius)~\cite{PierreAuger:2014yba}. In any case, it will be interesting to follow the evolution of these excesses with future data.

\begin{figure}[t]
%\sidecaption[b]
\begin{center}
\includegraphics[scale=0.16]{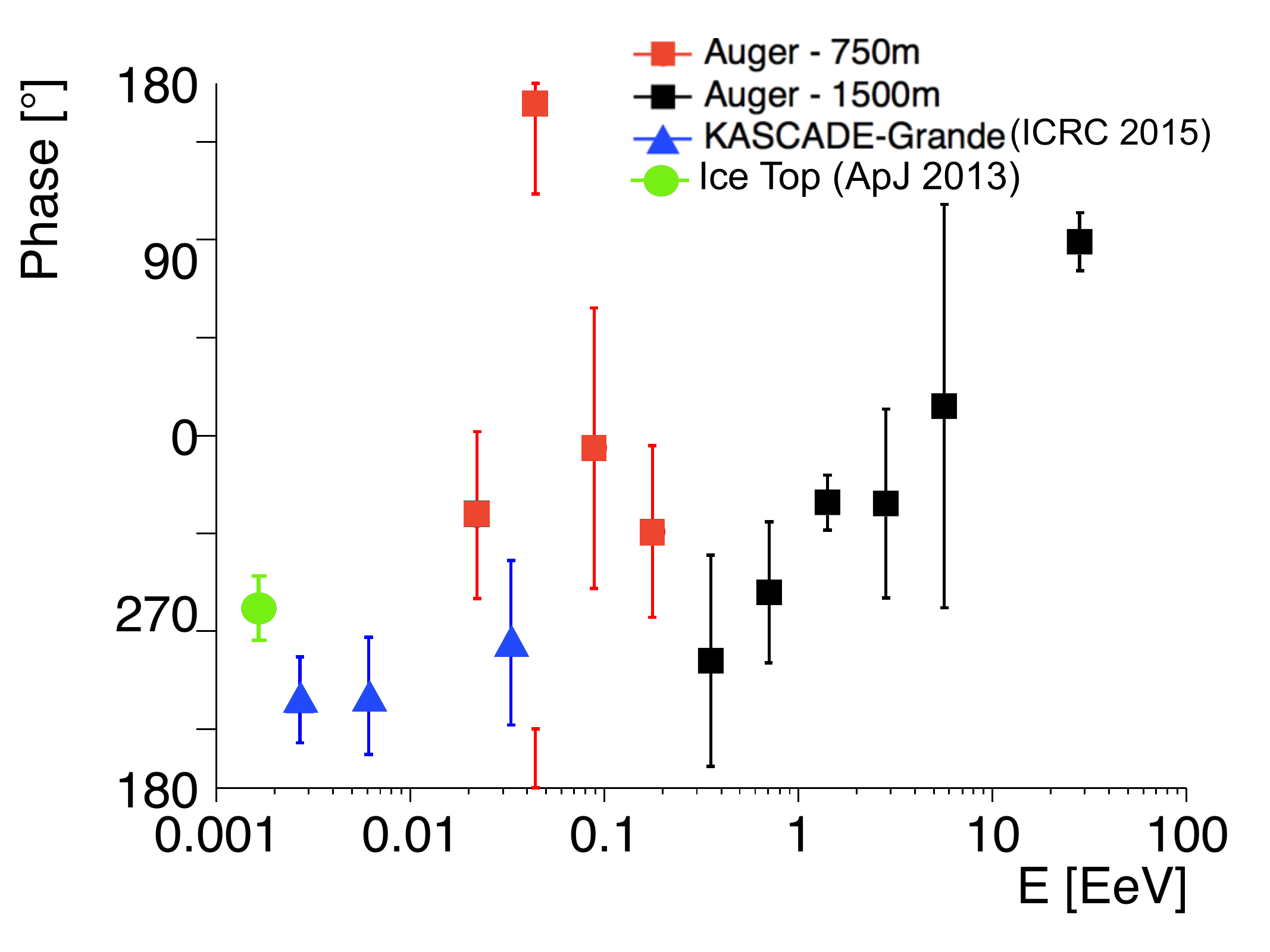}
\includegraphics[scale=0.16]{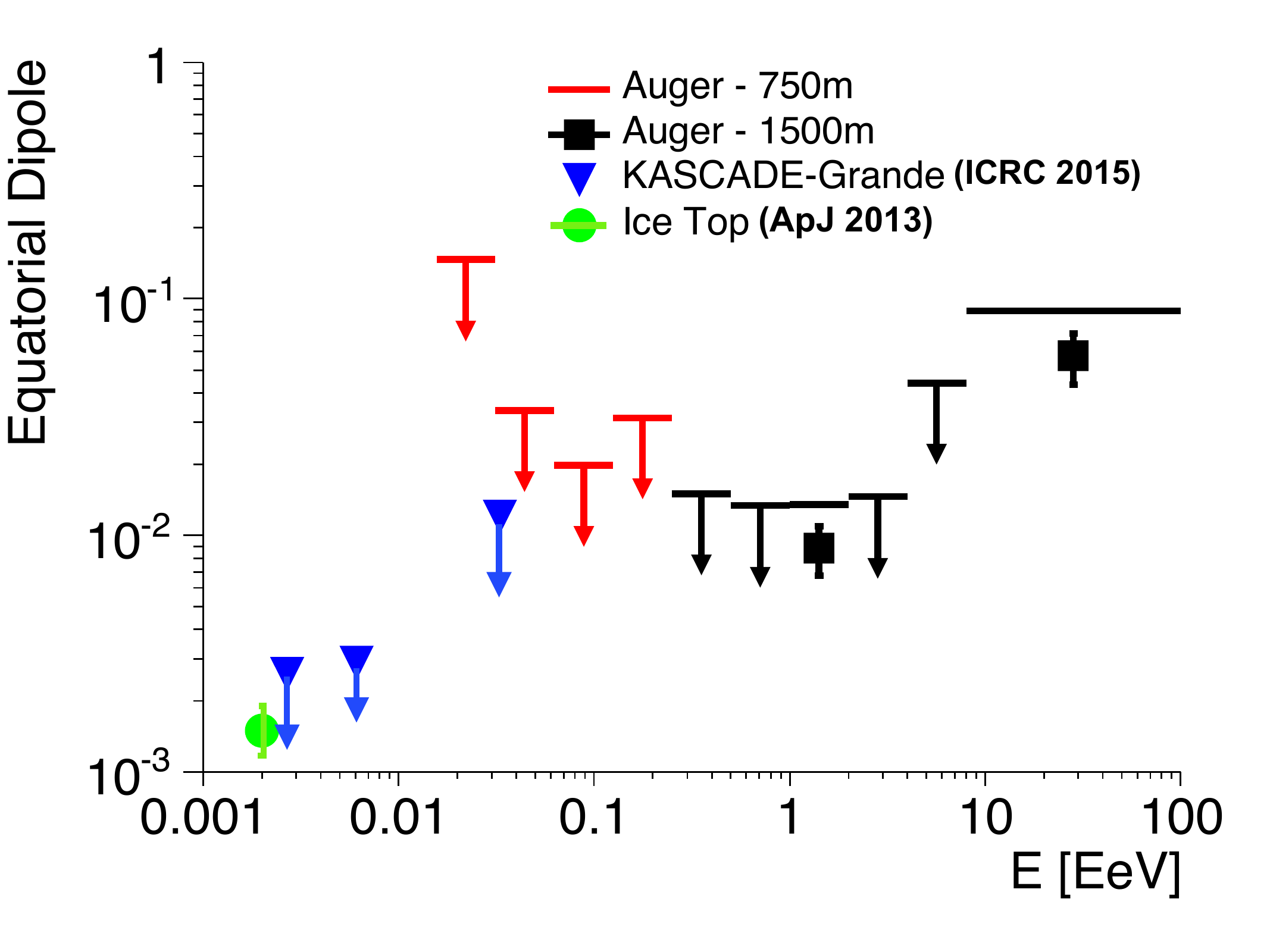}
\end{center}
\caption{Auger first harmonic analysis. Left: Measured phases of the first harmonic modulation in RA. Right: Upper limits of the dipole equatorial component. Amplitudes are also reported in the two energy bins when the corresponding p-value expected from isotropy is below 10$^{-3}$.}
\label{fig:UHEphtrans}   
\end{figure}

Another recent correlation study is based on the attempt to exploit IceCube neutrino observations to identify the sources of UHECRs~\cite{Aartsen:2015dml}. The study is a common effort by the IceCube, Auger and TA collaborations and is based on the fact that neutrinos can be related to charged particles both at astrophysical sources, through their interaction with the ambient matter and radiation, and along their propagation through the cosmic background radiation. The UHECRs comprise about 300 events above 50 EeV collected by Auger and TA. The neutrino events used are the most energetic (30 TeV to 2 PeV), which provided evidence for a neutrino flux of astrophysical origin. These comprise 39 {\it cascades} (signatures of charged-current $\nu_e$ interactions as well as neutral-current interactions of all flavors) and 16 high-energy {\it tracks} (signatures of charged-current $\nu_\mu$ interactions).  Another correlation study involves the so-called IceCube point-source sample  of about 400,000 tracks with a sub-degree angular resolution. No indications of correlations at discovery level are obtained for any of the searches performed. The smallest of the p-values comes from the search for correlation between UHECRs with IceCube high-energy cascades, a result that should continue to be monitored.

Large scale anisotropies are not suited to correlate directly cosmic rays to sources, but can be used to infer aspects of the global distribution of sources. Some large scale anisotropy (below the percent level) is expected as well because of the relative motion of cosmic rays with respect to the rest frame of background radiation~\cite{Kachelriess:2006aq}. Furthermore, a large scale analysis as a function of energy is important since it could provide information about the transition from galactic to extragalactic dominance in the cosmic ray flux.

The Auger collaboration has published evidence \cite{Abreu:2011ve} of a smooth phase transition of the first harmonic modulation in right ascension distribution from 270$^\circ$ to 100$^\circ$ around 1 EeV. This fact might correspond to a transition from dipole direction pointing to the galactic centre, at low energies, to one which is rotated by about 180$^\circ$ above a few EeV. This study is based on both Rayleigh and East-West analyses of the counting rates. This analysis has been recently updated and a prescribed test is running to confirm it~\cite{Aab:2015bza}. Fig. \ref{fig:UHEphtrans} shows the results from the latest data. The amplitude is still given as upper limits, but for two energy bins, between 1 and 2 EeV and above 8 EeV, the statistical significance is high enough to provide measured amplitudes.

\begin{figure}[t]
\begin{center}
\includegraphics[scale=0.25]{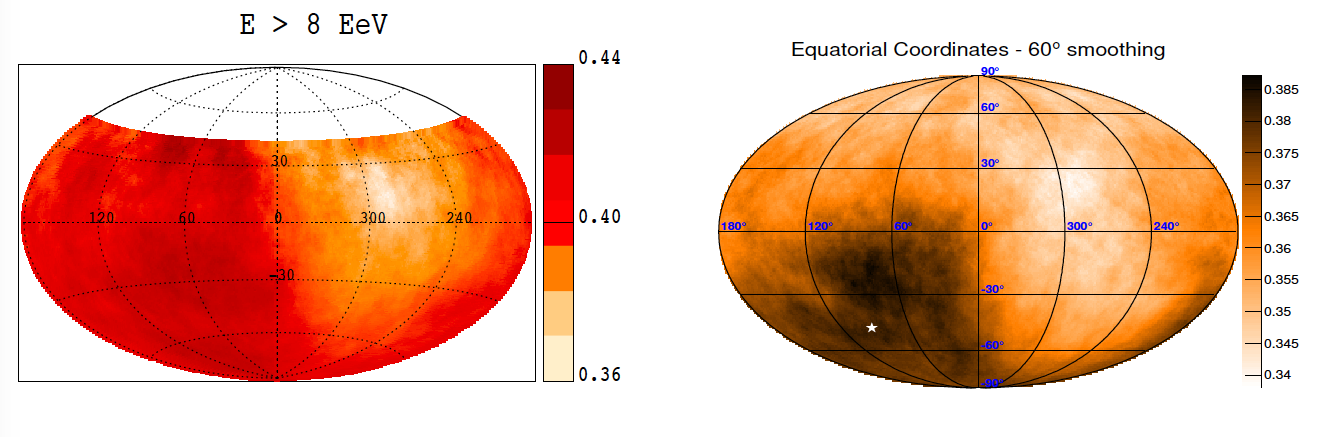}
\end{center}
\caption{Sky maps in equatorial coordinates of flux,  smoothed in angular windows of 45$^\circ$
(60$^\circ$) radius, for Auger (Auger and Telescope Array) events with 
$E > 8$ EeV (10 EeV), left (right) panel.}
\label{fig:dipoles8EeV}
\end{figure}

The large scale distribution of arrival directions has been also studied by combining the data of Auger and Telescope Array. Thanks  to the full-sky  coverage,  the measurement of the angular power spectrum does not rely on any assumption on the underlying flux of cosmic rays. The study, carried out using a spherical harmonic analysis for cosmic rays above 10$^{19}$ eV, has been published in~\cite{Aab:2014ila} and updated at ICRC 2015~\cite{OlivierICRC2015}. No deviation at discovery level from isotropy is found at any multipoles. The largest deviation, with a p-value of $5 \times 10^{-3}$, occurs for the dipole, with an amplitude of (6.5 $\pm$ 1.9)\%, pointing to 93$^\circ$ $\pm$ 24$^\circ$ in right ascension and -46$^\circ$ $\pm$ 18$^\circ$ in declination (see Fig. \ref{fig:dipoles8EeV}, right). It is worth noting that this result agrees with the one found with Auger-only data~\cite{ThePierreAuger:2014nja}, assuming pure dipolar or dipolar-quadrupolar distributions (see Fig. \ref{fig:dipoles8EeV}, left). 

\subsubsection{Future developments}
UHECR data provide other interesting outcomes on several aspects of cosmic ray physics (e.g. muon component in the shower events, photon and neutrino limits) and particle physics (e.g. proton cross section, hadronic interactions at energies higher than LHC) that are not included in this section and will be partly considered in the following. Considering all data, it appears difficult to build a consistent picture of the origin of UHECRs in the presence of so many unknowns about source distribution, composition, galactic and extragalactic magnetic fields, etc. To make further progress in this direction more accurate and extended information on the nature of the primaries is required: mass composition is currently unavailable above 40 EeV due to the intrinsic duty cycle of the FD and the scarce accuracy of the composition sensitive methods based on the surface array data. 

The  AugerPrime~\cite{Aab:2016vlz}  upgrade  of  the  Pierre Auger Observatory  has  been  specifically  designed  to  improve  mass composition in the whole energy range.  Along  the  line of a hybrid design, each SD will be equipped  with  a  top  scintillator  layer. Shower  particles  will  be  sampled  by  two detectors  (scintillators and water-\v{C}erenkov  stations) having different responses to the muonic and electromagnetic components, thus allowing to reconstruct each of them separately.  The muonic component will be derived in each station by subtracting the signal observed in the scintillator from that seen in the water \v{C}erenkov  tank. The upgraded array will provide data with no duty cycle limitation and then the access to the highest energies will be made possible.

The Telescope Array collaboration plan to extend the SD array by a factor four, TAx4~\cite{Sagawa:2016bsh}, by adding 500 surface detectors on a square grid of about 2 km spacing. With this new design the overall area will be approximately 3000 km$^2$. With enhanced statistics they expect to improve the investigation of the observed hotspot and possibly the correlation with other sources. TA are also designing new muon detectors~\cite{Nonaka:2016mco} to enable the simultaneous detection of electromagnetic and muonic components in shower events.

A possible jump in integrated statistics, even though with limited resolution on $X_\mathrm{max}$, would be provided by the observation of EAS fluorescence light 
from a space-based detector looking towards Earth's surface. The Extreme Universe Space Observatory on board the Japanese Experiment Module of the 
International Space Station, JEM-EUSO, is being designed for such a mission. It is based on a wide field of view (60$^{\circ}$) near-UV telescope with 
a diameter of 2.5 m, orbiting at an altitude of about 400 km, that would provide an annual exposure larger 
than $50,000\,$km$^2$ sr yr / yr, above $6 \times 10^{19}\,$eV \cite{Fenu:2017xct,Ebisuzaki:2014wka}.

\section{Transport of CRs in the Galaxy}
\label{sec:transport}

The transport of CRs in the Galaxy has been subject of active investigation for quite some time. The main handle we have on the propagation of galactic CRs comes from measurements of secondary-to-primary ratios, such as Boron/Carbon (B/C), which provide us with an estimate of the grammage that CRs traverse during propagation. In fact, this indicator is the main reason why we describe the CR transport as mainly diffusive in nature: a ballistic propagation of CRs would make CRs escape the Galaxy in times which are much shorter than the one necessary to explain the observed B/C ratio. 

The B/C ratio is related to the grammage traversed by CRs, $X(E)=\bar n \mu v \tau_{esc}(E)$, where $\bar n$ is the mean gas density in the confinement volume of the Galaxy (disc plus halo), $\mu$ is the mean mass of the gas, $v$ is the speed of particles. For particles with energy per nucleon of 10 GeV/n the measured B/C corresponds to $X\sim 10 g~cm^{-2}$. If the sources are located in the thin disc of the Galaxy with half thickness $h=150$ pc and the halo extends to a height $H$, the mean density can be estimated as $\bar n=n_{disc} h/H = 5\times 10^{-2}\left(\frac{n_{disc}}{1cm^{-3}}\right)\left(\frac{H}{3 kpc}\right)^{-1} cm^{-3}$. For a standard chemical composition of the ISM ($n_{He}\approx 0.15 n_{H}$) the mean mass is $\mu = (n_{H}+4n_{He})/(n_{H}+n_{He})\approx 1.4 m_{p}$. It follows that for a proton with energy $E_{*}=10$ GeV the typical escape time is 
\be
\tau_{*} \sim \frac{X(E_{*})}{\bar n \mu c} = 90 \left(\frac{H}{3 kpc}\right) Myr,
\label{eq:taustar}
\ee
which exceeds the ballistic propagation time scale by at least three orders of magnitude. This remains the strongest evidence so far for diffusive motion of CRs in the Galaxy. A diffusion coefficient can be introduced as $\tau_{esc}(E)=H^{2}/D(E)=\tau_{*}(E/E_{*})^{-\delta}$, so that at 10 GeV $D(E)\simeq 3\times 10^{28} \left(\frac{H}{3 kpc}\right) cm^{2} s^{-1}$. The grammage (and therefore the escape time) decreases with energy (or rather with rigidity) as inferred from the B/C ratio, illustrated in Fig. \ref{fig:BoverC}, which shows a collection of data points on the ratio of fluxes of boron and carbon. Fig. \ref{fig:BoverC} illustrates the level of uncertainty in the determination of the slope of the B/C ratio at high energies, which reflects on the uncertainty in the high energy behaviour of the diffusion coefficient. At low energies the uncertainty is due to the effects of solar modulation which suppresses CR fluxes in a different way during different phases of the solar activity (see \cite{Potgieter:2013pdj} for a recent review). The effect of modulation is more pronounced on the spectra of individual elements than on the B/C ratio. The high rigidity behavior of the B/C ratio is compatible with a power law grammage $X(R)\propto R^{-\delta}$ with $\delta = 0.3-0.6$.

While at high energy CR transport is most likely diffusive, at low energies other processes may become important or even dominant. For instance, particles can be advected with a galactic wind. This phenomenon leads to CR spectra at the position of the Sun that are much harder than the ones observed at high energies, a phenomenon that is apparent in the recent Voyager I data \cite{2013Sci...341..150S}. At energies below $\sim GeV$, energy losses due to ionisation also become important and result in spectral hardenings as compared with the high energy trend. For nuclei, spallation energy losses also become important and may harden nuclear spectra, as compared with the proton spectrum. 

Much can be learned in favour and against current models of CR propagation from the understanding of the microphysics of CR transport: spatial diffusion results from pitch angle diffusion of charged particles propagating in a background of Alfv\'en waves with random phases and a given power spectrum. This process requires resonance between the gyration radius of the particle in the background magnetic field and the wavelength of the relevant Alfv\'en waves. If the waves are assumed to propagate in both directions along the magnetic field, as it is usually implicitly assumed in standard propagation calculations, then CRs are scattered to reach diffusion, and gain energy through second order Fermi acceleration (diffusion in momentum space). 

The debate on whether the background Alfv\'en waves are the result of environmental processes (for instance SN explosions or other types of stirring of the ISM) or rather produced non-linearly by CRs themselves remains a hot topic in the field. Below we discuss the implications of some recent ideas concerning this point and some possible observational consequences. 

\subsection{Self-generation of waves}
\label{sec:self}

It was first shown by \cite{1975MNRAS.173..255S} that the super-Alfvenic streaming of charged particles may result in the excitation of a streaming instability, namely generation of weakly modified Alfv\'en waves with wavenumber $k\sim 1/r_{L}(p)$, where $r_{L}(p)$ is the Larmor radius of particles with momentum $p$ that are responsible for the instability. It is important to realize that these waves have the correct wavenumber $k$ to be effective for particle pitch angle scattering. Hence the question arises of whether it is possible to think of particle diffusion in self-generated waves. The problem of particle transport, that is usually solved in a test-particle approximation (the diffusion coefficient is pre-assigned and independent upon the particles that are being propagated), becomes non-linear when self-generation is included. The growth of the instability is limited by different mechanisms of wave damping. In the Galaxy, the two main damping mechanisms are ion-neutral damping \cite{1971ApL.....8..189K} and non-linear Landau damping (NLLD) \cite{1990JGR....9514881Z,1990JGR....9510291Z}, which are important in partially ionized and in totally ionized plasmas respectively. 

As recognized by \cite{1971ApJ...170..265S,1975MNRAS.170..251H}, the effect of ion-neutral damping in the ISM is expected to be so strong that the growth of Alfv\'en waves through streaming instability is strongly hindered. In these conditions the wave excitation is possible only far from the galactic disc, where the density of neutral hydrogen drops to very low values. On the other hand, while the average value of such density along a line of sight is well determined by observations \cite{Ferriere:2001rg}, it is possible that neutrals may be spatially segregated, so that dense regions of high density may be surrounded by vast regions where the gas is mostly ionized and ion-neutral damping is not very important. In this case, Alfv\'en waves can be excited by CRs and the growth of the instability occurs at a rate
\begin{equation} \label{eq:Gamma_cr}
  \Gamma_{\rm cr} = \frac{16 \pi^2}{3} \frac{v_A}{\mathcal{F}(k) B_0^2} 
  				\left[ p^4 v(p) \frac{\partial f}{\partial z} \right]_{p= e B_0/kc}  \,,
\end{equation}
where the gradient in the CR distribution has been assumed to be only along the $z$ direction perpendicular to the galactic plane. Here ${\cal F}(z,k)$ is the fractional power $\delta B^{2}(k)/B_{0}^{2}$ per unit logarithmic interval of $k$, which also determines the diffusion coefficient through 
\be
D(z,p)=\frac{1}{3}r_{L}(p) v(p) \frac{1}{{\cal F}(k)|_{k=r_{L}(p)}}.
\label{eq:Diff}
\ee
Eq. \ref{eq:Diff} returns Bohm diffusion coefficient only if there is the same power on all scales and $\delta B/B_{0}=1$. In general the real diffusion coefficient is larger than Bohm diffusion, as one can see from Eq. \ref{eq:Diff}, since ${\cal F}(z,k)\ll 1$ (Eq. \ref{eq:Diff} is strictly valid only in this limit).  The rate of NLLD can be written as \cite{2003A&A...403....1P}:
\begin{equation} \label{eq:Gamma_damp}
  \Gamma_{\rm nlld} = (2 c_k)^{-3/2} \, k v_A  \, {\mathcal{F}(k)}^{1/2} \,,
\end{equation}
with $c_k= 3.6$. 

Imposing that growth and damping balance each other locally, one can determine the power spectrum ${\cal F}(z,k)$. At high energy, where transport is dominated by diffusion, one can see that $\partial f/\partial z=f_{0}/H$, where $f_{0}(p)\sim p^{-\gamma}$ is the CR spectrum in the disc and $H$ is the size of the galactic halo. Hence ${\cal F}(z,k)\sim k^{-\frac{2}{3}(5-\gamma)}$, and consequently $D(p)\propto p^{-\frac{7}{3}+\frac{2}{3}\gamma}$. For $\gamma\approx 4.7$, one easily infers $D(p)\sim p^{0.8}$. The fast momentum dependence of the diffusion coefficient implies that the CR confinement due to self-generated waves, at least in purely diffusive models, is bound to be effective only at relatively low energies (typically below $\sim \rm few 100$ GeV). 

\begin{figure}[t]
%\sidecaption[b]
\begin{center}
\includegraphics[scale=.9]{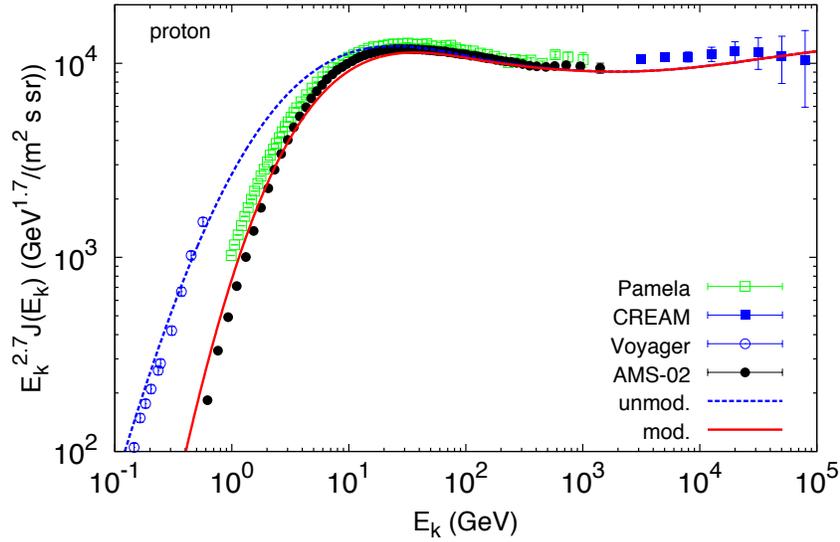}
\end{center}
\caption{Proton spectrum as measured by some experiments (see labels) compared with the flux as predicted by \cite{Aloisio:2015rsa}, with and without accounting for solar modulation (solid and dashed lines respectively). The flux without modulation fits well the recent Voyager data \cite{2013Sci...341..150S}.}
\label{fig:voyager}       
\end{figure}

After the discovery of the spectral breaks by PAMELA \cite{Adriani:2014xoa} and AMS-02 \cite{Aguilar:2015ooa,Aguilar:2015ctt} (see discussion in \S \ref{sec:hard}), models of self-generation have been reconsidered \cite{Blasi:2012yr,Aloisio:2013tda,Aloisio:2015rsa}: this work showed that the combination of self-generated waves and waves produced by cascading of large scale pre-existing turbulence may explain the spectral breaking: at high energies turbulence injected at large scale (possibly due to SN explosions) and cascading towards smaller scales dominates CR scattering; at low energies the self-generated waves provide the necessary scattering. This change of regime naturally leads to a spectral break at a few hundred GV rigidity. Moreover \cite{Aloisio:2015rsa} pointed out that, since the self-generated waves all move on average away from the disc and along the CR gradient, advection of such CRs with the waves may become important at low energies (below 10 GeV/n) and be responsible for the spectral hardening at such energies that seems to fit well the recent Voyager I data \cite{2013Sci...341..150S}. These phenomena are illustrated in Fig. \ref{fig:voyager} (from \cite{Aloisio:2015rsa}): the flux of protons as measured by PAMELA, AMS-02, CREAM and Voyager is shown, together with the prediction \cite{Aloisio:2015rsa} with (red solid line) and without (blue dotted line) the effect of solar modulation. A combination of advection with the self-generated waves and proton energy losses seems to provide a good description of the Voyager data in the ISM. 

\begin{figure}[t]
%\sidecaption[b]
\begin{center}
\includegraphics[scale=.45]{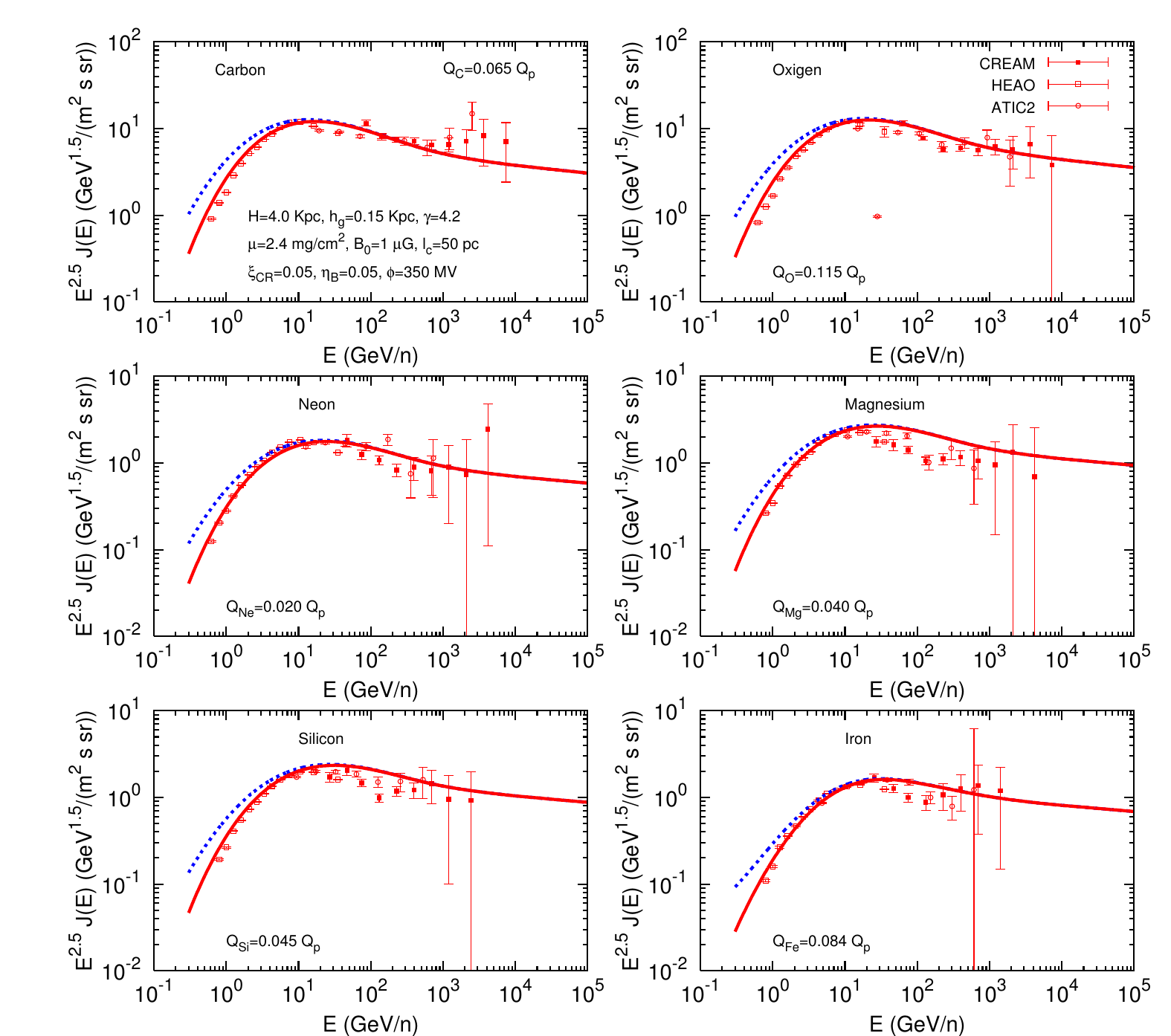}
\end{center}
\caption{Spectra of nuclei \cite{Aloisio:2013tda} in a model with self-generated waves and pre-existing turbulence, compared with available data.}
\label{fig:nuclei}       
\end{figure}

The calculations of \cite{Aloisio:2013tda,Aloisio:2015rsa} were extended to the spectra of all nuclei and provided a good description of their spectra, as illustrated in Fig. \ref{fig:nuclei}.

\begin{figure}[t]
%\sidecaption[b]
\begin{center}
\includegraphics[scale=.85]{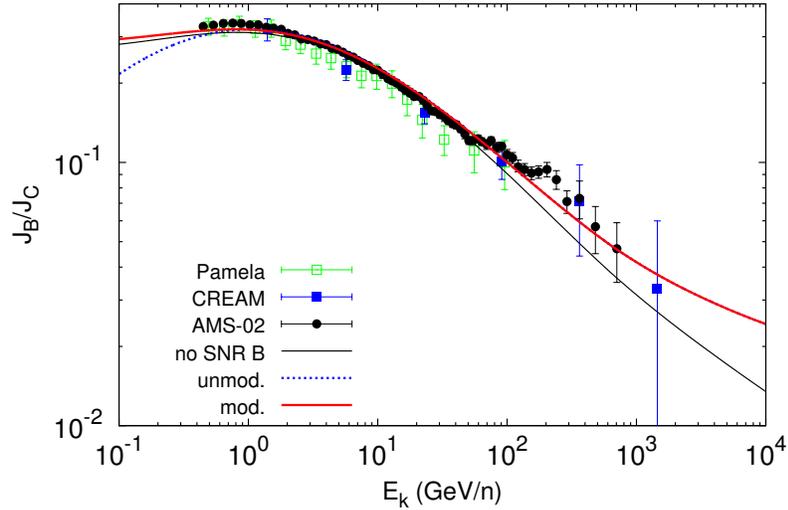}
\end{center}
\caption{B/C ratio (from \cite{Aloisio:2015rsa}) in a model with self-generated waves and pre-existing turbulence (black solid line), compared with available data. The red line is obtained by adding an energy-independent grammage of $\sim 0.17 \rm g cm^{-2}$ accumulated by CRs while being downstream of a supernova shock for $\sim 10^{4}$ years.}
\label{fig:BC}       
\end{figure}

Interestingly, despite the good agreement with the spectra of primary nuclei, the B/C ratio as calculated in Ref. \cite{Aloisio:2015rsa}, shows a slight excess at high energies (black solid line in Fig. \ref{fig:BC}). The red solid line in Fig. \ref{fig:BC} shows the B/C ratio obtained by accounting for a grammage of $\sim 0.17 \rm g cm^{-2}$ accumulated by CRs while being downstream of a supernova shock for $\sim 10^{4}$ years. This irreducible contribution seems to be required if to improve the fit to the observed B/C data (this conclusion, that was found by \cite{Aloisio:2015rsa} using preliminary AMS-02 data, seems to apply also to the recently AMS-02 data \cite{2016PhRvL.117w1102A} on the B/C ratio).

A spectral breaking at $\sim 200$ GV similar to the one discussed above may also be induced by a spatial dependence of the diffusion coefficient in the halo, as discussed by \cite{Tomassetti:2012ga}: if the diffusion coefficient is $D_{1}(p)\propto p^{\delta_{1}}$ closer to the disc, and $D_{2}(p)\propto p^{\delta_{2}}$ farther away, then the spectrum at the Earth is $\propto p^{-\alpha-\delta_{2}}$ at low energies, and $\propto p^{-\alpha-\delta_{1}}$ at high energy, if injection is $\propto p^{-\alpha}$. 

The propagation of CRs in the Galaxy leaves an imprint in measurable quantities other than the B/C ratio. The galactic emission of gamma rays reflects the density of hadronic and leptonic CRs and the density of gas and photons in the environment. In the disc of the Galaxy, most gamma ray emission is due to pp inelastic collisions, which result in neutral pion production and decays. If the density of gas is reliably traced, the detection of gamma radiation from a given line of sight results in a measurement of the local density of CRs. This analysis has been recently carried out by the \fermilat collaboration \cite{Acero:2016qlg} and revealed an interesting trend: 1) the density of CRs in the inner Galaxy (within a few kpc from the galactic center) is rather peaked where the density of SNRs is also observed to be peaked, while the CR density decreases very slowly with Galactocentric distance in the outer Galaxy; 2) the spectrum of CRs with energies $\leq 100$ GeV in the inner Galaxy is somewhat harder than in the outer regions. These findings were qualitatively confirmed by an independent analysis of the \fermilat data \cite{Yang:2016jda}. 

The weak dependence of the CR density on Galactocentric distance $R$ for $R>5$ kpc is the well known CR gradient problem \cite{Stecker:1977ph,Collaboration:2010cm,FermiLAT:2011lax}: the CR density drops much slower than proportional to the density of sources. These findings are difficult to reconcile with the standard approach to CR propagation, which is based upon solving the transport equation under the assumption that the diffusive properties are the same in the whole propagation volume \cite{Ginzburg:1990sk}. Within the context of this approach, several proposals have been put forward to explain the radial gradient problem. Among them:
{\it a)} assuming a larger halo size or 
{\it b)} a flatter distribution of sources in the outer Galaxy \cite{Collaboration:2010cm};
{\it c)} accounting for advection effects due to the presence of a galactic wind  \cite{1993A&A...267..372B};
{\it d)} assuming a sharp rise of the CO-to-H$_2$ ratio in the external Galaxy \cite{Strong:2004td};
{\it e)} speculating on a possible radial dependence of the injected spectrum \cite{Erlykin:2015sca}.
None of these ideas, taken individually, can simultaneously account for both the spatial gradient and the spectral behavior of CR protons. Moreover, many of them have issues in accounting for other observables \cite{Evoli:2012ha}.

A different class of solutions invoke the breakdown of the hypothesis of a spatially constant diffusion coefficient. For instance, \cite{Evoli:2012ha} proposed a correlation between the diffusion coefficient parallel to the galactic plane and the source density in order to account for both the CR density gradient and the small observed anisotropy of CR arrival directions. Ref. \cite{Gaggero:2014xla} followed the same lines of thought and showed that a phenomenological scenario where the transport properties (both diffusion and convection) are position-dependent can account for the observed gradient in the CR density. It is however unsatisfactory that these approaches do not provide a convincing physical motivation for the assumed space properties of the transport parameters.

In the context of models of CR transport with self-generated diffusion and advection the CR accumulation and the harder spectra in the inner Galaxy find a relatively simple explanation \cite{Recchia:2016bnd}: waves are excited more easily where there is more injection, so that the diffusion coefficient is correspondingly smaller and CRs are accumulated there for longer times resulting in a higher CR density. At the same time, advection with self-generated waves remains dominant up to higher energies, thereby implying harder CR spectra. One should keep in mind that so far the evidence for harder spectra in the inner Galaxy derives from low energy gamma ray observations. It is not clear if observations will show that this phenomenon holds even at higher energies. In that case, self-generated waves alone would not provide the full explanation of observations. 

\begin{figure}[t]
%\sidecaption[b]
\begin{center}
\includegraphics[scale=.28]{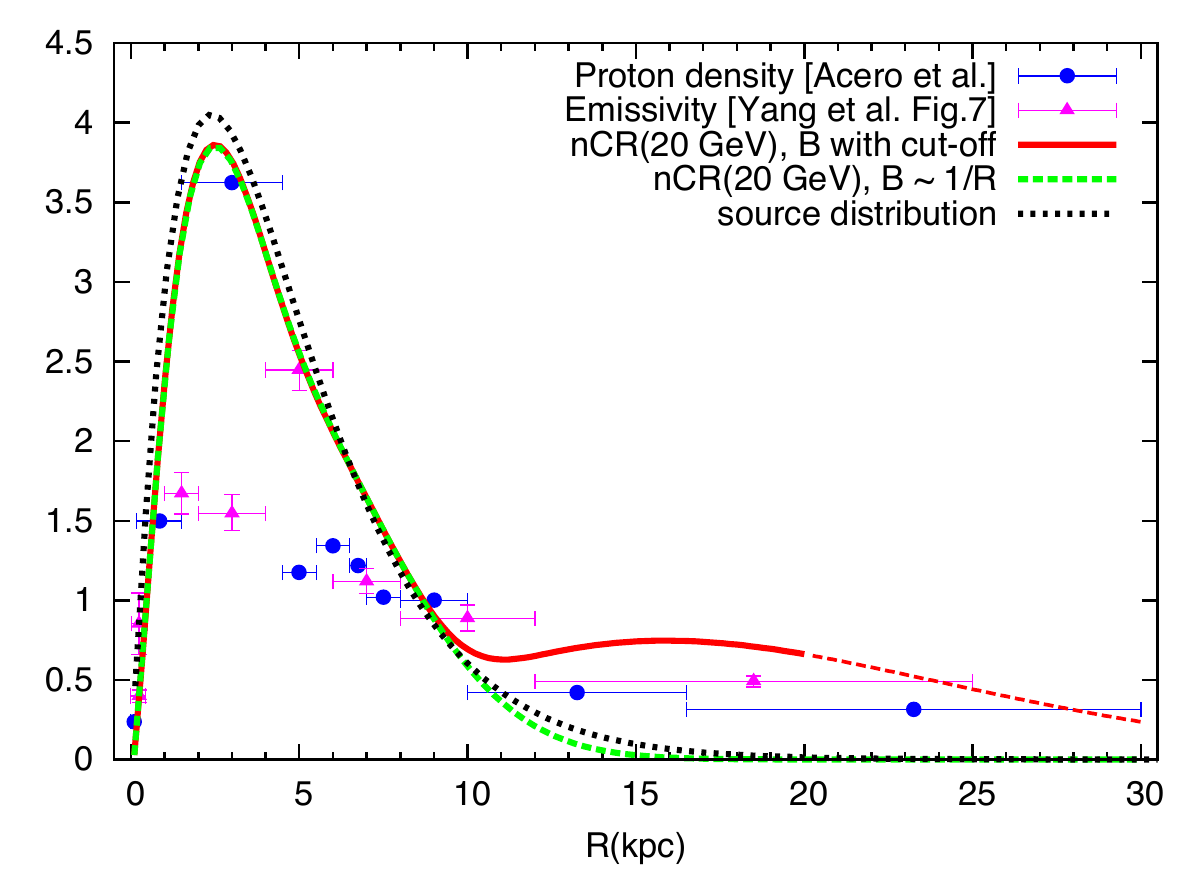}
\includegraphics[scale=.28]{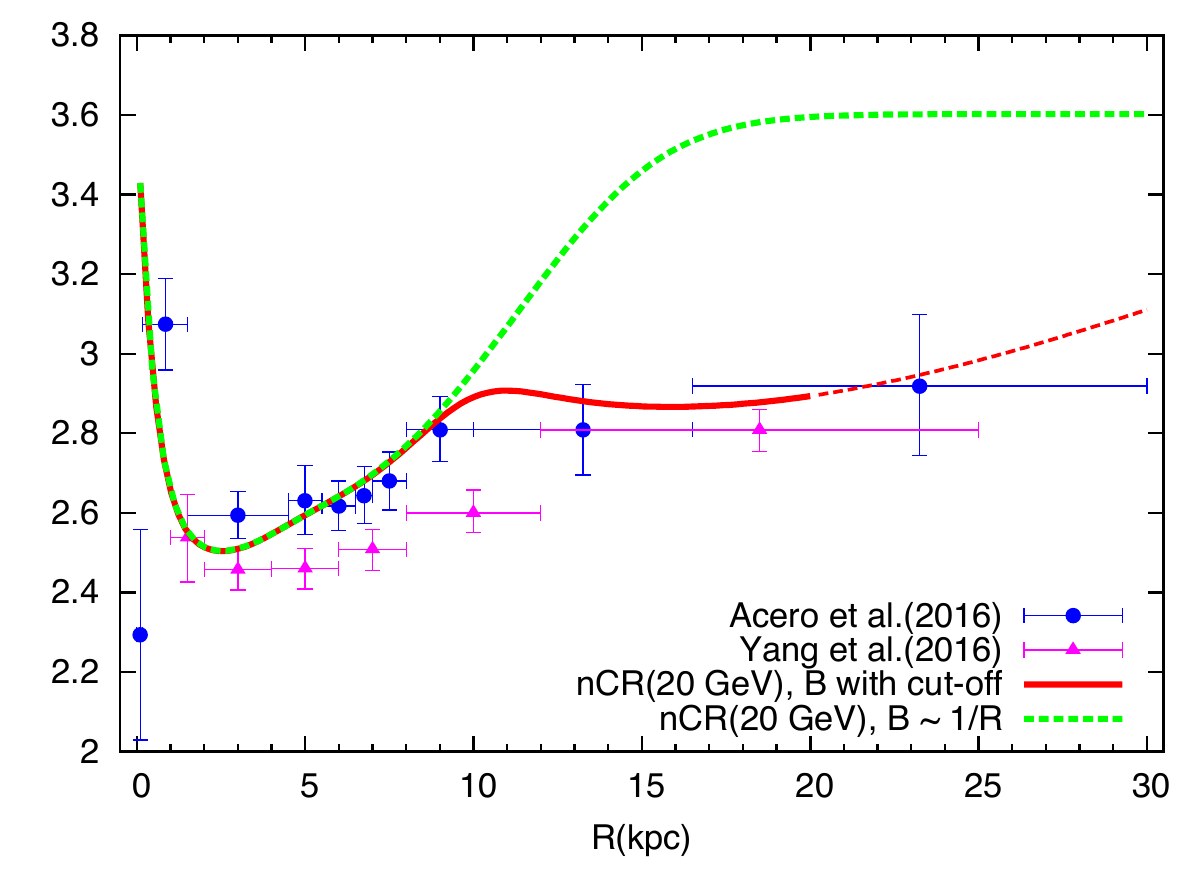}
\end{center}
\caption{Density (left panel) and spectral slope (right panel) of CRs with energy $\sim 20$ GeV as a function of the Galactocentric distance. The data points are from \cite{Acero:2016qlg} and \cite{Yang:2016jda}, while the curves refer to the results of \cite{Recchia:2016bnd}.}
\label{fig:slope}
\end{figure}

The CR density and spectral slope as found by \cite{Recchia:2016bnd} in the context of self-generated diffusion are shown in Fig. \ref{fig:slope}, together with the findings of the analysis of the \fermilat data. Both the density and slope are well fitted if an exponential cutoff is assumed in the spatial distribution of the background magnetic field at distances beyond $10$ kpc from the galactic center. In the absence of this magnetic field drop at large galactocentric distances, the CR gradient problem becomes even more severe in non-linear modes than it is in the standard model, because the density of CRs at large $R$ drops even faster as a result of the smaller density of sources and larger values of the diffusion coefficient.

Finally, we stress that, as discussed above, if the effect of an R-dependent slope is confirmed to exist at higher energies as well then an alternative explanation of the CR gradient problem should be sought.

\subsection{CR driven galactic Winds}
\label{sec:winds}

The possibility that a galaxy may launch winds has attracted attention for many different reasons. For instance, star formation is regulated by the amount of gas available, and winds modify the availability of such gas. In fact, galactic models that do not include feedback processes suffer from over-predicting the amount of baryons and star formation rates \cite{Crain:2006sb,Stinson:2012uh}. 
Winds also pollute galactic halos with hot dilute plasma that may provide an important contribution to the number of baryons in the Universe \cite{Kalberla:1998A&A...332L..61K,Kalberla:2008uu,Miller:2013nza}. Such gas might in fact have already been detected \cite{1995ApJ...454..643S} in the form of X-ray emitting plasma with temperature of several million degrees, and possibly associated with a galactic wind \cite{1994Natur.371..774B,Breitschwerdt:1999yd} (see also \cite{Everett:2007dw}). Finally winds can affect the transport of cosmic rays (CRs) in a galaxy, by advecting them away from their sources. 

Galactic winds may be thermally-driven, namely powered by core-collapse SNe \cite[]{Chevalier:1985pc} or momentum-driven, powered by starburst radiation \cite{Scoville:2003JKAS...36..167S,Murray:2004dd}. These two mechanisms of wind launching are thought to be at work in starburst galaxies and galaxies with active nuclei \cite{Veilleux:2005ia}. On the other hand in a galaxy like the Milky Way, winds are unlikely to be due to such processes because thermal and radiation pressure gradients are expected to be too small. A possible exception is the innermost part of the galactic Center region where the recently discovered Fermi Bubbles may originate from direct bursting activity of Sgr A$^*$ \cite{Cheng:2011xd,Zubovas:2011py} or past starburst activities \cite{Lacki:2013zsa}. On the other hand, CRs can play an important role in launching winds because of the gradient that their pressure develops as a consequence of the gradual escape of CRs from the Galaxy. The force $-\nabla P_{\rm CR}$ associated with such gradient is directed opposite to the gravitational force, and in certain conditions the plasma above and below the disc can be lifted off to form a CR driven wind. Notice that the gravitational force may be dominated by the dark matter component or the baryonic (gas and stars) components depending on the location. The force exerted by CRs depends in a complicated manner on the density of sources of CRs but also on non-linear processes of excitation of Alfv\'en waves through streaming instability (see discussion above). Both the force induced by CRs on the background plasma and the streaming instability induced by CRs depend on the gradient in CR density. In turn the distribution function of CRs is affected by their transport: diffusion is self-regulated through the production of Alfv\'en waves, and advection is determined by the velocity of the wind, if any is launched, and by the Alfv\'en waves' velocity, directed away from the sources of CRs. This complex interplay makes the problem non-linear. The first pioneering attempt to describe the hydrodynamics of a CR driven wind was described in Ref. \cite{Ipavich:1975p3566}, where the author used a spherically symmetric model of the Galaxy and considered only baryons and stars for the calculation of the gravitational potential. Later \cite{Breitschwerdt:1991A&A...245...79B} presented an extensive discussion of the hydrodynamics of CR driven winds: dark matter was included and a realistic geometry of the wind was considered, in which the launch takes place at some distance from the galactic disc and proceeds in a roughly cylindrical symmetry out to a distance of about $\sim 15$ kpc, where the flow opens up into a spherical shape. The calculations of \cite{Breitschwerdt:1991A&A...245...79B} treated CRs as a fluid, hence no information on the spectrum of CRs was retained. The important role of wave damping in the wind region was also discussed in \cite{Breitschwerdt:1991A&A...245...79B}, although only in the simplified case of a spherical outflow. As mentioned above, \cite{Breitschwerdt:1991A&A...245...79B} assumed that the wind is launched some distance away from the disc of the Galaxy and this assumption raises the issue of what happens in the region between the disc and the base of the wind, a problem of both mathematical and physical importance \cite{Breitschwerdt:1993p3640}. 

The dynamical role of CRs in launching winds was also studied via purely hydrodynamical simulations \cite{Uhlig:2012rt,Booth:2013asa,Salem:2013npa} and through MHD simulations \cite{Girichidis:2016ApJ...816L..19G,Peters:2015yaa,Ruszkowski:2016jzu}. These simulations, with their progressive level of sophistication, demonstrated that CRs play an important role in wind launching. Nevertheless, all this bulk of work treated CRs as a fluid, thereby not providing any information on the CR spectrum.

At present, the only two attempts at calculating the spectrum of galactic CRs in the presence of CR driven winds in a self-consistent manner were made by \cite{Ptuskin:1997A&A...321..434P} and \cite{Recchia:2016ylf}. The analytical approach put forward by \cite{Ptuskin:1997A&A...321..434P} is illuminating in terms of understanding the basic physical aspects of CR driven winds and CR transport in such words, although the conclusions may be rather different in more realistic winds, as discussed by \cite{Recchia:2016ylf}. 

In the cases studied in Ref. \cite{Ptuskin:1997A&A...321..434P}, the Alfv\'en velocity at the base of the wind is larger than the wind launching speed $v_{w}$, and increases linearly with the distance $z$ from the galactic disc. There is a distance $s_{*}(p)$ where the time scale for advection and diffusion are equal:
\be
\frac{s_{*}^{2}}{D(p)}=\frac{s_{*}(p)}{v_{A}+u_{w}}\approx \frac{s_{*}(p)}{v_{A}}\propto \sim\rm constant,
\ee
since $v_{A}\propto z\sim s_{*}$. This implies that the critical distance $s_{*}(p)$, as a function of the momentum $p$ scales as $s_{*}(p)\propto D^{1/2}(p)$. In CR driven wind models of the origin of CRs, the distance $s_{*}$ plays the role of the halo size $H$ in the standard leaky box model. By analogy, the spectrum of CRs in the disc can be written as $f(p)\propto Q(p)s_{*}(p)/D(p)\propto Q(p)/D^{1/2}(p)$, which is quite different from the standard result $f(p) \propto Q(p)/D(p)$. 

Although these scalings are very useful to pin down the essential physical ingredients of the problem, they do not fully reflect the complexity of the CR transport in winds: as recently discussed in \cite{Recchia:2016ylf}, most hydrodynamical wind solutions lead to CR spectra at the Earth which are quite unlike the observed ones. The main reason for this result is that the advection velocity at the base of the wind is usually found to be around $\sim 100$ km/s, which leads to advection dominated transport even at relatively high energies, say in the TeV range, at odds with the observed CR spectra. Moreover, standard wind solutions are characterised by a spectral softening at energies above $\sim $TeV, rather than the observed hardening found by PAMELA and AMS-02. On the other hand, \cite{Recchia:2016ylf} pointed out that the region between the base of the wind and the disc, where the sources are assumed to be localised, is crucial for the determination of the spectrum: for instance, assuming that the wind is actually launched at $z_{0}\sim 1$ kpc from the disc and that for $|z|<z_{0}$ the diffusion coefficient is fixed (not self-generated) with a Kolmogorov-like shape, the overall structure of the CR spectrum at the Earth may be recovered: a low energy hardening due to advection is visible, a steep spectrum at energies $20\leq E\leq 1000$ GeV is produced by the self-generation of waves in the wind region, and finally at high energy a hardening is produced as due to the dominance of the Kolmogorov spectrum upon the spectrum of self-generated waves. 

\subsection{Positrons and antiprotons}
\label{eq:positrons}

The ratios $e^{+}/(e^{-}+e^{+})$ and $\bar p/p$ are often used to infer the propagation properties of galactic CRs. If positrons and antiprotons are solely produced in hadronic interactions of primary CRs with ISM gas, it is easy to demonstrate that both ratios should be decreasing functions of energy, at least for energies high enough that radiative energy losses of electrons dominate their propagation and solar modulation can be neglected. These conditions are typically satisfied for energies above $\sim 10$ GeV. For both ratios, it is expected that they decrease with energy as $\sim 1/D(E)$, where $D(E)$ is the diffusion coefficient of particles with energy $E$ (see, for instance, Ref. \cite{Serpico:2011wg} and \cite{Serpico:2015caa} for a recent review), although the uncertainties in the cross section of antiproton production \cite{2014PhRvD..90h5017D} may affect such conclusion as far as the $\bar p/p$ ratio is concerned.

The PAMELA experiment measured the positron ratio and found that it grows with energy for $E>10$ GeV \cite{Adriani:2008zr}, at least up to $\sim 100$ GeV. A similar trend was also obtained by analyzing lepton fluxes from the Fermi-LAT telescope \cite{Ackerman:2012ph}. These results were later confirmed and extended to higher energies by AMS-02 \cite{Aguilar:2013qda}. The PAMELA \cite{Adriani:2011xv,Adriani:2013uda} and AMS-02 \cite{2014PhRvL.113l1102A} measurements of the separate fluxes of electrons and positrons showed that the increasing trend in the ratio $e^{+}/(e^{-}+e^{+})$ is due to an excess of positrons rather than a deficit of electrons, at odds with the simplest interpretation of positrons as secondary products of hadronic interactions.

As discussed in \S \ref{sec:observ} (see also \cite{Serpico:2015caa}), the $\bar p/p$ does not show any rise with energy, thereby suggesting that whatever the sources of positrons, they should not produce appreciable amounts of antiprotons. However, it has been suggested that the recent AMS-02 measurement of the flux of antiprotons \cite{2016PhRvL.117i1103A}, which extends the previous measurements to higher energies, indicate that the spectrum of $\bar p$ is harder than expected based on the standard model of CR transport. This conclusion is, at present, questionable, in that the uncertainties in the cross sections of $\bar p$ production and in the parameters of CR transport to be adopted do not allow to reach a definite conclusion in this matter. Nevertheless, several authors noticed the intriguing similarity between the spectrum of antiprotons and those of positrons and protons (see for instance \cite{Lipari:2016vqk} and references therein) and postulated that both positrons and antiprotons might still be the result of CR interactions provided the model of CR transport is changed in a suitable way. We will comment further on these models below.

The discovery of an increasing positron ratio stimulated much interest in dark matter annihilation as the source of the {\it excess} positrons. Such phenomenon typically leads to copious production of pions and finally gamma rays, electrons and positrons and, to a lesser extent, hadrons. The explanation of the positron excess in terms of dark matter annihilation requires the dark matter candidate to have peculiar properties: it should be leptophylic (otherwise the $\bar p/p$ would be affected), and should be characterized by Sommerfeld enhancement of the cross section (in order to explain the normalization of the positron spectrum). In addition, the contribution of clumps of dark matter in the Milky Way's halo should be prominent, in order to account for the observed flux of positrons at the Earth. As discussed in detail in Ref. \cite{Serpico:2011wg}, these conditions appear to be rather {\it ad hoc}, and an explanation of the positron excess in terms of dark matter annihilation is, at present, considered as disfavored, at least by these authors. 

Several astrophysical explanations of the excess of CR positrons have also been put forward. It has been suggested that the positron excess may be due to old SNRs \cite{Blasi:2009hv}: the idea is that electrons and positrons are also produced as secondary products of hadronic interactions inside accelerators, such as SNR shocks. The peculiarity of these electrons and positrons is that, since they are produced inside the acceleration region, they also take part in the acceleration process which leads to particularly hard spectra, required to explain the observed  $e^{+}/(e^{-}+e^{+})$ ratio. It was soon realized that the same fate would occur to antiprotons \cite{Blasi:2009bd} and to nuclear secondaries, such as Boron \cite{Mertsch:2009ph}: in both the $\bar p/p$ ratio and B/C ratio one would expect a rising trend at sufficiently high energies. The measurements of the B/C ratio and of the $\bar p/p$ ratio by PAMELA and AMS-02 experiments have shown no sign of such rise, thereby providing strong constraints on the applicability of such model.

Electron-positron pairs are also copiously produced in pulsar magnetospheres. In the vicinity of pulsars the electromagnetic fields are so intense that pair cascades develop with very high multiplicity: each electron extracted from the star surface typically produces $\sim 10^{4}-10^{6}$ $e^+$-$e^-$ pairs. Aside from being theoretically predicted, direct evidence of this phenomenon is provided by multi-wavelength observations of Pulsar Wind Nebulae (PWNe), bright synchrotron and IC nebulae surrounding many young pulsars (see {\it e.g.} \cite{amato14} for a review). Since these particles will have to be released in the ISM at some point, their contribution to CR leptons is unavoidable and must be taken into account in any model aimed at explaining the positron excess.  In fact, it has been argued \cite{Hooper:2008kg,2012CEJPh..10....1P} that PWNe could be the most important sources of the excess positrons observed by PAMELA and AMS-02 (the possibility that pulsars could be sources of CR positrons had been put forward long before the recent developments in the field \cite{1995A&A...294L..41A}).

The $e^+$-$e^-$ pairs created in the pulsar magnetosphere become part of the relativistic wind into which pulsars convert most of their rotational energy. The interaction between the wind and the surrounding medium, the SNR during early stages and the ISM later on, is what makes the PWN shine: a shock develops from this interaction and propagates towards the pulsar down to a distance that guarantees pressure equilibrium between the unshocked wind and the downstream nebula. Extremely efficient particle acceleration occurs at this shock: long power-law spectra extending from about 1 GeV to even 1 PeV are formed and the radiation of these particles in the ambient magnetic field directly reveals the accelerated particle spectrum. The $e^+$-$e^-$ pairs are seen to be described by a flat spectrum ($N(E)\propto E^{-\gamma}$ with $1<\gamma<2$), at low energies, which then steepens to $\gamma>2$ beyond a few hundred GeV. The hard lepton spectrum at low energies is extremely appealing in terms of explaining the CR positron spectrum if, as expected, at some point in the history of the PWN these particles are released into the ISM. What needs to be assessed is the effective rate of release of the pairs and their potential contribution to the CR spectrum.  

Most of the positrons produced in a PWN are likely to be confined in the nebula for long times and perhaps lose their energy there, but pairs produced after the pulsar escapes the parent remnant and forms a bow shock nebula could leak into the ISM and account for the positron excess \cite{Blasi:2010de}. In this model, no antiprotons are expected to accompany the pairs, hence the absence of a rise in the $\bar p/p$ ratio is easily accounted for. The pulsar explanation of the positron excess leads to expecting several interesting and potentially observable effects, that are currently being investigated (see for instance \cite{Linden:2013mqa}).

As anticipated above, the approximate similarity of the high energy spectra of positrons, antiprotons and protons has stimulated some speculations on the possibility that $e^{+}$ and $\bar p$ could be solely secondary products of CR interactions \cite{2013PhRvL.111u1101B,2010MNRAS.405.1458K,Lipari:2016vqk}, by invoking modifications of the standard model of CR transport. 

In Ref. \cite{Lipari:2016vqk}, it is noticed that the observed $e^{+}$ and $\bar p$ production rates are in perfect accord with those calculated from standard CR interactions in the ISM and that their spectral shape is consistent with being the same as that of protons at energies above $\sim 300$ GeV, thereby supporting the idea that the observed flux of antiprotons and positrons may be interpreted in terms of CR interactions. On the other hand, in the standard model of CR transport, where the grammage is inferred from the B/C ratio, the propagation of leptons (electrons and positrons) is dominated by radiative losses for energies above $\sim 10$ GeV, which steepen the spectrum of leptons and lead to a spectral difference with antiprotons. It follows that the conclusion of Ref. \cite{Lipari:2016vqk} can only be considered as potentially viable if energy losses are negligible, which requires a residence time in the Galaxy much smaller than usually assumed. This also implies that in this model the boron production must be decoupled from the production of other secondaries, which is not very appealing from the theoretical point of view, although one implementation of this scenario is already present in the literature, the so-called nested leaky box model \cite{2010PhRvD..82b3009C,2016ApJ...827..119C} (NLB). 

The basic assumption of the NLB model is that CRs accumulate most grammage in {\it cocoons} around the sources, while lesser grammage is accumulated during propagation throughout the Galaxy. The former grammage is assumed to be energy dependent while the latter is assumed to be energy independent. By construction, the two values of the grammage become comparable around few hundred GeV/n. Since secondary Boron nuclei are produced by primary CRs (mainly Carbon and Oxygen) at the same energy per nucleon, the B/C ratio reflects the energy dependence of the grammage at the same energy, hence one should expect that the B/C should be decreasing with energy (reflecting the near source grammage) below $100$ GeV/n and become energy independent at higher energies. Since the $e^{+}$ and $\bar p$ production is characterized by a large inelasticity ($E_{e^{+},\bar p}\sim 0.1 E_{p}$ with $E_{p}$ the proton energy), the production of $e^{+}$ and $\bar p$ at energy $\sim 10-100$ GeV reflects the grammage  traversed by CR at $0.1-1$ TeV, where the grammage is assumed to be flat in the context of the NLB model. One should keep in mind that the recent paper by AMS-02 on the measurement of the B/C ratio up to $\sim 1$ TeV/n did not find evidence for a flattening of the ratio at the highest energies. The idea of cocoons around sources, presented in Ref. \cite{2010PhRvD..82b3009C} as a speculative possibility, might find a theoretical justification in the mechanism of CR self-confinement near sources discussed in Ref. \cite{marta}).

\section{Acceleration of galactic CRs}
\label{sec:accelera}

There are several aspects of the problem of the origin of CRs that are tightly linked to the physical processes responsible for particle energization in astrophysical sources: 1) why are there non-thermal particles in the first place? 2) what are the physical mechanisms through which nature energizes a small fraction of the particles in a plasma to non-thermal energies? 3) What is the spectrum of the accelerated particles? 4) What is the maximum energy that particles can be accelerated to? 

In terms of energetics of CR injection in the Galaxy, once the confinement time of CRs has been normalized to the B/C ratio, the only class of sources that are left as plausible sources of CRs are supernovae, with typical efficiency $\sim 5-20\%$. The question of which types of SNe contribute the most is very complex, and it is probably dependent upon the energy of CRs we are interested in. 

The mechanism of particle acceleration that is expected to account for the required acceleration efficiency is diffusive shock acceleration (DSA) at the forward shock that accompanies the supersonic motion of the plasma associated with the SN explosions. The theory of DSA was initially developed in \cite{1977DoSSR2341306K,Blandford:1978ky,1977ICRC11.132A,Bell:1978zc,Bell:1978fj} in the so-called test particle regime. 

Let us consider a shock front characterized by a Mach number $M_{s}$. The compression factor at the shock is $r=u_{1}/u_{2}$ where $u_{1}$ and $u_{2}$ are the plasma velocities upstream and downstream of the shock respectively. The compression factor can be expressed in terms of the Mach number using conservation of mass, momentum and energy at the shock:
\be
r=\frac{4 M_{s}^{2}}{M_{s}^{2}+3},
\ee
which tends to 4 in the limit of strong shocks, $M_{s}\to \infty$. A test particle diffusing in the upstream or downstream plasma alone does not gain or lose energy (although the second order Fermi process discussed above may be at work). 

For a stationary parallel shock, namely a shock for which the normal to the shock is parallel to the orientation of the background magnetic field, the transport of particles is described by the diffusion-convection equation (see for instance \cite{Blandford:1978ky}), which in the shock frame reads:
\be
u \frac{\partial f}{\partial z} = \frac{\partial}{\partial z}\left[ D \frac{\partial f}{\partial z}\right] + \frac{1}{3} \frac{du}{dz} p \frac{\partial f}{\partial p} + Q,
\label{eq:transport}
\ee
where $f(z,p)$ is the distribution function of accelerated particles, normalized in a way that the number of particles with momentum $p$ at location $z$ is $\int dp 4\pi p^{2} f(p,z)$. In Eq. \ref{eq:transport} the LHS is the convection term, the first term of the RHS is the spatial diffusion term. The second term on the RHS describes the effect of fluid compression on the accelerated particles, while $Q(x,p)$ is the injection term. 

A few comments on Eq. \ref{eq:transport} are in order: 1) the shock will appear in this equation only in terms of a boundary condition at $z=0$, and the shock is assumed to have infinitely small size along $z$. This implies that this equation cannot properly describe the thermal particles in the fluid. The distribution function of accelerated particles is continuous across the shock. 2) In a self-consistent treatment in which the acceleration process is an integral part of the processes that lead to the formation of the shock one would not need to specify an injection term. Injection would result from the microphysics of the particle motions at the shock. 

For the purpose of the present discussion we assume that injection only takes place at the shock surface, immediately downstream of the shock, and that it only consists of particles with given momentum $p_{inj}$:
\be
Q(p,x)=\frac{\eta n_{1}u_{1}}{4\pi p_{inj}^{2}}\delta(p-p_{inj})\delta(z)=q_{0} \delta(z),
\label{eq:inj}
\ee
where $n_{1}$ is the fluid density upstream of the shock and $\eta$ is the acceleration efficiency, defined here as the fraction of the incoming number flux across the shock surface that takes part in the acceleration process. 

The compression term vanishes everywhere but at the shock since $du/dz=(u_{2}-u_{1})\delta(z)$. Integration of Eq. \ref{eq:transport} around the shock surface (between $z=0^{-}$ and $z=0^{+}$) leads to:
\be
\left[D\frac{\partial f}{\partial z}\right]_{2} - \left[D\frac{\partial f}{\partial z}\right]_{1} + \frac{1}{3} (u_{2}-u_{1}) p \frac{df_{0}}{dp} + q_{0}(p) = 0,
\label{eq:boundary}
\ee
where $f_{0}(p)$ is now the distribution function of accelerated particles at the shock surface. Particle scattering downstream leads to a homogeneous distribution of particles, at least for the case of a parallel shock, so that $\left[\partial f/\partial z\right]_{2}=0$. In the upstream region, where $du/dz=0$ the transport equation reduces to:
\be
\frac{\partial}{\partial z}\left[ u f - D\frac{\partial f}{\partial z} \right]=0,
\ee
and since the quantity in parenthesis vanishes at upstream infinity, it follows that 
\be
\left[D\frac{\partial f}{\partial z}\right]_{1} = u_{1} f_{0}.
\ee
Using this result in Eq. \ref{eq:boundary} we obtain an equation for $f_{0}(p)$
\be
u_{1} f_{0} = \frac{1}{3}(u_{2}-u_{1}) p \frac{d f_{0}}{dp} + \frac{\eta n_{1}u_{1}}{4\pi p_{inj}^{2}}\delta(p-p_{inj}),
\ee
which is easily solved to give:
\be
f_{0}(p) = \frac{3r}{r-1} \frac{\eta n_{1}}{4\pi p_{inj}^{2}} \left( \frac{p}{p_{inj}}\right)^{-\frac{3r}{r-1}}.
\ee
The spectrum of accelerated particles is a power law in momentum (not in kinetic energy) with a slope $\alpha$ that only depends on the compression ratio $r$:
\be
\alpha=\frac{3r}{r-1}.
\ee
The slope tends asymptotically to $\alpha=4$ in the limit $M_{s}\to \infty$ of an infinitely strong shock front. The number of particles with energy $\epsilon$ is $n(\epsilon)d\epsilon = 4\pi p^{2} f_{0}(p)(dp/d\epsilon) d\epsilon$, therefore $n(\epsilon)\propto \epsilon^{-\alpha}$ for relativistic particles and $n(\epsilon) \propto \epsilon^{(1-\alpha)/2}$ for non-relativistic particles. In the limit of strong shocks, $n(\epsilon)\propto \epsilon^{-2}$ ($n(\epsilon)\propto \epsilon^{-3/2}$) in the relativistic (non-relativistic) regime. 

Even in the pioneering work of Refs. \cite{Bell:1978zc,Bell:1978fj} it was already recognized that for acceleration efficiency $\sim 10\%$ non-linear effects would become important. A full non-linear theory of DSA was developed later, first in the hydrodynamical limit (two fluid models) and then in kinetic models (see \cite{Malkov:2001kya} for a complete review of these approaches). 

The main effects of this non-linearity can be summarised as follows:

\begin{itemize}
\item[1)] {\it Dynamical reaction of accelerated particles}

For typical efficiencies of CR acceleration at a SN shock, $\sim 10\%$, the pressure exerted by accelerated particles on the plasma around the shock affects the shock dynamics as well as the acceleration process. The dynamical reaction that accelerated particles exert on the shock is due to two different effects: {\it a}) the pressure in accelerated particles slows down the incoming upstream plasma as seen in the shock reference frame, thereby creating a precursor. In terms of dynamics of the plasma, this leads to a compression factor that depends on the location upstream of the shock \cite{1981ApJ.248.344D,1982A&A.111.317A}. {\it b}) The escape of the highest energy particles from the shock region makes the shock {\it radiative-like} \cite{1999ApJ.526.385B}, thereby inducing an increase of the compression factor between upstream infinity and downstream. Both these effects result in a modification of the spectrum of accelerated particles, which turns out to be no longer a perfect power law \cite{Berezhko:1994cj,Berezhko:1997yw,1999ApJ.526.385B,Malkov:1998nq,Blasi:2001wi}. Moreover, the fact that a sizeable fraction of the ram pressure at the shock is converted to accelerated particles implies that the temperature of the shocked gas is lower than in the absence of particle acceleration. 

\item[2)] {\it Plasma instabilities induced by accelerated particles}

As discussed above, SNRs can be the source of the bulk of CRs in the Galaxy, up to rigidities of order $\sim 10^{6}$ GV only if substantial magnetic field amplification takes place at the shock surface. Since this process must take place upstream of the shock in order to reduce the acceleration time, it is likely that it is driven by the same accelerated particles, which would therefore determine the diffusion coefficient that describes their motion. The existence of magnetic field amplification is also the most likely explanation of the observed bright, narrow X-ray rims of non-thermal emission observed in virtually all young SNRs (see \cite{Vink:2011ei,Ballet:2005jn} for recent reviews). The non-linearity here reflects in the fact that the diffusion coefficient becomes dependent upon the distribution function of accelerated particles, which is in turn determined by the diffusion coefficient in the acceleration region.

\item[3)] {\it Dynamical reaction of the amplified magnetic field}

The magnetic fields required to explain the X-ray filaments are of order $100-1000 \mu G$. The magnetic pressure is therefore still a fraction of order $10^{-2}-10^{-3}$ of the ram pressure $\rho v_{s}^{2}$ for typical values of the parameters. However, the magnetic pressure may easily become larger than the upstream thermal pressure of the incoming plasma, so as to affect the compression factor at the shock. A change in the compression factor affects the spectrum of accelerated particles which in turn determines the level of magnetic field amplification, another non-linear aspect of DSA.

\end{itemize}

The dynamical reaction of accelerated particles leads to concave spectra, and above $\sim 10-100$ GeV are harder than $p^{-4}$, the standard prediction of DSA in the test particle approximation. This simple finding refers to the instantaneous spectrum at a given time in the evolution of a SN shock. The time integrated spectrum may be somewhat different, but in general it is not easy to make it steeper than $p^{-4}$. In addition, one should notice that the CR spectrum released into the ISM by an individual SNR, integrated over the injection history, is made of two contributions: the spectrum of particles accelerated at the shock and advected downstream, and eventually released after the remnant ends its evolution, and the spectrum of particles that escape at the maximum momentum at any given time. This point was discussed in detail in \cite{Caprioli:2009fv}. 

The hardness of the spectra of injected CRs into the ISM is somewhat of an issue for the SNR paradigm: in fact the rule of thumb that the spectrum observed at the Earth should be $\sim E^{-\gamma-\delta}$, where $\delta$ identifies the energy dependence of the diffusion coefficient, would imply that in order to have, at Earth, a slope $\sim 2.6$, a diffusion coefficient $\propto E^{0.7\div 0.8}$ would be required. In turn, this would, most likely cause problems with anisotropy \cite{Blasi:2011fm}, although some solutions may be devised to mitigate such a problem \cite{Mertsch:2014cua}.

Observations of gamma ray emission from young SNRs also suggests that the spectra of accelerated particles are somewhat steeper than predicted by diffusive shock acceleration \cite{Damiano:2011tp,Acero:2015prw}. However, it should be stressed that such spectra may also reflect a complex morphology of the emission region (see \cite{Berezhko:2012av} for an implementation of this idea to the case of the Tycho SNR) or the presence of neutral material in the acceleration region \cite{Blasi:2012xe,Morlino:2015jvq}. Finally this phenomenon might be due to subtle aspects of DSA, such as the role played by the velocity of scattering centers in the acceleration process, as discussed in Refs. \cite{Caprioli:2009fv,Caprioli:2012cd} and in Ref. \cite{Morlino:2011di} for the case of the Tycho SNR.

\subsection{The quest for PeVatrons}
\label{sec:pevatrons}

The name {\it PeVatron} is used to indicate an astrophysical source able to accelerate protons to energies of order $\sim 1$ PeV (nuclei with charge $Z$ would therefore be accelerated to Z times larger energies). The need for the existence of PeVatrons mainly derives from the empirical fact that at the knee the mass composition of CRs is dominated by light nuclei. As discussed in \S \ref{sec:observ}, this piece of observation has recently been questioned, in that some experiments show evidence of a knee-like structure in the spectrum of light nuclei at $\sim 700$ TeV, well below the knee. This finding is at odds with the KASCADE measurement  of the spectrum in the same energy region, and the nature of this difference is not clear at present. 

As discussed by many authors, the assumption that CR protons may accelerate protons up to a few PeV may lead to a satisfactory description of both the knee and the transition region between galactic and extragalactic CRs. Most members of the community would agree that at least a class of sources must accelerate CRs up to PeV energies, while it is less clear whether such sources are also the ones that provide the main contribution to the CR spectrum below the knee. These issues were discussed, for instance, in Refs. \cite{Schure:2013kya,Bell:2013kq,Cardillo:2015zda}. 

What are the conditions needed for particle acceleration to $\sim$PeV energies? As discussed early on \cite{Lagage:1983p1347}, even in case of magnetic field amplification $\delta B/B\sim 1$ and Bohm diffusion, the maximum energy in a typical SNR is unlikely to be higher than $\sim 100$ TeV, thereby failing to reach the knee by more than one order of magnitude.  

The situation evolved quite abruptly in the last decade or so, mainly because of two findings: 1) X-ray observations of young SNRs led to the discovery that virtually all of them are characterised by thin filaments of non-thermal emission, to be interpreted as the result of synchrotron emission of high energy electrons \cite{Vink:2011ei}. In the assumption of Bohm diffusion, the thickness of such filaments can be used to infer the strength of the magnetic field in the emission region, and in all cases this estimate returns $B\sim 100-1000\mu G$, about $\sim 100$ times larger than the typical magnetic field in the ISM. Such finding can be only explained as a result of efficient magnetic field amplification, which may be due to hydrodynamical instabilities \cite{Giacalone:2007p962} and/or to CR induced instabilities. 2) Investigation of the streaming instability induced by CRs at a SNR shock led \cite{Bell:2004p737} to find a non-resonant branch of quasi-purely growing modes with a growth rate much higher than the resonant modes studied in \cite{Lagage:1983p1347}. These modes could explain the large magnetic fields at SNR shocks, and possibly lead to higher values of the maximum energy of accelerated particles. 

There are however some subtle features concerning these fastly growing modes that need to be taken into account if to assess their importance for particle acceleration \cite{Bell:2004p737,Schure:2013kya,Bell:2013kq}. From the physical point of view, the instability is due to the return current generated by electrons in the background plasma in order to compensate the CR current $J_{CR}$ upstream of the shock. Hence the scales that grow the fastest are the ones on very small spatial scales (large wavenumber $k$). The growth rate of the fastest modes can be written as:
\begin{equation}
\gamma_{M} = k_{M} v_{A},
\label{eq:gmax}
\end{equation}
where $v_{A}$ is the Alfv\'en speed in the unperturbed magnetic field $B_{0}$. The wavenumber where the growth is the fastest can also be easily estimated using the condition 
\begin{equation}
k_{M} B_{0}\cong\frac{4\pi}{c}j_{CR},
\label{eq:Btension}
\end{equation}
which corresponds to balance between current and magnetic tension. For a spectrum of particles accelerated at the shock $\propto p^{-4}$, the above expression can be rewritten as:
\begin{equation}
k_{M}r_{L}=\xi_{CR}\frac{1}{\Lambda}\left(\frac{v_{sh}}{V_{A}}\right)^{2}\left(\frac{v_{sh}}{c}\right)\gg 1,
\label{eq:non_resonance}
\end{equation}
where $r_{L}$ is the gyration radius of the particles dominating the current, $\Lambda=\ln\left(\frac{p_{max}}{m_{p}c}\right)$ and $\xi_{CR}$ is the CR acceleration efficiency. Eq. \ref{eq:non_resonance} illustrates well the fact that the fast growing modes are the ones that grow on scales much smaller than the Larmor radius of the particles. Since the resonance condition $k r_{L}^{-1}=1$ cannot be achieved, scattering of particles off these waves is not effective, hence it might seem at first sight that the excitation of these modes should not appreciably impact the scattering of particles near the shock, and not help achieving higher values of the maximum momentum $p_{max}$. However, this conclusion only applies to the linear growth of the modes, while at later stages the situation becomes more interesting. 

An element of background plasma is subject to a force $\sim \frac{1}{c}j_{CR}\delta B$ due to the exponentially growing magnetic field $\delta B$, so that within the growth time $\gamma_{M}^{-1}$ the fluid element is displaced by 
\begin{equation}
\delta x \approx \frac{J_{CR}\delta B}{c \rho \gamma_{M}^{2}}.
\end{equation}
When the loops of magnetic fields get stretched by $\delta x\sim r_{L}$ (where now the Larmor radius is calculated in the amplified magnetic field), the spatial scale in the magnetic field becomes sufficient to cause particle scattering and the current gets destroyed. Hence, one can envision the condition $\delta x\sim r_{L}$ as the saturation condition for the instability. The corresponding value of the magnetic field is given by 
\begin{equation}
\frac{\delta B^{2}}{4\pi} \approx \frac{\xi_{CR}}{\Lambda} \rho v_{s}^{2} \frac{v_{s}}{c}.
\label{eq:equi}
\end{equation}
The right end side of Eq. \ref{eq:equi} represents the energy density of accelerated particles escaping the accelerator from upstream, so that Eq. \ref{eq:equi} suggests that the saturation level is reached when equipartition between magnetic energy density (LHS) and energy density of escaping particles (RHS) is reached. The implications of this finding are potentially very impressive: particles at a given maximum energy $E_{M}$ escape the accelerator because there are no scattering centers able to scatter them back to the shock; however, the growth of the waves that they excite leads to particles of the same energy at a later time to be confined to the shock region, thereby reaching higher energy. 

The maximum energy that can be reached at a SNR shock as due to the mechanism above has been calculated by several authors \cite{Schure:2013kya,Bell:2013kq,Cardillo:2015zda}. For a type Ia SN with ejecta mass $M_{ej}$ and total kinetic energy $E_{SN}$, exploding in the ISM with density $n_{ISM}$, the maximum energy can be written as 
\be
E_{M}\cong 130\ \left(\frac{\xi_{CR}}{0.1}\right) \left(\frac{M_{ej}}{M_\odot}\right)^{-\frac{2}{3}} \left(\frac{E_{SN}}{10^{51}\rm{erg}}\right) \left(\frac{n_{ISM}}{\rm{cm^{-3}}}\right)^{\frac{1}{6}}\ TeV.
\label{eq:emaxI}
\ee
Despite the efficient magnetic field amplification due to CR streaming, the maximum energy falls short of the knee by about one order of magnitude, thereby leading to the conclusion that type Ia SNe are unlikely to act as PeVatrons.

Most core collapse SNe on the other hand explode in the wind produced by their pre-supernova red giant progenitor. For a reference value of the mass loss rate $\dot M\sim 10^{-5} M_\odot \rm{yr^{-1}}$ and wind speed $V_{w}\sim 10$ km/s, the maximum energy can be estimated as
\be
E_{M}\approx 1\ \left(\frac{\xi_{CR}}{0.1}\right)\left(\frac{M_{ej}}{M_\odot}\right)^{-1} \left(\frac{E_{SN}}{10^{51}\rm{erg}}\right) \left(\frac{\dot M}{10^{-5} M_\odot \rm{yr^{-1}}}\right)^{\frac{1}{2}} \left(\frac{V_w}{10\,\rm{km\, s^{-1}}}\right)^{-\frac{1}{2}}\ PeV\ .
\label{eq:emaxII}
\ee
In principle SNe of this type can accelerate particles to PeV energies, although one should notice that the values of the parameters have been rather optimised and while it is easy to lower this estimate, it is not that easy to raise it to higher values. In this sense, acceleration to PeV energies remains challenging although for the first time, at least theoretically, it appears possible to explain how SNe can accelerate particles to these very high energies. In both cases of type Ia and type II SNe, the spectrum of accelerated particles is expected to steepen at $E_{M}$, reached at the beginning of the Sedov-Taylor phase. Notice however that while such stage is reached a few hundred years after the explosion for type Ia SNe, a type II SN enters its adiabatic stage a few tens of years after the explosion. This implies that catching a PeVatron of this type in our Galaxy in the act of accelerating to the highest energies, for instance using gamma ray observations, is and will remain in the near future rather challenging. Notice that in the case of particle acceleration in type II SNe exploding in the wind of the presupernova red giant, the maximum energy $E_{M}$ is defined as the highest energy of accelerated particles at the beginning of the Sedov-Taylor phase. However even higher energies can be reached at earlier times, when the mass processed by the shock is appreciably smaller than the ejecta mass. This phenomenon reflects in a steepening in the CR spectrum at energy $E_{M}$ that is not exponential, but rather a steeper power law. This finding has important implication for the description of the transition from galactic to extragalactic CRs \cite{Cardillo:2015zda}. 

The investigation of the effects of streaming instability in the regime discussed in Ref. \cite{Bell:2004p737} has recently received a strong boost thanks to the adoption of hybrid and Particle-in-cell (PIC) codes to describe particle acceleration at non-relativistic shocks (see \cite{Caprioli:2015cda} for a recent review). 

These simulations aim at describing the basic physics of the formation of a collisionless shock wave: particle acceleration seems to be a by-product of the formation of such shocks. Hybrid simulations of diffusive shock acceleration at parallel shocks \cite{Caprioli:2013dca} have shown that the mechanism works and that CR acceleration efficiency of $10-20\%$ can be achieved. The spectrum of accelerated particles is confirmed to be consistent with the standard prediction, $\sim p^{-4}$. On the other hand, the temperature of the downstream plasma is measured to be lower by the amount predicted by the conservation relations at the shock and accounted for in non-linear theories of DSA. This is the first time, to our knowledge, that non-linear effects induced by particle acceleration are detected in simulations based on basic physical principles of the formation of collisionless shocks. These simulations also show magnetic field amplification \cite{Caprioli:2014tva}, especially through the non-resonant channel proposed in \cite{Bell:2004p737}. The instability is seen to grow on small scales as predicted by quasi-linear treatments \cite{Bell:2004p737,Amato:2008vj}. On the other hand the saturation level of the magnetic field is found to depend upon the strength of the pre-existing magnetic field $B_{0}$, contrary to what one would expect based on Eq. \ref{eq:equi}, and to be somewhat lower than the value returned by Eq. \ref{eq:equi}.

One of the concepts that are most central to the investigation of DSA at collisionless shocks is that of injection, namely of establishing what are the physical criteria that differentiate thermal from non-thermal particles. A substantial advancement was recently achieved in this direction in Ref.  \cite{Caprioli:2014dwa}, where injection was studied as a function of the orientation of the pre-existing magnetic field and a semi-analytical theory of injection, supported by hybrid simulations, was proposed. This study revealed that ions are effectively injected for quasi-parallel shocks (inclination angle of $\leq 45^{o}$ of the magnetic field with respect to the shock normal) with the help of shock drift acceleration that serves as a pre-energisation process. For more inclined shocks, \cite{Caprioli:2014dwa} found a strong suppression of injection, namely the acceleration process does not get bootstrapped. It is not clear at present whether this conclusion may change by assuming the presence of some level of pre-existing turbulence on small scales, independent of CR induced instabilities. 

Recently, PIC simulations have been used to investigate the problem of electron acceleration at collisionless shocks \cite{Park:2014lqa}. For the first time, these simulations have demonstrated that at quasi-parallel shocks both ions and electrons can be accelerated through DSA, with a ratio of fluxes at given momentum $K_{ep}\sim 10^{-3}-10^{-2}$, that compares well with the ratio observed at the Earth between the electron and proton flux in the GeV range, where both protons and electrons lose a negligible fraction of their energy during propagation. 

Finally, it is worth recalling that recently gamma ray observations carried out with the HESS telescope provided the very first evidence of a PeVatron in our Galaxy \cite{Abramowski:2016dhk}. The source is spatially located at the galactic center, coincident with the position of Sagittarius $A^{*}$. At the present time, it is not clear as yet whether the black hole at the galactic center or some other type of source could be responsible for this emission. 

\subsection{DSA in the presence of neutral hydrogen}
\label{sec:neutrals}

The presence of neutral atomic hydrogen in the acceleration region may potentially lead to modifications in the spectrum of accelerated particles \cite{Blasi:2012xe} for shock velocities below $\sim 3000$ km/s. The theory of DSA in the presence of neutrals was recently developed in \cite{Blasi:2012xe} in the test-particle regime and in \cite{Morlino:2012nb} in its non-linear version. The spectrum of accelerated particles is modified because of the so-called neutral return flux: neutral hydrogen atoms crossing the (collisionless) shock towards downstream can suffer a charge exchange reaction with a hot ion downstream of the shock. There is a finite probability that the resulting neutral atom moves in the direction of the upstream of the shock, where it can damp its energy and momentum onto the plasma, thereby heating it and slowing it down. This phenomenon leads to a reduction of the Mach number and to a correspondingly steeper spectrum of accelerated particles for momenta such that the diffusion length $D(p)/v_{s}<\lambda_{n}$, where $\lambda_{n}$ is the pathlength for reactions of charge exchange and ionization. For typical values of the parameters, the effect of neutrals on the spectrum of accelerated particles is limited to energies below $\sim 1-10$ TeV. The neutral return flux requires that charge exchange reactions downstream occur faster than ionization, a condition that restricts the importance of this phenomenon to shocks moving with velocity below $\sim (3-4)\times 10^{3}$ km/s. 

The presence of neutrals in the acceleration region also leads to Balmer line emission, which represents a powerful diagnostic tool for testing particle acceleration. In fact the presence of CRs at the shock leads to an enhanced width of the narrow Balmer line (due to charge exchange reactions in the CR induced precursor discussed above) and to a reduced width of the broad Balmer line (due to the lower gas temperature downstream of the shock that follows from non-linear DSA). These effects are described quantitatively in \cite{Morlino:2012ik,Morlino:2012nb,Morlino:2013gka}, where the appearance of a Balmer line component with intermediate width $\sim$ few hundred km/s was also discussed, as resulting from the phenomenon of neutral return flux. 

In principle, the simultaneous measurement of the width of the narrow and broad component of the Balmer line at a SNR shock may allow us to measure the CR acceleration efficiency and to infer the existence of a CR induced precursor, due to the pressure of accelerated particles slowing down the plasma incoming into the shock from upstream. The phenomenon of neutral return flux discussed above also gives rise to an intermediate component of the Balmer line, with typical width of $100-300$ km/s. The cases of SNRs RCW 86 (G315.4 - 2.3) and SNR 0509-67.5 were recently studied in \cite{Morlino:2013uaa} and \cite{Morlino:2013vza} respectively. Unfortunately, at the present time it is difficult to extract unambiguous information about CRs from these observations, due to uncertainties in the level of thermalisation between electrons and ions in the system and poor knowledge of the shock velocity (see \cite{Ghavamian:2013aka} for a recent review of these issues). On the other hand, if this information becomes available, the observation of Balmer line emission has the potential to provide precious insights in the acceleration process at SNR shocks.

\section{Transport of extragalactic CRs}
\label{sec:prop}

The extragalactic origin of UHECRs, at least at energies above the ankle $E>10^{19}$ eV, is widely accepted \cite{Aloisio:2012ba}. The propagation of UHECRs across intergalactic space is conditioned primarily by astrophysical photons backgrounds and, if any, by the presence of extragalactic magnetic fields. 

The astrophysical photon backgrounds relevant for the propagation of UHECRs are the Cosmic Microwave Background (CMB) and the Extragalactic Background Light (EBL). The former background, with the highest density, is the well known relic radiation from the big bang, while the latter is composed of infrared, optical and ultraviolet photons produced and reprocessed by astrophysical sources at present and past cosmological epochs. While the cosmological evolution of the CMB is analytically known, the case of the EBL is model dependent. In the past years several models for the cosmological evolution of the EBL have been proposed \cite{Franceschini:2008tp,Stecker:2006eh,Stecker:2005qs,Kneiske:2003tx}. These models show sizeable differences only at high redshift ($z > 4$), not actually relevant in the propagation of UHECRs but affecting the production of secondary neutrinos \cite{Allard:2011aa,Aloisio:2015ega}, as we discuss in section \S \ref{sec:sec}.

In the forthcoming sections \S \ref{sec:protons-int}, \S \ref{sec:prospec}, \S \ref{sec:sec} and \S \ref{sec:mag} we discuss the details of UHECRs propagation, through astrophysical backgrounds and intergalactic magnetic fields, and the production of secondary particles. 

\subsection{Interactions of ultra-high energy cosmic rays}
\label{sec:protons-int}

Ultra-high energy protons\footnote{Here we do not consider the case of neutrons because their decay time is much shorter than all other scales involved in the propagation of UHECR \cite{Aloisio:2008pp,Aloisio:2010he}.} or nuclei propagating in the intergalactic space interact with CMB and EBL photons through the processes of pair-production, photo-pion production and, only in the case of nuclei heavier than protons, photo-disintegration. Given the distribution of background photons and the energies involved, the propagation of protons is substantially affected only by the CMB radiation field, while in the case of nuclei, and only for the photo-disintegration process, also the EBL field is important \cite{Aloisio:2008pp,Aloisio:2010he}. As we discuss below, the effect of the EBL on proton propagation plays a role only for the production of secondary particles, but it negligibly affects the expected proton flux. 

These mechanisms of energy losses and their relevance for the propagation of UHECRs were discussed soon after the discovery of the CMB radiation field \cite{Penzias:1965wn}. Greisen \cite{Greisen:1966jv} and, independently, Zatsepin and Kuzmin \cite{Zatsepin:1966jv} realised that, due to photo-pion production, the interaction of UHE protons with CMB would produce a sharp suppression in the expected UHECR spectrum at $E\ge 10^{20}$ eV, the so-called GZK feature\footnote{Also referred as GZK cut-off or suppression.}. In the same period, Hillas \cite{Hillas:1967} and Blumenthal \cite{Blumenthal:1970nn} studied the effect of pair production of UHE protons with $E>10^{18}$ eV on the CMB radiation field and Berezinsky and Zatsepin \cite{Beresinsky:1969qj,Stecker:1973sy,Strong:1973,Berezinsky:1975zz} realised that the propagation of UHE protons would give rise to the production of secondary cosmogenic particles, generated by the decay of photo-produced mesons, such as neutrinos and gamma rays. 

The interaction rate associated to processes involving UHECRs can be written in a very general form as \cite{Stecker:1968uc,Aloisio:2008pp}:
\begin{equation}
\frac{1}{\tau} = \frac{c}{\Gamma^2}\int_{\epsilon'_{\min}}^{+\infty} \epsilon'\sigma(\epsilon')\int_{{\epsilon'}/{2\Gamma}}^{+\infty} \frac{n_\gamma(\epsilon)}{2\epsilon^2} d\epsilon \,d\epsilon' \label{eq:tau}~,
\end{equation} 
where $\Gamma$ is the Lorentz factor of the particle, $\sigma(\epsilon')$ is the total cross-section associated to the particle interactions, $\epsilon'$ is the background photon energy in the particle rest frame, $\epsilon'_{\min}$ is the lowest value of~$\epsilon'$ above which the interaction is kinematically possible (threshold), and $n_\gamma(\epsilon)\,d\epsilon$ is the number per unit volume of background photons with energy between $\epsilon$ and $\epsilon + d\epsilon$ in the laboratory reference frame. The photon energy in the particle rest frame is related to that in the laboratory frame by $\epsilon' = \Gamma\epsilon(1-\cos\theta)$, where $\theta$ is the angle between the particle and photon momenta $(0 \le \epsilon' \le 2\Gamma\epsilon)$.

Nucleons ($N$), whether free or bound in nuclei, with Lorentz factor larger than $\Gamma\ge m_\pi/2\epsilon(1-\cos\theta)\simeq 10^{10}$ interacting with the CMB photons give rise to the photo-pion production process: 

\begin{equation}
N+\gamma \to N + \pi^0   \qquad N+\gamma \to N + \pi^{\pm}.
\end{equation}

At lower energies ($\Gamma< 10^{10}$) the same processes can occur on the EBL field \cite{Aloisio:2015ega}, although with a lower probability. 

The photo-pion production process involves a sizeable energy loss for protons resulting in the GZK feature \cite{Greisen:1966jv,Zatsepin:1966jv} which arises at the threshold for photo-pion production, which in the nucleon rest frame reads $\epsilon'_{\min} = m_{\pi} + m_{\pi}^2/2m_N \approx 145$ MeV. The photo-pion production cross-section has a complex behavior with a number of peaks corresponding to different hadronic resonances, the largest one being the $\Delta$ resonance placed at $\epsilon' =\epsilon_{\Delta}\approx 340$ MeV \cite{Berezinsky:2002nc}. At energies much larger than $\epsilon_{\Delta}$ the cross-section is approximately constant \cite{Berezinsky:2002nc}. The photo-pion production process holds also for nucleons bound within UHE nuclei, being the interacting nucleon ejected from the parent nucleus. This process is subdominant if compared with photo-disintegration except at extremely high energies \cite{Allard:2011aa} and, as we discuss later, has some relevance only in the case of production of secondary cosmogenic particles. 

UHE nuclei propagating through astrophysical backgrounds can be stripped of one or more nucleons through the interactions with the CMB and EBL photons, a process named photo-disintegration:
\begin{equation}
(A,Z) + \gamma \to (A-n, Z-n') + nN
\end{equation}
being $A$ and $Z$ the atomic mass number and atomic number of the nucleus, $n$ ($n'$) the number of stripped nucleons. In the nucleus rest frame the energy involved in such processes is usually much less than the rest mass of the nucleus itself, hence in the laboratory frame all fragments approximately inherit the same Lorentz factor of the parent nucleus, i.e. we can neglect nucleus recoil \cite{Aloisio:2008pp,Aloisio:2010he}.

For photon energies close to the threshold ($8~{\rm MeV} \approx \epsilon'_{\min} < \epsilon' \le 30~{\rm MeV}$), the cross-section is dominated by a smooth peak, the giant dipole resonance, that corresponds to the extraction of one nucleon and it is the dominating process in UHE nuclei propagation \cite{Puget:1976nz,Allard:2005ha,Aloisio:2008pp,Aloisio:2010he}. At larger energies $\epsilon'>30~{\rm MeV}$ the quasi-deuteron process dominates with the extraction of two or more nucleons. This regime corresponds to an almost constant cross-section and has a small impact on the propagation of UHE nuclei \cite{Puget:1976nz,Allard:2005ha,Aloisio:2008pp,Aloisio:2010he}. 

The process of photo-disintegration is responsible for the production of secondary, hadronic, particles that will compose a sizeable fraction of the flux of UHECRs observed at the Earth.  

\begin{figure}[t]
%\sidecaption[b]
\begin{center}
\includegraphics[scale=.45]{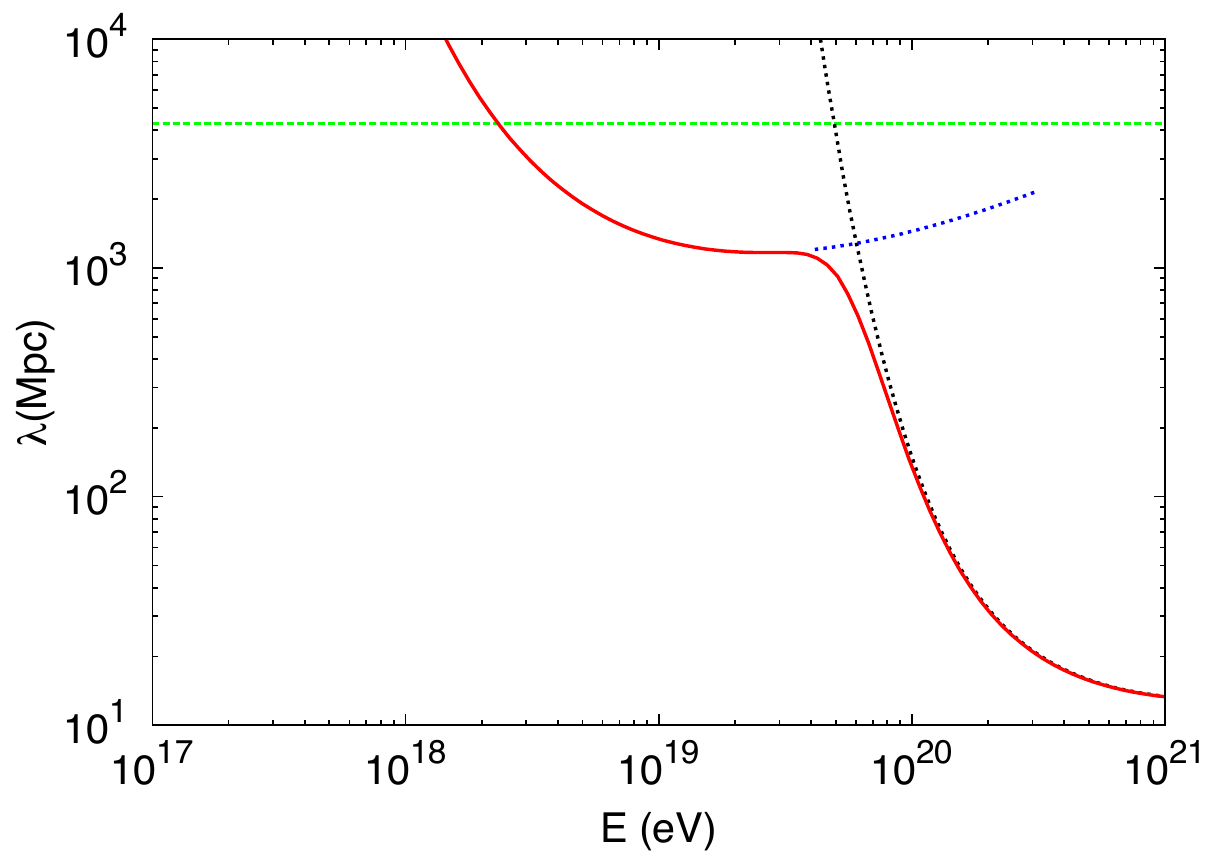}
\end{center}
\caption{Loss length of UHE protons (red solid line total energy losses). Photo-pion production dotted black line, pair production dotted blue line. The size of the visible universe is seen by the dashed green line.}
\label{fig1}   
\end{figure}

The interaction rate associated to the processes of photo-pion production and photo-disintegration can be written using Eq. (\ref{eq:tau}) specifying the cross-section, the background photon density and all relevant kinematical thresholds of the process. It is interesting that, while the first process changes the particle's Lorentz factor, leaving the nature of the particle unchanged, the second process changes the particle's nature leaving the Lorentz factor unchanged. 

Protons and nuclei with Lorentz factor $\Gamma\ge 2m_e/\epsilon(1-\cos\theta)\simeq 10^{9}$ can undergo the process of pair production $p + \gamma \to p + e^{+} + e^{-}$. The mean free path associated to pair production is relatively short compared with all other length scales of UHECR propagation, with a very small amount of energy lost by the propagating particle in each interaction \cite{Berezinsky:2002nc} so that frequently this process is treated in the approximation of continuum energy losses. In this case the rate of energy losses due to pair production is given by Eq. (\ref{eq:tau}) substituting $\sigma\to \sigma f$ being $f$ the inelasticity of the process, i.e. the average fraction of energy lost by the particle in one interaction \cite{Ginzburg:1990sk}. In the case of nuclei the rate of pair-production energy losses can be computed starting from that of protons and taking into account that $f^A=f^p/A$ and $\sigma^A=Z^2\sigma^p$. 

\begin{figure}[t]
%\sidecaption[b]
\begin{center}
\includegraphics[scale=.45]{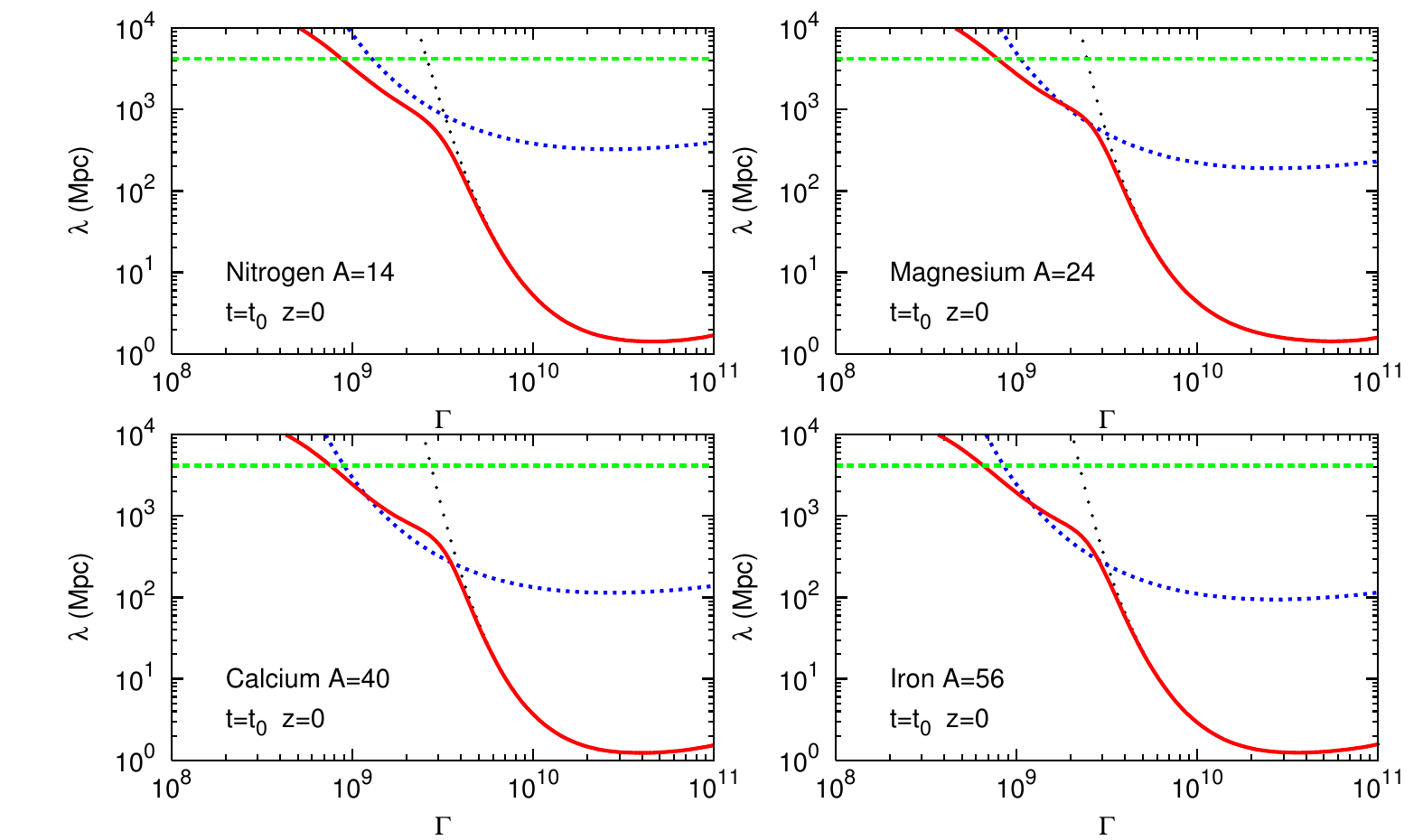}
\end{center}
\caption{Loss length of UHE nuclei. Red solid line photo-disintegration and blue dotted line pair-production. The effect of EBL on photo-disintegration is seen by the black dotted line that shows photo-disintegration due to the sole CMB field. The size of the visible universe is seen by the dashed green line.}
\label{fig2}   
\end{figure}

UHECRs can propagate through cosmological distances, in this case the expansion of the universe produces adiabatic energy losses. Assuming standard $\Lambda$CDM cosmology, we can write the energy lost per unit time by UHECRs (protons or nuclei) as 
\begin{equation}
\left(-\frac{1}{\Gamma}\frac{d\Gamma}{dt}\right)_\textrm{ad} = H(z) = H_0\sqrt{(1+z)^3\Omega_\textrm{m} + \Omega_\Lambda}
\label{adiabatic}
\end{equation}
where $z$ is the redshift at time~$t$, $H_0 \simeq 70~\mathrm{km}/\mathrm{s}/\mathrm{Mpc}$ is the Hubble constant, $\Omega_\textrm{m} \simeq 0.26$ is the matter density, and $\Omega_\Lambda \simeq 0.74$ is the dark energy density \cite{Ade:2015xua}.

In figures \ref{fig1}, \ref{fig2} we plot the interaction path length of protons and nuclei computed at zero red-shift as a function of energy.

In the case of protons (figure \ref{fig1}), at low energies, up to few $10^{18}$ eV, energy losses are dominated by the expansion of the universe. The pair production process starts to be relevant at the energy $2\times 10^{18}$ eV; pion photo-production becomes important at $5\times 10^{19}$ eV where the loss length drops to very low values and the large scale universe becomes opaque to UHECRs. 

In figure \ref{fig2}, the loss length of nuclei is plotted as function of the Lorentz factor $\Gamma$, assuming the EBL model presented in \cite{Stecker:2005qs}. Two sharp drops can be seen in the pathlength of nuclei: the first drop, at energy $A\times 10^{18}$ eV ($\Gamma\simeq 10^{9}$), is due to the combined effect of photo-disintegration on far infra-red photons (low energy EBL) and pair-production on the CMB; the second drop, even more pronounced, is due to photo-disintegration on CMB photons and arises at energies $A\times 4\times 10^{18}$ eV ($\Gamma\simeq 4\times 10^{9}$). In particular, the position of this last drop is less model dependent, being related to photo-disintegration on CMB photons only, and fixes the highest energy behaviour of the fluxes of nuclei expected at the Earth. 

\subsection{Propagated spectra}
\label{sec:prospec}
As discussed in section \ref{sec:UHEdet}, the main spectral features of UHECRs observed at the Earth are: (i) the ankle, a flattening of the spectrum at energy around $\simeq 5\times 10^{18}$ eV, observed since 1960s (Volcano Ranch experiment \cite{Linsley:1963}) and confirmed by all observations \cite{Aab:2015bza,Abbasi:2015bha}; (ii) a sharp suppression of the spectrum at the highest energies. The energy of such suppression is not clearly identified, due to some tension between the observations of Auger and TA (see figure \ref{fig:CompSpectra} and section \ref{sec:UHEdet} for a detailed discussion). 

Once energy losses of UHECRs are specified one can determine the spectra expected at the Earth by assuming an injection spectrum at the sources and their cosmological evolution\footnote{We will not discuss here the case of the possible presence of extragalactic magnetic fields, we will come back to this point in the forthcoming section \S \ref{sec:mag}.}. In the following we will assume that the energy per unit volume injected by the sources in the form of UHECRs only depends on red-shift, that sources share the same injection spectral index $\gamma_g$ (with a power law injection $\propto E^{-\gamma_g}$) and the same maximum energy $E_{max}$ at the sources. 

As discussed in section \ref{sec:UHEdet}, Auger and TA are currently not providing an unambiguous measurement of the mass composition. Hence, in what follows, we discuss separately the two cases of a pure proton composition (according to TA data) and a mixed composition with heavy nuclei contributing to the UHECR flux (according to Auger data). 

\subsubsection{Protons and the dip model} 
\label{sec:dip}

In the case of a pure proton composition the only relevant astrophysical background is the CMB \cite{Aloisio:2008pp,Aloisio:2010he}. This fact makes the propagation of UHE protons free from the uncertainties related to the background, being the CMB exactly known as a pure black body spectrum that evolves with redshift through its temperature. In this case, therefore, any signature of the propagation in the observed spectrum will depend only on the assumptions made at the source and the details of the interactions suffered by propagating protons. 

\begin{figure}[t]
\begin{center}
%\sidecaption[b]
\includegraphics[scale=.4]{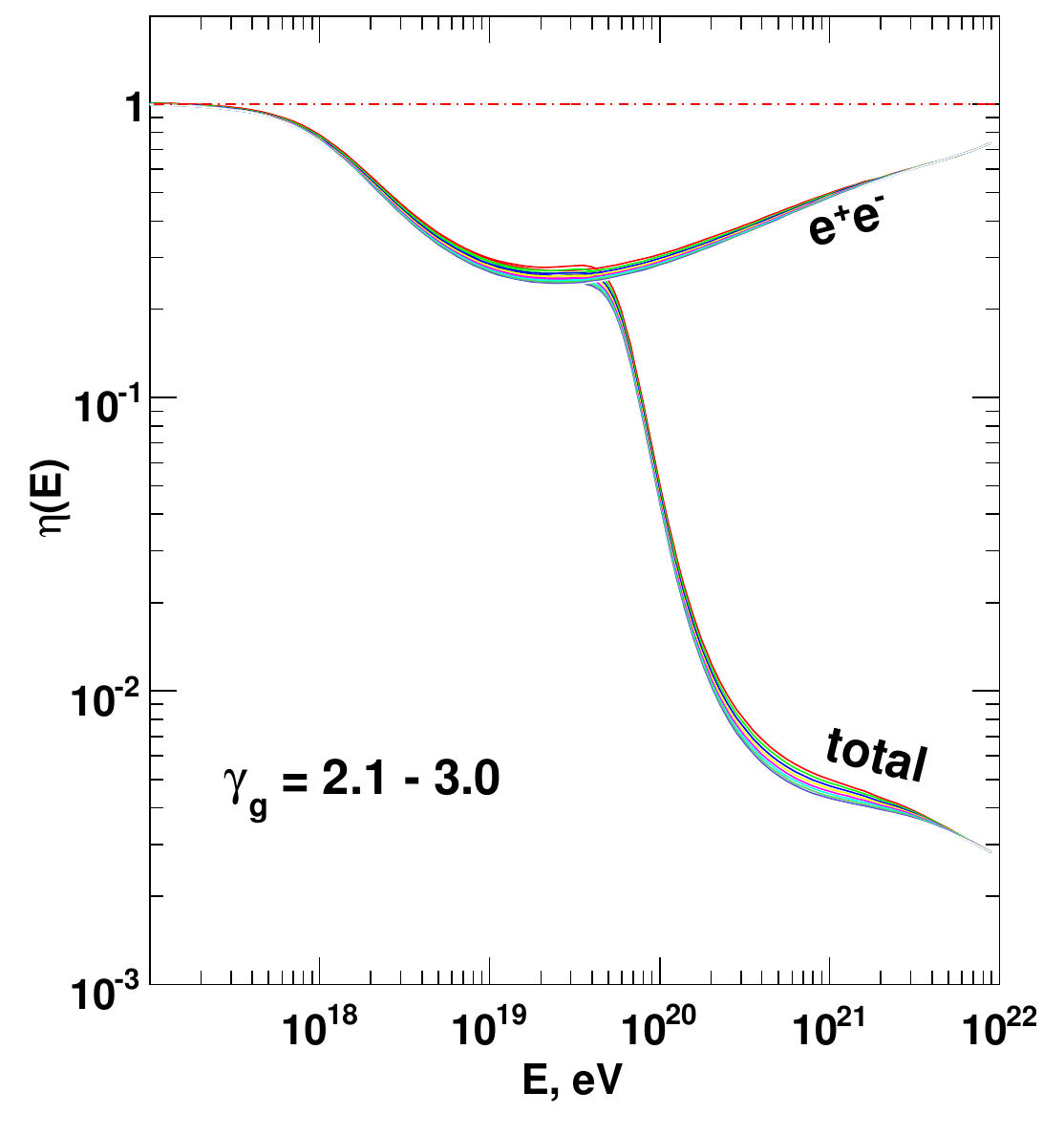}
\end{center}
\caption{Theoretical modification factor computed for different values of the injection power law index as labeled.}
\label{fig3}   
\end{figure}

In order to isolate the effects of energy losses in the propagated proton spectrum it is useful to use the so-called modification factor $\eta(E)$ defined as the ratio: 
\begin{equation}
\eta(E)=\frac{J_p(E)}{J_{unm}(E)}
\label{eq:modfact}
\end{equation}
where $J_p$ is the proton spectrum, computed with all energy losses taken into account, and $J_{unm}(E)$ is the unmodified spectrum computed taking into account only adiabatic energy losses due to the expansion of the universe. 

\begin{figure}[t]
\sidecaption[b]
\includegraphics[scale=.85]{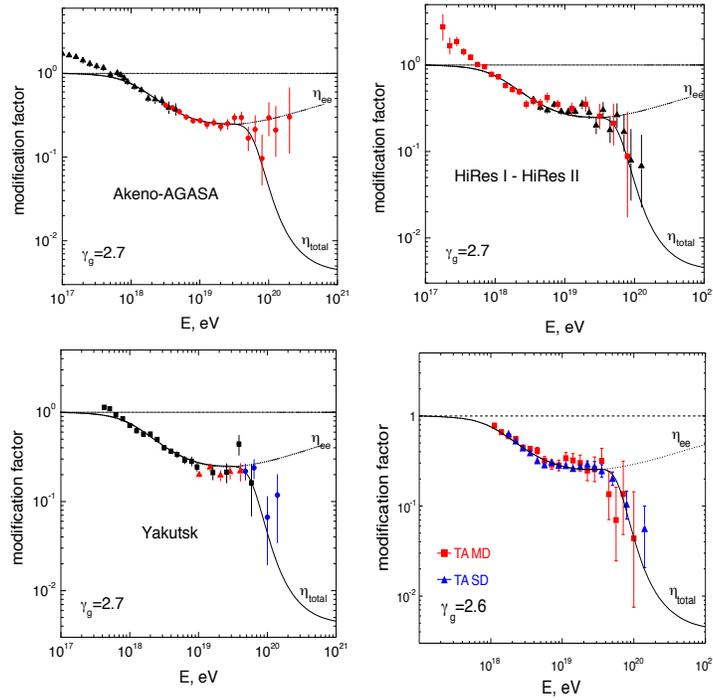}
\caption{Comparison of the modification factor with data \cite{Abbasi:2002ta,Egorova:2004,Shinozaki:2006,Honda:1993,Jui:2011vm} of the UHECR experiments favouring a proton dominated composition.}
\label{fig4}   
\end{figure}

Assuming an injection spectrum at the source with a power law form $\propto E^{-\gamma_g}$, in figure \ref{fig3} we plot the modification factor $\eta(E)$ as computed for different values of the injection power law $\gamma_g$ as labeled. From this figure it is evident that $\eta(E)$ clearly shows the signatures of energy losses suffered by protons, being almost independent of the injection. At low energy, losses are dominated by the adiabatic expansion of the universe, the spectrum at the  Earth keeps the injection shape, and the modification factor is $\eta(E)=1$; above the pair production threshold (around $2\times 10^{18}$ eV) pair production energy losses become important and the propagated spectrum hardens (the curves labelled with $e^{+}e^{-}$ take into account only pair-production); above the photo-pion production threshold (around $5\times 10^{19}$ eV) the propagated spectrum experiences an abrupt steepening which corresponds to the GZK feature (curves labeled with ${\it total}$ take into account all relevant energy losses). 

Particularly relevant is the feature associated to the pair-production energy losses, named "dip" \cite{Berezinsky:2002nc,Aloisio:2006wv}, that reproduces quite well the ankle observed in the UHECR spectrum, provided that the injection power law at the source is around $\gamma_g=2.6\div 2.7$. In figure \ref{fig4} we plot the theoretical modification factor together with the experimental data of several detectors as labeled, which all claim a pure proton composition \cite{Abbasi:2002ta,Egorova:2004,Shinozaki:2006,Honda:1993,Jui:2011vm}. From this figure it is evident that the behaviour of the pair production dip reproduces quite well the observations. 

The results presented in figure \ref{fig4} refer to the case without a cosmological evolution of the sources, i.e. density and luminosity of sources are independent of red-shift. Assuming a cosmological evolution, which typically gives a larger weight to distant sources, the transition between adiabatic and pair production energy losses arises at lower energy and the pair production dip will be deeper and slightly shifted to lower energies. As a consequence of this fact the spectral index at the sources needed to reproduce the observations will be lowered with respect to the best fit value (see figure \ref{fig4}) obtained without evolution \cite{Berezinsky:2002nc,Aloisio:2006wv}.

Remarkably, the dip model explains the observed flux with only one extragalactic component of pure protons, directly linking the flux behaviour to energy losses. Hence, in the case of the dip model, the transition between galactic and extragalactic cosmic rays occurs at energies below the pair-production threshold, i.e. $E_{tr}<2\times 10^{18}$ eV \cite{Aloisio:2012ba,Berezinsky:2002nc,Aloisio:2006wv,Berezinsky:2005cq,Aloisio:2007rc,DeMarco:2005ia}.

The source parameters that can be fitted by the comparison of figure \ref{fig4} are basically only two: the injection power law index $\gamma_g$ and the source emissivity ${\cal L}_S$, i.e. the energy emitted (in the form of UHECRs) per unit time and volume by sources\footnote{Given a distribution of sources with a number density $n_S$ each with the same luminosity $L_S$, the energy emitted per unit time and volume (emissivity) is given by ${\cal L}_S=n_SL_S$.}. The value of the required emissivity depends on the power law index, with a value that ranges from $\gamma_g=2.5$ (for strong cosmological evolution) up to $\gamma_g=2.7$ (without evolution) \cite{Berezinsky:2002nc,Aloisio:2006wv}. Using these values of $\gamma_g$ and assuming a single power law injection down to the lowest energies (GeV) results in a prohibitive energy budget for any astrophysical source. To avoid this problem, in the original papers introducing the dip model, a change in the spectral index at injection was assumed \cite{Berezinsky:2002nc}: at energies below $10^{18}$ eV, $\gamma_g=2.0$ while at larger energies $\gamma_g$ takes the best fit values quoted above. Under this assumption the required emissivity (at red-shift $z=0$) necessary to reproduce UHECR data is around ${\cal L}_S=10^{45}\div 10^{46}$ erg/Mpc$^3$/yr \cite{Berezinsky:2002nc,Aloisio:2006wv,Aloisio:2015ega}.  

In figure \ref{fig4} the same maximum acceleration energy $E_{max}=10^{21}$ eV for all sources is adopted. Releasing this hypothesis and taking into account that sources can be distributed over different values of the maximum energy we can assume an injection power law index as $\gamma_g=2.0$ for all sources \cite{Kachelriess:2005xh,Aloisio:2006wv}. At energies above the minimal maximum energy, a softer spectral index arises from the convolution of the distribution function over $E_{max}$ ($dN(E_{max})/dE_{max}\propto E_{max}^{-\alpha}$ with $\alpha=1.6\div 1.7$) leading to an "effective" spectral index $\gamma_{g}^{eff}=\gamma_g+\alpha-1$ at the highest energies \cite{Kachelriess:2005xh,Aloisio:2006wv}. 

As discussed above, we stress that in the framework of the dip model, a suppression of the flux at low energy ($<10^{18}$ eV) is needed not only to avoid too high luminosity of the sources but also not to overshoot the observed flux at energies below $10^{18}$ eV. As we discuss later (see section \ref{sec:mag}), magnetic horizon effects can also be invoked to reduce the proton fraction at $E\le10^{18}$ \cite{Aloisio:2004fz} even if such effects would leave the energy budget of the sources unaffected. 

\subsubsection{Mixed composition model} 
\label{sec:mix}

The discussion in the previous section was centred around the hypothesis of a pure proton composition of UHECRs. However, as discussed in section \ref{sec:UHEdet}, a somewhat different picture arises from the Auger observations that claim a mixed composition of UHECRs characterised by light nuclei at low energies ($\le 5\times 10^{18}$ eV) and heavier nuclei at the highest. The qualitative new finding that mass composition of UHECRs might be mixed has served as a stimulus to build models that can potentially explain the phenomenology of Auger data. These models all show that the Auger spectrum and mass composition at $E\ge 5\times 10^{18}$ eV can be fitted at the same time only at the price of requiring very hard injection spectra for all nuclei ($\propto E^{-\gamma_g}$ with $\gamma_g=1\div 1.6$) and a maximum acceleration energy of $E_{max}\le 5 Z\times 10^{18}$ eV \cite{Aloisio:2013hya,Aloisio:2009sj,Taylor:2013gga,Aab:2016zth}. 

The need for hard spectra can be understood taking into account that the low energy tail of the flux of UHECRs reproduces the injection power law. Therefore, taking $\gamma\ge 2$ causes the low energy part of the spectrum to be polluted by heavy nuclei thereby producing a disagreement with the light composition observed at low energy. 

One should appreciate here the change of paradigm that these findings imply: while in the case of a pure proton composition it is needed to find sources and acceleration mechanisms able to energise CR protons up to energies larger than $10^{20}$ eV with steep injection ($\gamma_g\simeq 2.5\div 2.7$), the Auger data require that the highest energy part of the spectrum ($E>5\times 10^{18}$ eV) has a flat injection ($\gamma_g\simeq 1.0\div 1.6$) being dominated by heavy nuclei with protons' maximum energy not exceeding few$\times 10^{18}$ eV. 

By accepting the new paradigm, it follows that the Auger spectrum at energies below $5\times 10^{18}$ eV requires an additional component, with a steep injection spectrum ($\gamma_g\simeq 2.5\div 2.7$), composed by protons and helium nuclei that in principle could be both of galactic or extragalactic origin \cite{Aloisio:2013hya,Taylor:2013gga,Globus:2015xga}. However, the anisotropy expected for a galactic light component extending up to $10^{18}$ eV exceeds by more than one order of magnitude the upper limit measured by Auger \cite{Abreu:2012ybu}. This observation, just restricting the analysis to Auger data, would constrain the transition between galactic and extragalactic CRs to energies below $10^{18}$ eV \cite{Giacinti:2011ww,Aloisio:2012ba}.

The Auger data can be modelled essentially in two ways: (i) assuming the presence of two classes of sources: one injecting heavy nuclei with a hard injection and the other only protons and helium nuclei with a soft spectrum \cite{Aloisio:2013hya,Taylor:2013gga} or (ii) identifying a peculiar class of sources that could provide at the same time a steep light component and a flat heavy one \cite{Globus:2014fka,Globus:2015xga,Unger:2015laa}. These findings were recently confirmed by the Auger collaboration through a combined fit of the flux and mass composition data at energies $E>5\times 10^{18}$ eV \cite{Aab:2016zth}.

\begin{figure}[t]
\sidecaption[b]
\includegraphics[width=0.45\textwidth,height=0.335\textwidth]{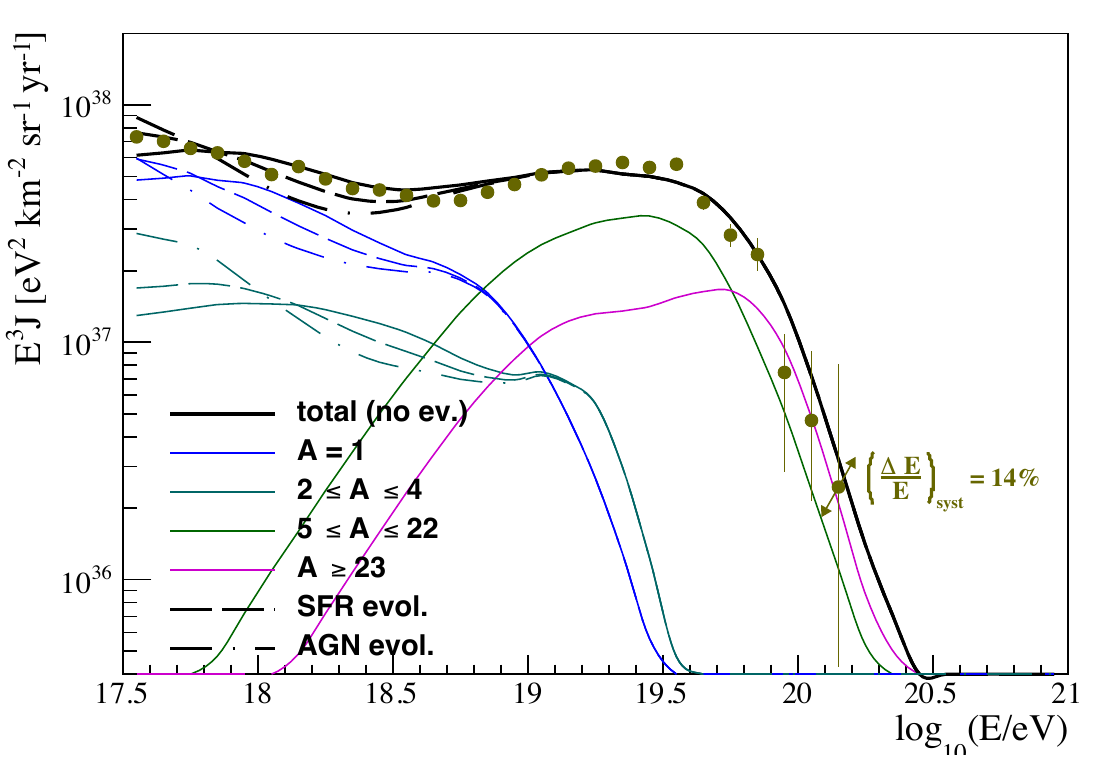}
\includegraphics[width=0.45\textwidth,height=0.335\textwidth]{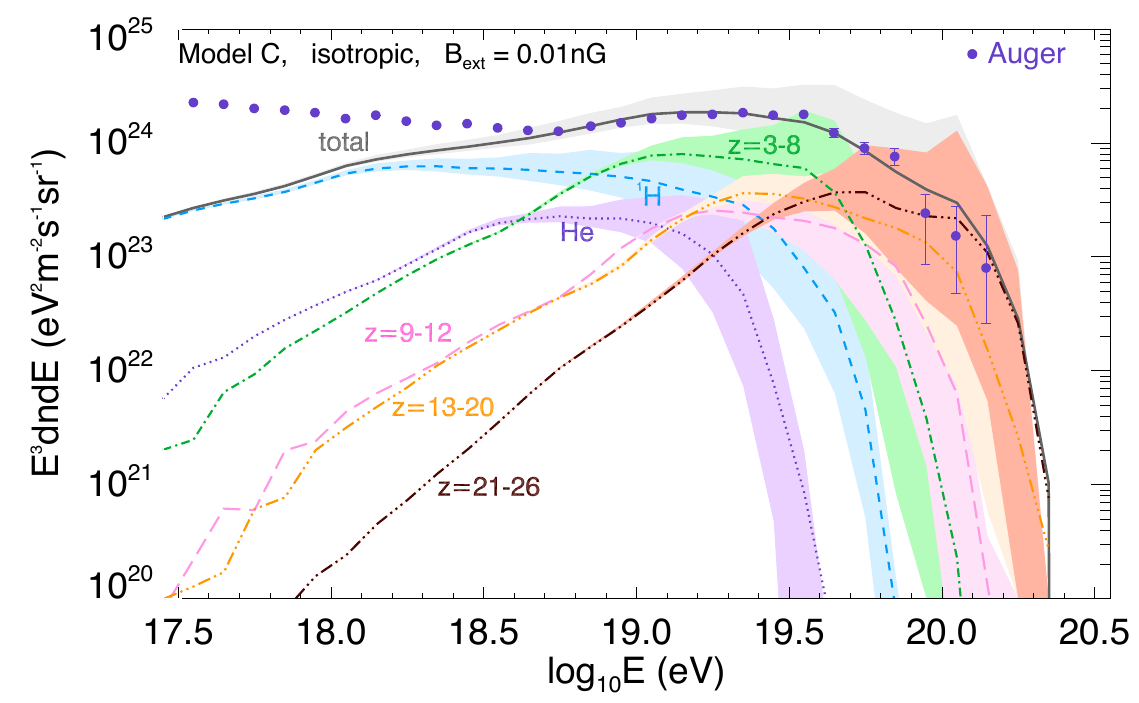}
\caption{Comparison of the Auger spectrum with theoretical expectations in the case of models with mixed composition. [Left Panel] Model with two classes of sources as in \cite{Aloisio:2015ega}. Continuous, dashed and dot-dashed lines correspond respectively to the cases of: no cosmological evolution of sources, evolution as the star formation rate and as active galactic nuclei. [Right Panel] Model with UHECR production in the internal shock of GRB as discussed in \cite{Globus:2014fka}. }
\label{fig5}   
\end{figure}

\begin{figure}[t]
\sidecaption[b]
\includegraphics[width=0.45\textwidth,height=0.335\textwidth]{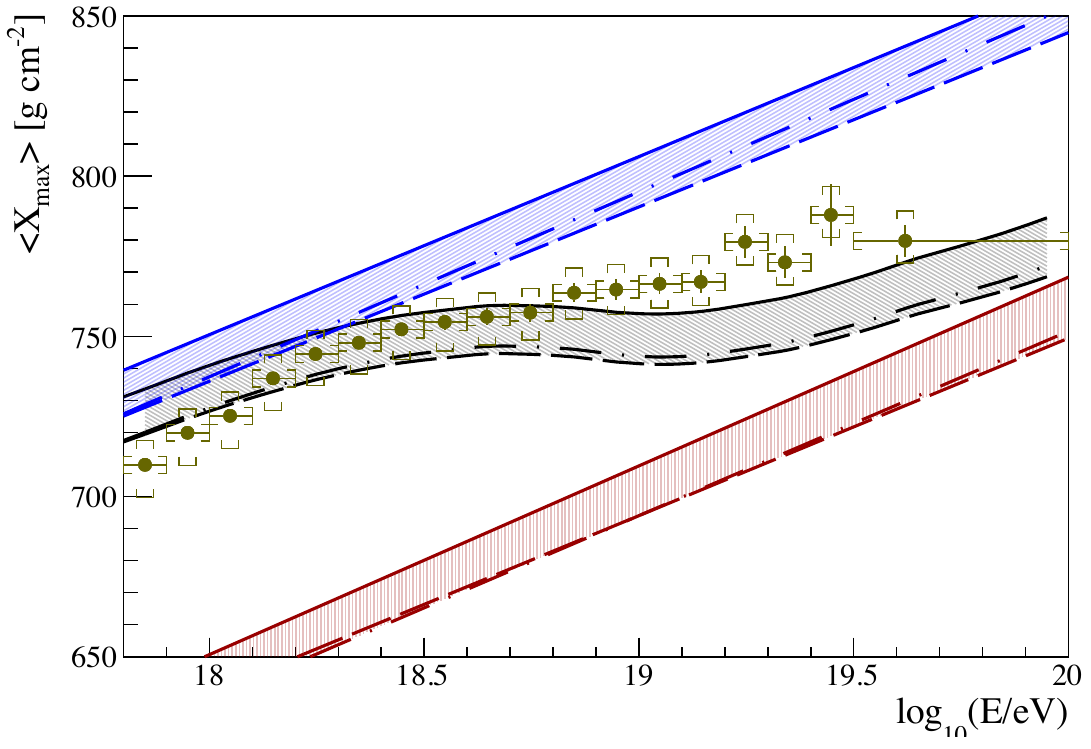}
\includegraphics[width=0.45\textwidth,height=0.335\textwidth]{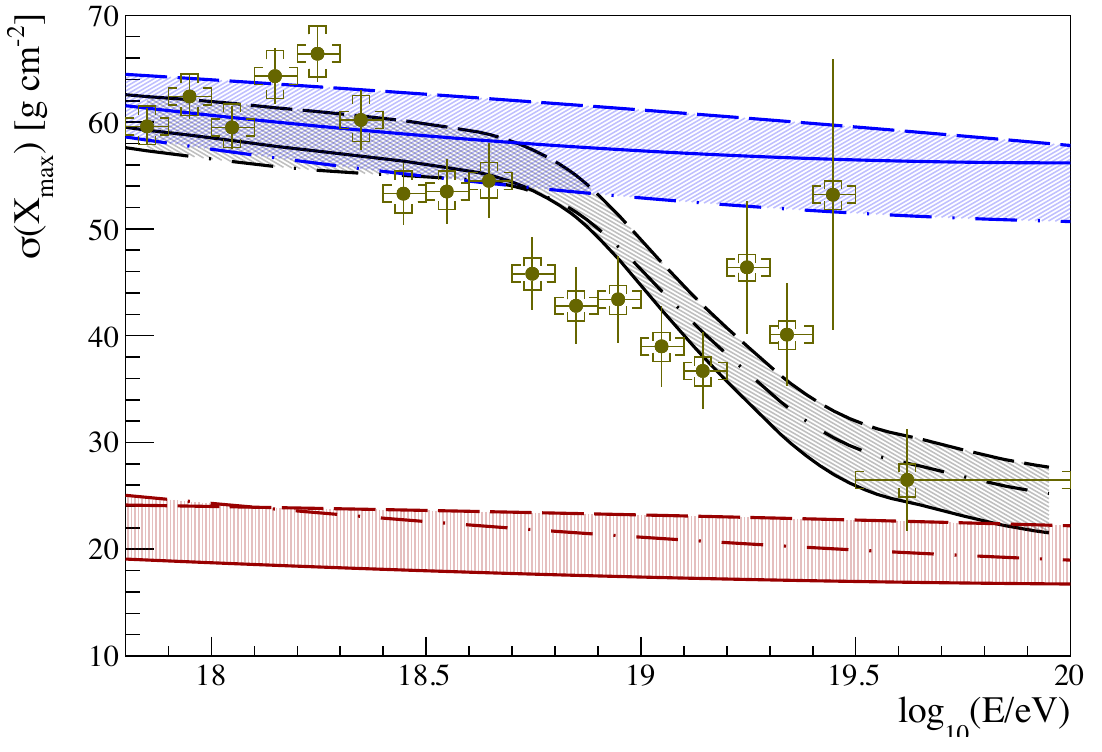}
\caption{Comparison of the elongation rate and its root mean square computed assuming the model with two classes of sources as discussed in \cite{Aloisio:2015ega}.}
\label{fig6}   
\end{figure}

\begin{figure}[t]
\sidecaption[b]
\includegraphics[width=0.45\textwidth,height=0.335\textwidth]{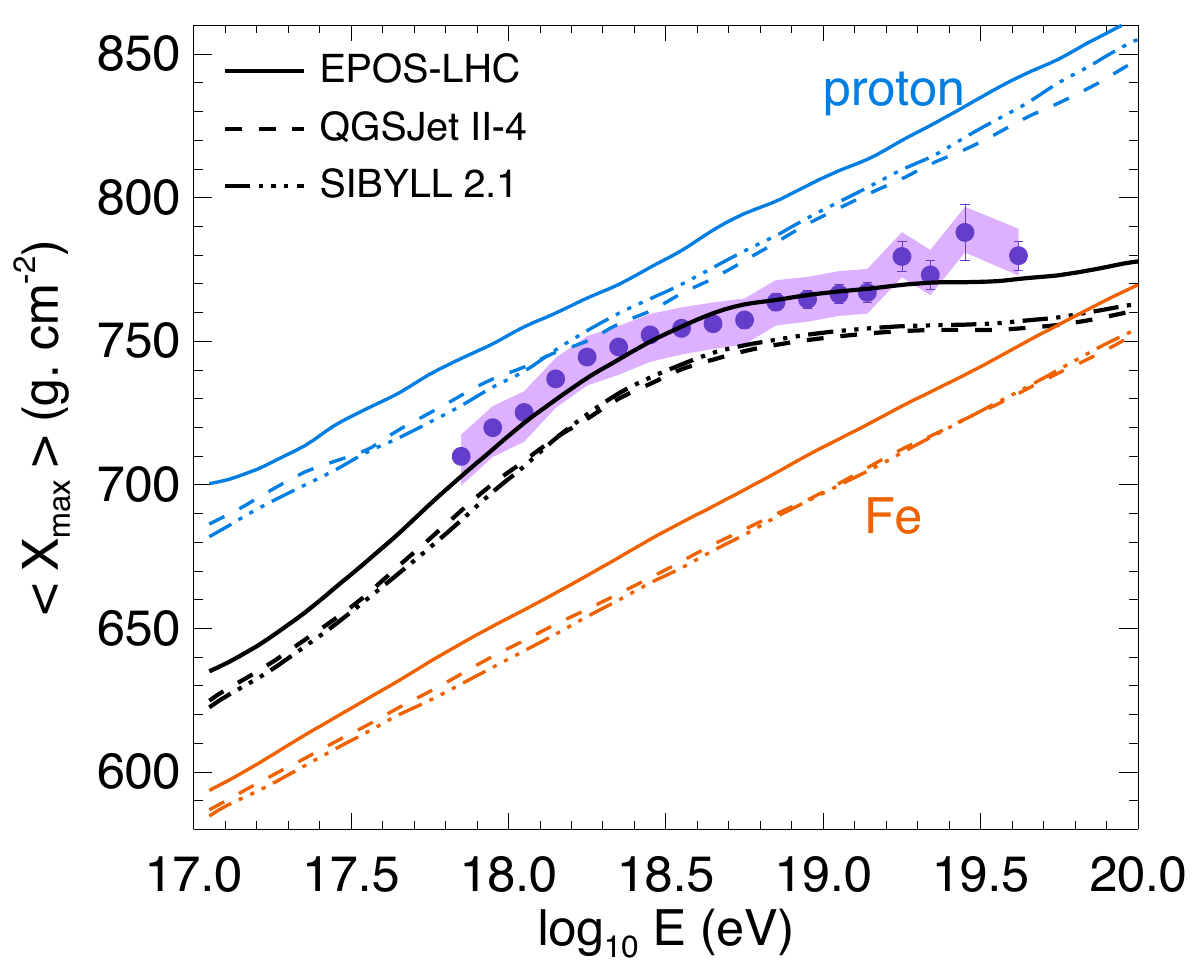}
\includegraphics[width=0.45\textwidth,height=0.335\textwidth]{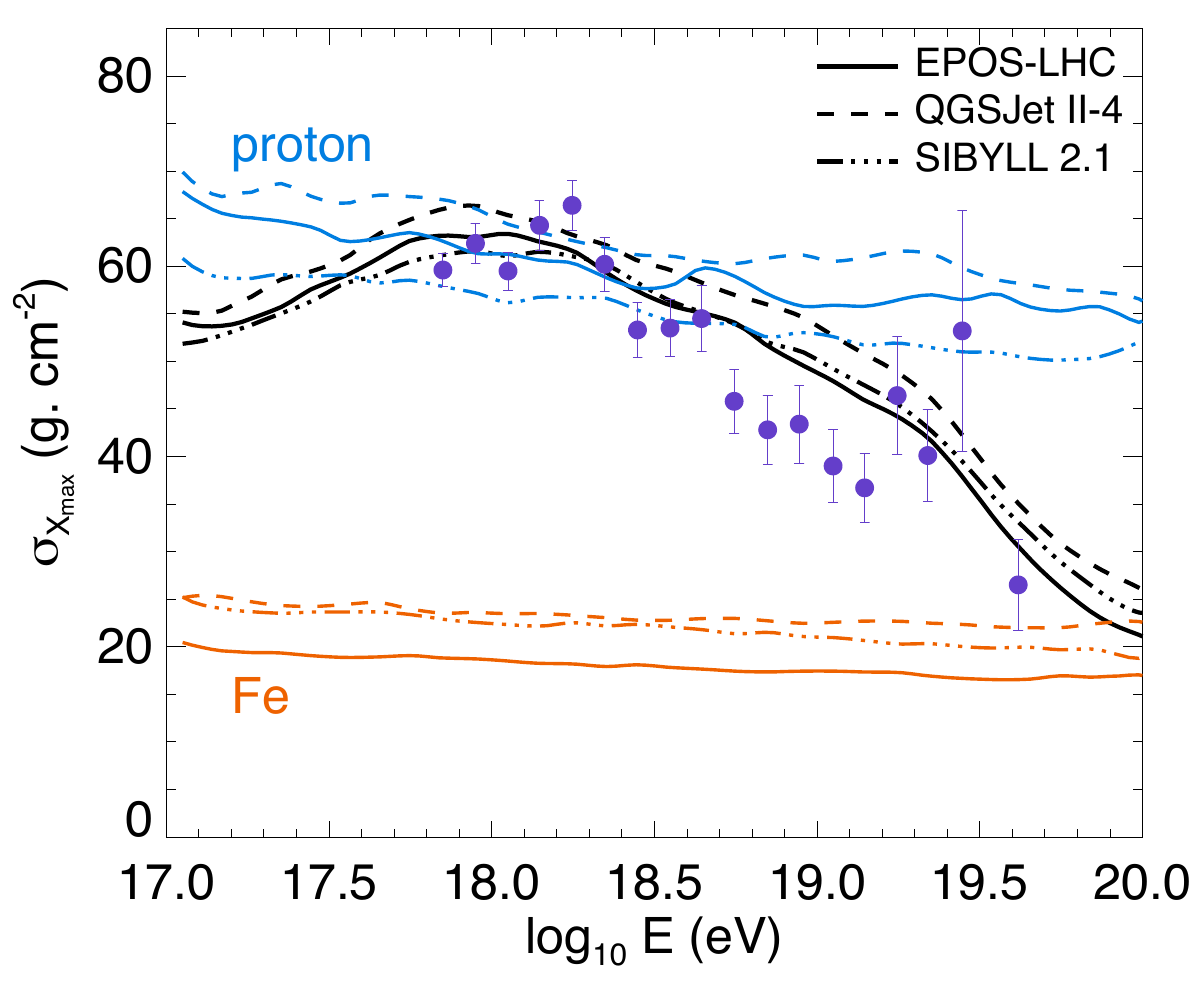}
\caption{Comparison of the elongation rate and its root mean square in the case of UHECR production in the internal shock of GRB as discussed in \cite{Globus:2014fka}.}
\label{fig7}   
\end{figure}

In figures \ref{fig5},\ref{fig6},\ref{fig7} we plot the comparison of Auger data on flux and chemical composition with the theoretical expectation in the case of two classes of generic sources discussed in \cite{Aloisio:2015ega} (left panel figure \ref{fig5} and figure \ref{fig6}) and in the case of a single class of sources (internal shocks in Gamma Ray Bursts, GRBs) discussed in \cite{Globus:2014fka} (right panel of figure \ref{fig5} and figure \ref{fig7}). In the left panel of figure \ref{fig5} we also plot the spectra computed with different assumptions on the cosmological evolution of sources as discussed in \cite{Aloisio:2015ega}: no cosmological evolution, evolution following the star formation rate \cite{Gelmini:2011kg,Yuksel:2008cu,Wang:2011qc} and of Active Galactic Nuclei (AGN) \cite{Gelmini:2011kg,Hasinger:2005sb,Ahlers:2009rf}. 

In figures \ref{fig6} and \ref{fig7}, chemical composition is inferred from the mean value of the depth of shower maximum $\langle X_\mathrm{max} \rangle$ and its dispersion (RMS) $\sigma(X_\mathrm{max})$. As discussed in section \ref{sec:UHEdet} (see also \cite{Kampert:2012mx,Abreu:2013env,Aloisio:2007rc}), the combined analysis of $\langle X_\mathrm{max} \rangle$ and $\sigma(X_\mathrm{max})$ allows one to obtain less model dependent information on the mass composition of UHECRs. The main uncertainties in such a procedure are introduced by the dependence of $\langle X_\mathrm{max} \rangle$ and its fluctuations on the hadronic interaction model used to describe the shower development. Most of such models fit low energy accelerator data while providing somewhat different results when extrapolated to the energies of relevance for UHECRs (for a review see \cite{Engel:2011zzb} and references therein). In figure \ref{fig6}, to highlight the uncertainties in the atmospheric shower development, four different models of hadronic interaction were included in the coloured bands, namely EPOS 1.99 \cite{Pierog:2006qv},  Sibyll 2.1 \cite{Ahn:2009wx}, QGSJet 01 \cite{Kalmykov:1997te} and  QGSJet 02 \cite{Ostapchenko:2005nj}. In figure \ref{fig7} different lines correspond to different interaction models as labeled. 

\subsection{Secondary cosmogenic messengers} 
\label{sec:sec}

The propagation of UHECRs through intergalactic space gives rise to the production of several unstable particles, produced by photo-hadronic interactions with CMB and EBL photons, that in turn produce high energy photons, electrons and neutrinos. The possible detection of these signal carriers, as realised soon after the proposal of the existence of the GZK cut-off \cite{Beresinsky:1969qj,Stecker:1973sy,Berezinsky:1975zz,Stecker:1968uc}, is extremely important to constrain models for UHECR sources, mass composition and the details of propagation \cite{Aloisio:2015ega,Ahlers:2009rf,Engel:2001hd,Kalashev:2002kx,Hooper:2004jc,Seckel:2005cm,DeMarco:2005kt,Allard:2006mv,Anchordoqui:2007fi,Takami:2007pp,Kotera:2010yn,Berezinsky:2010xa,Stanev:2014asa,Berezinsky:2016feh}.

\subsubsection{Neutrinos}
\label{sec:nu}

There are two processes that lead to neutrino production in the propagation of UHECRs: (i) the decay of charged pions, produced by photo-pion production, $\pi^{\pm}\to \mu^{\pm}+\nu_{\mu}(\bar{\nu}_{\mu})$ and the subsequent muon decay $\mu^{\pm}\to e^{\pm}+\bar{\nu}_{\mu}(\nu_{\mu})+\nu_e(\bar{\nu}_e)$; (ii) the beta-decay of neutrons and nuclei produced by photo-disintegration: $n\to p+e^{-}+\bar{\nu}_e$, $(A,Z)\to (A,Z-1)+e^{+}+\nu_e$, or $(A,Z)\to (A,Z+1)+e^{-}+\bar{\nu}_e$. These processes produce neutrinos in different energy ranges: in the former the energy of each neutrino is around a few percent of that of the parent nucleon, whereas in the latter it is less than one part per thousand (in the case of neutron decay, larger for certain unstable nuclei). This means that in the interactions with CMB photons, which have a threshold Lorentz factor around $\Gamma\ge 10^{10}$, neutrinos are produced with energies of the order of $10^{18}$~eV and $10^{16}$~eV respectively. Interactions with EBL photons contribute with a much lower probability than CMB photons, affecting a small fraction of the propagating protons and nuclei. Neutrinos produced through interactions with EBL, characterised by lower thresholds, have energies of the order of $10^{15}$~eV in the case of photo-pion production and $10^{14}$~eV in the case of neutron decay (see \cite{Aloisio:2015ega} and references therein).   

\begin{figure}[t]
\sidecaption[b]
\includegraphics[width=0.45\textwidth,height=0.335\textwidth]{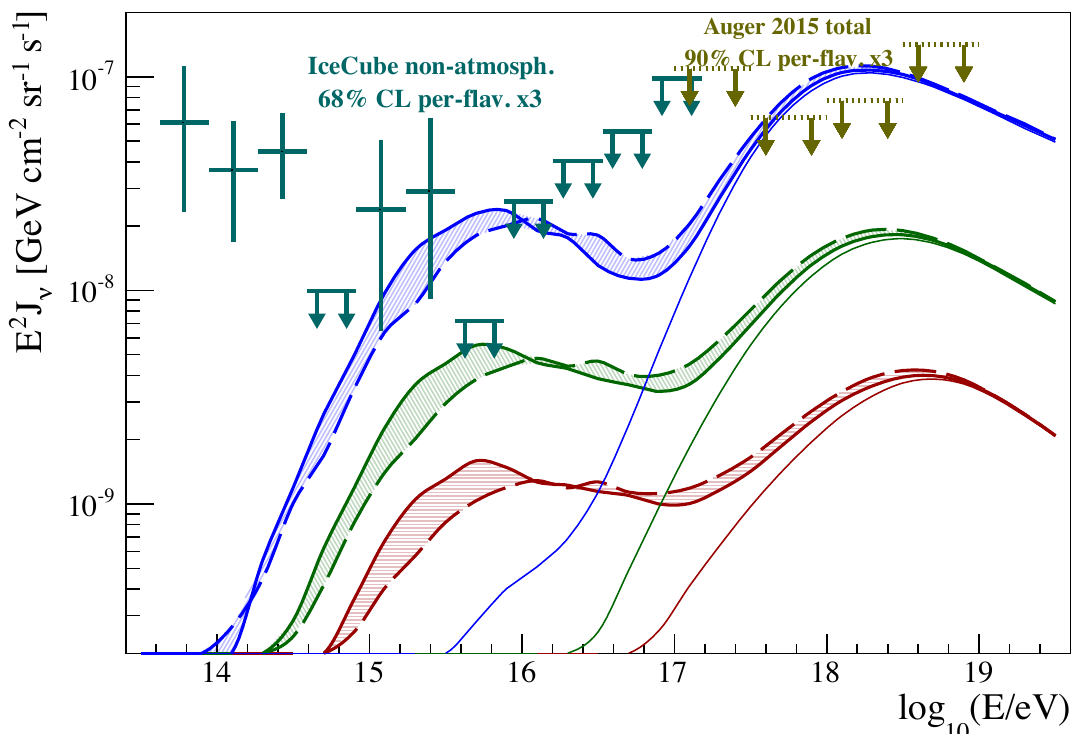}
\includegraphics[width=0.45\textwidth,height=0.335\textwidth]{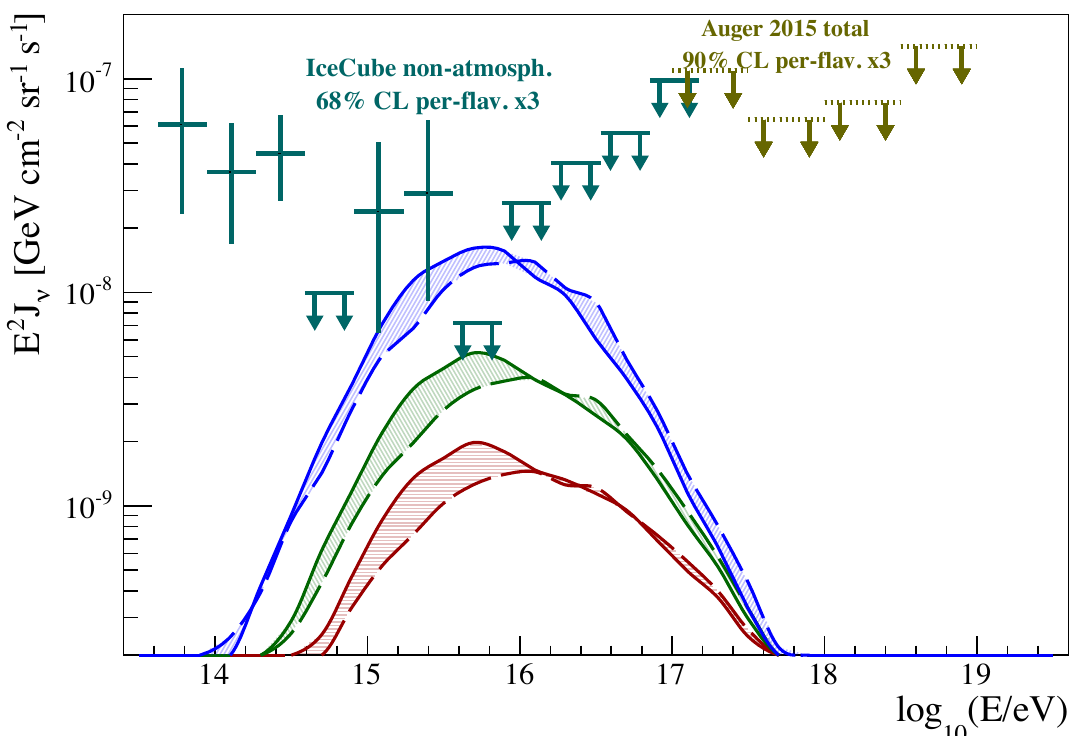}
\caption{[Left Panel] Fluxes of neutrinos in the case of the dip model. The three different fluxes correspond to different assumptions on the cosmological evolution of sources (from bottom to top): no evolution (red), SFR (green) and AGN (blue), coloured bands show the uncertainties due to the EBL model considered \cite{Stecker:2005qs,Stecker:2006eh,Kneiske:2003tx}. Thin solid lines are neutrino fluxes obtained taking into account the sole CMB field. [Right Panel] Neutrino fluxes in the case of mixed composition (as shown in figure \ref{fig5} left panel) with the same color code of left panel. Experimental points are the observation of IceCube on extraterrestrial neutrinos \cite{Aartsen:2013jdh,Aartsen:2013bka} and the Auger limits on neutrino fluxes \cite{Abreu:2013zbq}.} 
\label{fig8}   
\end{figure}

As discussed in the previous sections, theoretical models aiming at describing UHECR observations can be distinguished in two general scenarios depending on the mass composition: the dip model, based on a pure proton composition, and mixed composition models, with protons and heavy nuclei contributing to the flux of UHECRs. The flux of secondary neutrinos can be a powerful tool to investigate the actual composition of UHECRs. In figure \ref{fig8} we plot the flux of cosmogenic neutrinos expected in the case of the dip model (left panel) and in the case of mixed composition (right panel). Comparing the two panels of figure \ref{fig8} it is evident the huge impact of the composition on the expected neutrino flux: heavy nuclei provide a reduced flux of neutrinos because the photo-pion production process in this case is subdominant. 

The production of cosmogenic neutrinos during the propagation of UHECRs is not sensitive to variations in the distribution of sources, the whole universe contributes to the flux of neutrinos up to the maximum red-shift of astrophysical structures able to energise UHECRs. This red-shift is typically placed around $z_{max}\simeq 10$, which is the redshift of the first stars (pop III) \cite{Berezinsky:2011bb}. 

Once produced at cosmological distances neutrinos travel toward the observer almost freely, except for the adiabatic energy losses and flavour oscillations, being the opacity of the universe to neutrinos relevant only at $z \gg 10$ \cite{Weiler:1982qy,Gondolo:1991rn}. Hence, cosmogenic neutrinos are also a viable probe of the cosmological evolution of sources while UHE protons and nuclei, given their energy losses, can be observed only if produced at red-shifts $z<3\div 4$.

In figure \ref{fig8} three different hypotheses on the cosmological evolution of sources are taken into account, following the same assumptions used in figure \ref{fig5} (left) \cite{Gelmini:2011kg,Hasinger:2005sb,Ahlers:2009rf}.
 
There is a solid consensus about the light composition of UHECRs in the low energy part of the observed spectrum. This assures a flux of cosmogenic neutrinos in the PeV energy region produced by the protons' photo-pion production process on the EBL photons. Coloured bands in figure \ref{fig8} show the uncertainties connected with the EBL background \cite{Stecker:2005qs,Stecker:2006eh,Kneiske:2003tx}. 

\subsubsection{Gamma rays}
\label{sec:gamma}

While neutrinos reach the observer without being absorbed, high energy photons and electrons colliding with astrophysical photon backgrounds (CMB and EBL) produce electromagnetic cascades (EMC) through the processes of pair production (PP, $\gamma+\gamma_{CMB,EBL}\to e^{+}+e^{-}$) and Inverse Compton Scattering (ICS, $e+\gamma_{CMB,EBL}\to \gamma + e$). While PP is characterised by a kinematic threshold, the ICS occurs at all energies. From this simple observation follows that once a cascade is started by a primary photon/electron/positron it develops for as long as the energy of photons produced through ICS is above the PP threshold. The final output of the cascade, namely what is left behind when the cascade is completely developed, is a flux of low energy photons all with energies below the PP threshold.

The two astrophysical backgrounds, CMB and EBL, against which the EMC develops are characterised by typical energies $\epsilon_{CMB}\simeq 10^{-3}$ eV and $\epsilon_{EBL}\simeq 1$ eV. Hence, the typical threshold energy scale for pair-production will be respectively ${\mathcal E}_{CMB}=m_e^2/\epsilon_{CMB}=2.5\times 10^{14}$ eV and ${\mathcal E}_{EBL}=m_e^2/\epsilon_{EBL}=2.5\times 10^{11}$ eV; the radiation left behind by the cascade will be restricted to energies below ${\mathcal E}_{EBL}$. Clearly, numerical values quoted here should be intended as reference values being background photons distributed over energy and not monochromatic.

The cascade development has a universal nature independent of the spectrum of the initial photon/pair. It can be proven\footnote{For a recent detailed discussion of EMC development on CMB and EBL see \cite{Berezinsky:2016feh} and references therein.} that the spectrum of photons produced in the cascade, those left behind with energy below threshold, is always of the type: 
\begin{equation}
n_{\gamma}(E_{\gamma})\propto \left \{ \begin{array}{ll}
E_{\gamma}^{-3/2} & E_{\gamma} < {\mathcal E}_X \\
&\\
E_{\gamma}^{-2} & {\mathcal E}_X \le E_{\gamma} \le {\mathcal E}_{EBL}
\end{array}
\right.
\label{eq:EMC-spectrum}
\end{equation} 
being ${\mathcal E}_X=(1/3) {\mathcal E}_{EBL} \epsilon_{CMB}/\epsilon_{EBL}$ the (average) minimum energy of a photon produced through the ICS mechanism by an electron with the minimum allowed energy ${\mathcal E}_{EBL}/2$ \cite{Blumenthal:1970nn,Blumenthal:1970gc,Ginzburg:1990sk}. The normalisation of the spectrum (\ref{eq:EMC-spectrum}) can be easily determined imposing energy conservation, i.e. the total energy of cascading photons should correspond to the energy of the photon/pair that started the cascade. 

\begin{figure}[!h]
\centering
\includegraphics[scale=.30]{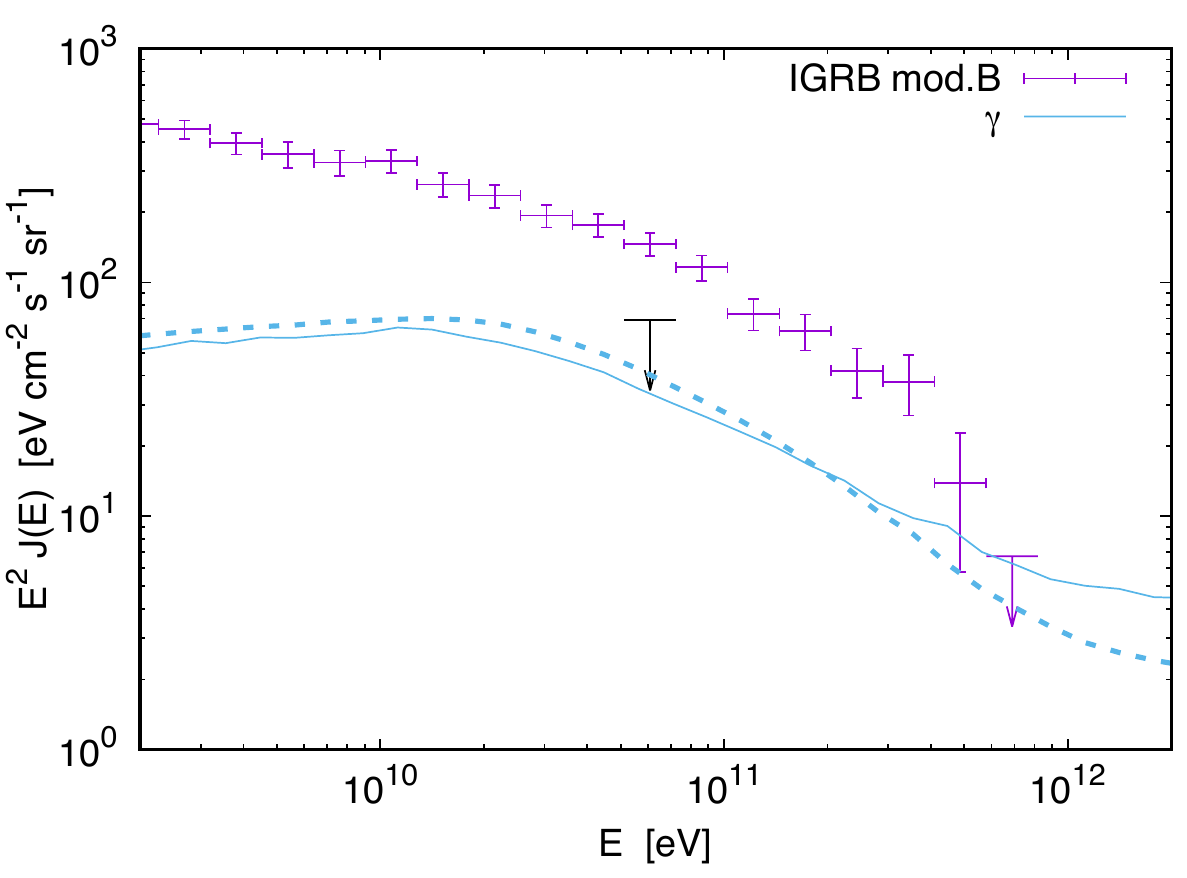}
\includegraphics[scale=.32]{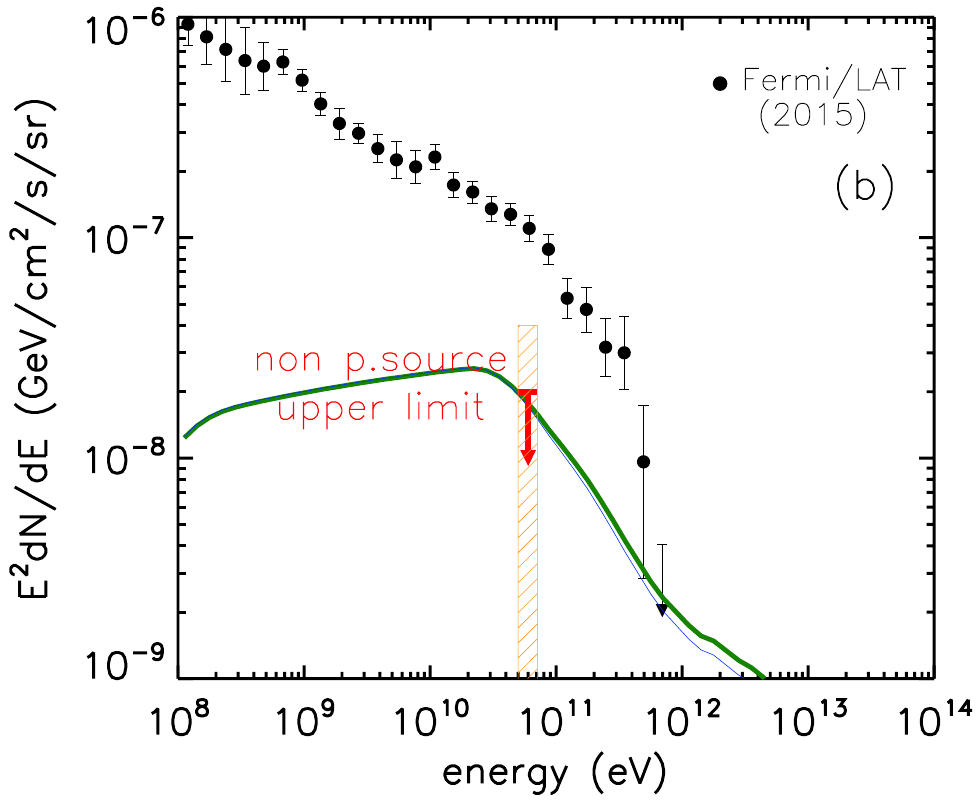}
\caption{Spectra of cosmogenic gamma rays obtained in the case of pure proton composition of UHECR without cosmological evolution of sources, as computed in \cite{Berezinsky:2010xa} (left panel) and in \cite{Liu:2016brs} (right panel), together with the Fermi-LAT data on diffuse gamma ray background, as in model-B (left panel) and model-A (right panel) of the analysis presented in \cite{Ackermann:2014usa,TheFermi-LAT:2015ykq}.} 
\label{fig:gamma_lim}   
\end{figure}

The propagation of UHECRs produces EMCs started by pairs and photons produced in the processes of pair production and photo-pion production. These cascades transform the energy lost by UHECRs into low energy gamma-ray photons, with the characteristics discussed above, that in turn contribute to the diffuse gamma ray background \cite{Beresinsky:1969qj,Berezinsky:1975zz,Gelmini:2011kg,Berezinsky:2010xa,Berezinsky:2016feh,Ferrigno:2004am,Ahlers:2010fw,Liu:2016brs}. Hence, the observation of a diffuse extragalactic gamma-ray background by the Fermi-LAT satellite \cite{Abdo:2010nz,Ackermann:2014usa} can be used to constrain models of UHECRs. 

A pure proton composition maximises the production of secondary gamma rays. The fast decrease in energy ($\propto E^{-2.4}$) of the diffuse background as observed by the Fermi-LAT satellite already constrains versions of the dip model (see section \ref{sec:dip}) with strong red-shift evolution of sources \cite{Berezinsky:2010xa}. These constraints could become even more severe taking into account a recent analysis of the Fermi-LAT collaboration \cite{TheFermi-LAT:2015ykq} that shows how the 'true' diffuse extragalactic gamma-ray background could be lower than what considered in the past if the contribution of unresolved sources is taken into account \cite{Liu:2016brs,Berezinsky:2016jys}. In figure \ref{fig:gamma_lim} we plot the expected gamma ray background in the case of the dip model without cosmological evolution of sources in comparison with the experimental data of Fermi-LAT in two cases of models for unresolved sources as discussed in \cite{TheFermi-LAT:2015ykq}.

At extreme energies ($>10^{19}$ eV), given the behaviour of PP cross-section, the universe becomes more and more transparent to photons \cite{Blumenthal:1970nn,Blumenthal:1970gc,Ginzburg:1990sk}, so that very high energy gamma-rays can propagate few tens of Mpc without being absorbed. This high energy gamma radiation was discussed in the literature either as a signature of exotic models for the production of UHECRs \cite{Berezinsky:1997hy,Berezinsky:1998ft,Sigl:1998vz,Aloisio:2003xj,Aloisio:2006yi,Aloisio:2015lva} (see section \ref{sec:td}) or as a probe of the astrophysical acceleration of UHECRs in the local universe \cite{1993PThPh..89..833Y,Taylor:2009we,Gelmini:2011kg}.

The detection of point-like gamma-ray sources at GeV-TeV energies could also be a promising way to reveal powerful astrophysical accelerators of UHECRs \cite{Ferrigno:2004am,Essey:2009ju,Essey:2010er}. This possibility critically depends on the magnetic field in the intergalactic space as it could modify the spatial development of the cascade. Hence, to reveal a point-like source of this kind it is needed to confine the cascade within small angular size around the source line of sight and the corresponding magnetic field should be quite low ($B<10^{-14}$ G) \cite{Ferrigno:2004am,Essey:2009ju,Essey:2010er}. The detection of such effect and its firm correlation with a source of UHECRs is anyway very difficult as the electromagnetic cascading gamma-rays have a universal spectrum independent of the primary particle type and energy (provided it is large enough, see above). 

Let us conclude stressing the importance of the magnetic field in the physics of EMC. Cascades can be sustained only if the process of ICS dominates over electron interactions, nevertheless increasing the magnetic field synchrotron interaction becomes more and more important with the net result of producing low energy ($\le$MeV) photons thus damping the cascade development. 

\subsection{Intergalactic magnetic fields} 
\label{sec:mag}

The propagation of UHECRs can be heavily affected by the presence of intergalactic magnetic fields (IMF). Here we will not discuss the effect of the galactic magnetic field which affects only the arrival distribution of particles leaving unchanged spectrum and mass composition \cite{Stanev:1996qj,Harari:2000az,Prouza:2003yf,Yoshiguchi:2003mc,Kachelriess:2005qm,Takami:2005ij}. 

Our experimental knowledge of intergalactic magnetic fields is still poor and fragmented, even if several important constraining observations were achieved (see \cite{Kronberg:1993vk,Grasso:2000wj,Carilli:2001hj,Kulsrud:2007an,Beck:2011he,2012SSRv..166....1R,Durrer:2013pga} and references therein). In certain environments of the universe such as galaxy clusters, which could harbour sources of UHECR, the magnetic field is better known with typical observed values in the range of $1 ~\mu$G \cite{Kronberg:1993vk,Grasso:2000wj,Carilli:2001hj,Kulsrud:2007an,Beck:2011he,2012SSRv..166....1R,Durrer:2013pga}. 

It is outside clusters, far the largest space traversed by extragalactic CR, in filaments and voids, that the value of the magnetic field is not known and, untill now, no convincing mechanism to produce strong fields over very large (supra-cluster) scales has been clearly found. The most reliable observations of the IMF are those of synchrotron emission, its polarisation and Faraday rotation at radio frequencies ($0.1\div 10$ GHz) \cite{Kronberg:1993vk,Grasso:2000wj,Carilli:2001hj,Kulsrud:2007an,Beck:2011he,2012SSRv..166....1R,Durrer:2013pga}. These measurements imply an upper limit for the IMF that depends on the coherence length of the field itself. For instance, according to \cite{Blasi:1999hu}, in the case of an inhomogeneous universe, $B < 4$ nG with a coherence scale of about $l_c = 50$ Mpc. 

The detection of magnetic fields in voids or, more generally, along paths with very low field intensities can be achieved by observing cascading propagation of TeV gamma rays coming from a point-like source \cite{Elyiv:2009bx,Neronov:2009gh}. As discussed in \ref{sec:gamma}, for low enough magnetic fields secondary pairs produced in the cascade will be weakly deflected producing a narrow emission of ICS gamma rays around the source line of sight. The variation with energy of the emission size can be detected allowing for a measure of the magnetic field strength in the range $10^{-16}\div 10^{-12}$ G \cite{Elyiv:2009bx,Neronov:2009gh}.

Apart from observations, the IMF can be predicted, in principle, implementing Magneto-Hydrodynamics (MHD) evolution of magnetic fields in numerical simulation of Large Scale Structures (LSS) formation \cite{2002A&A...387..383D,Dolag:2004kp,Sigl:2004yk,Donnert:2008sn,Ryu:2008hi}. The main ambiguities in these simulations are related to the assumed seed magnetic fields, to the mechanism invoked in their growth and to the capability of reproducing the local density velocity field (constrained \cite{2002A&A...387..383D,Dolag:2004kp} and unconstrained simulations \cite{Sigl:2004yk}). Unfortunately, because of these uncertainties, MHD simulations are not completely conclusive. The volume filling factor for strong fields, of the order of $1$ nG, vary by several orders of magnitude from one simulation another. The predicted magnetic field in voids (filaments) vary from $10^{-3}$ nG ($10\div 1$ nG) \cite{Dolag:2004kp,Donnert:2008sn,Ryu:2008hi} up to $10^{-1}$ nG ($10$ nG) \cite{Sigl:2004yk}.

In order to discuss the effects of magnetic fields on the propagation of UHECRs let us consider the ideal configuration of a homogeneous turbulent magnetic field with strength $B$ on the coherence scale $l_c$. A charged particle with energy $E$ in a magnetic field $B$ has a Larmor radius given by:
\begin{equation}
r_L(E) \simeq 1 \left (\frac{E_{EeV}}{ZB_{nG}}\right ) Mpc
\label{eq:rL}
\end{equation}
being $E_{EeV}$ the energy in units of $10^{18}$ eV, $B_{nG}$ the magnetic field in units of nano-Gauss and $Z$ the charge of the particle. 

Depending on their energy, particles can feel the effect of the magnetic field in different ways: if the Larmor radius of the particles is smaller than $l_c$, then the particles can resonantly scatter against turbulence at the appropriate wavelength, a phenomenon that naturally leads to diffusion. This happens for $$E<E_c\simeq 10^{18} Z B_{nG} (l_c/Mpc) ~{\rm eV}.$$
The diffusion length $l_D$, i.e. the distance that corresponds to a typical particle deflection of $1$ rad, depends on the turbulent power encountered by particles at the Larmor radius scale, therefore it depends on the turbulent spectrum of the magnetic field. It can be easily shown that $l_D\simeq l_c (E/E_c)^{\alpha}$ \cite{Aloisio:2004jda,Lemoine:2004uw,Aloisio:2008tx,Mollerach:2013dza} being $\alpha$ related to the turbulent spectrum ($\alpha=1/3$ Kolmogorov, $\alpha=1/2$ Kraichnan and $\alpha=1$, i.e. $l_D=r_L$, in the case of Bohm diffusion). As discussed earlier in this review, the quasi-linear expression for the diffusion coefficient reads
$$D=\frac{1}{3} cl_D=\frac{1}{3} c l_c \left (\frac{E}{E_c}\right )^{\alpha}~.$$  

At higher energies, namely for $r_L>l_c$  ($E>E_c$), the scattering becomes non-resonant and particles' propagation results in a series of small deflections $\delta\theta\simeq l_c/r_L$ in each coherence length $l_c$. The diffusion length $l_D$ can be easily evaluated as the space traversed by a particle to suffer a deflection $\Delta\theta\simeq 1$ rad, one has $l_D\simeq l_c (E/E_c)^2$ with a diffusion length that strongly increases with energy \cite{Aloisio:2004jda,Aloisio:2008tx,Mollerach:2013dza}. A formal derivation of the overall transport of particles in a spectrum of turbulence in the different regimes was recently presented in \cite{2017ApJ...837..140S}.

In this case the propagation of particles can be described as diffusive or rectilinear depending on the distribution of sources, namely on the average distance between sources $r_s$. If $l_D<r_s$ the propagation is still diffusive, with a diffusion coefficient 
$$D=\frac{1}{3} cl_D=\frac{1}{3} c l_c \left (\frac{E}{E_c}\right )^{2}~,$$  
while at the highest energies when $l_D>r_S$ the rectilinear propagation regime is recovered and the magnetic field has no effects on propagation. 

On general grounds, the effect of the magnetic field on spectrum and mass composition of UHECRs is closely tied to the assumptions made about density and luminosity of sources. In other words, even a very strong field would have no effect on UHECRs if the mean separation between sources is smaller than all other propagation length scales, i.e. diffusion and energy losses lengths \cite{Aloisio:2004jda,Aloisio:2008tx}.   

Typical density expected for UHECR sources is in the range $n_s=10^{-4}\div 10^{-6}$ Mpc$^{-3}$, with a rather large uncertainty, as can be estimated based on detection (or non-detection) of small-scale clustering in the arrival directions \cite{Dubovsky:2000gv,Fodor:2000yi,Blasi:2003vx,Kachelriess:2004pc}. The typical separation distance between sources will be $r_s=(n_s)^{-1/3}\simeq 10\div 10^2$ Mpc.

As discussed above, assuming a non-resonant diffusion regime ($r_L > l_c$), cosmic rays from a source at distance $r_s$ will diffuse whenever $l_D\le r_s$, or in terms of energy:
\begin{equation}
E\le E_D\simeq 10^{18} Z B_{nG}\sqrt{\frac{r_s l_c}{{\rm Mpc}^2}} eV~.
\label{eq:ED}
\end{equation}
Particles with energies below $E_D$, traveling for longer times with respect to rectilinear propagation, lose more energy. The net effect of this kind of transport is to reduce the expected flux at energies below $E_D$. 

As was first realised in \cite{Aloisio:2004fz,Lemoine:2004uw}, the (IMF induced) flux suppression (in the energy range $10^{17}\div 10^{18}$ eV) is of paramount importance in tagging the transition between galactic and extragalactic CR. Such scenario can also be invoked to reduce the flux of protons below $10^{18}$ eV in the case of the dip model (see paragraph \ref{sec:dip}) or to allow for softer injection spectra in the case of mixed composition (see paragraph \ref{sec:mix}). As follows from Eq. (\ref{eq:ED}), the viability of these scenarios clearly depends on the assumptions made about the magnetic field configuration that should be at the $nG$ level. It is worth recalling here that, as follows from equipartition, a $nG$ field is an absolute upper limit of the expected magnetic field strength in voids. 

Considering more realistic configurations with inhomogeneous magnetic fields, i.e. taking into account different field intensities and coherence scales in voids and filaments, the suppression effect on the low energy flux will be less pronounced and shifted to lower energies \cite{Kotera:2007ca,Kotera:2008ae,Kalli:2011ud}.

Let us conclude by stressing that the simple description given above of the role of magnetic fields for UHECR propagation is likely to be an oversimplification, mainly because it fails to catch the easiest ingredients of the interaction between UHECRs and magnetic fields. For instance, within several Mpc from any source of UHECRs, the current induced by the escape of the accelerated particles from the source is likely to induce instabilities that are bound to affect both the local magnetic fields and the transport of UHECRs in that region. As discussed in \cite{Blasi:2015esa}, these phenomena may well induce a natural confinement of the lowest energy end of UHECRs close to sources because of the excitation of plasma instabilities. At present the phenomenological implications of these phenomena have not yet been investigated.

\section{Astrophysical sources} 
\label{sec:source}

\begin{figure}[t]
\sidecaption[b]
\includegraphics[scale=.35]{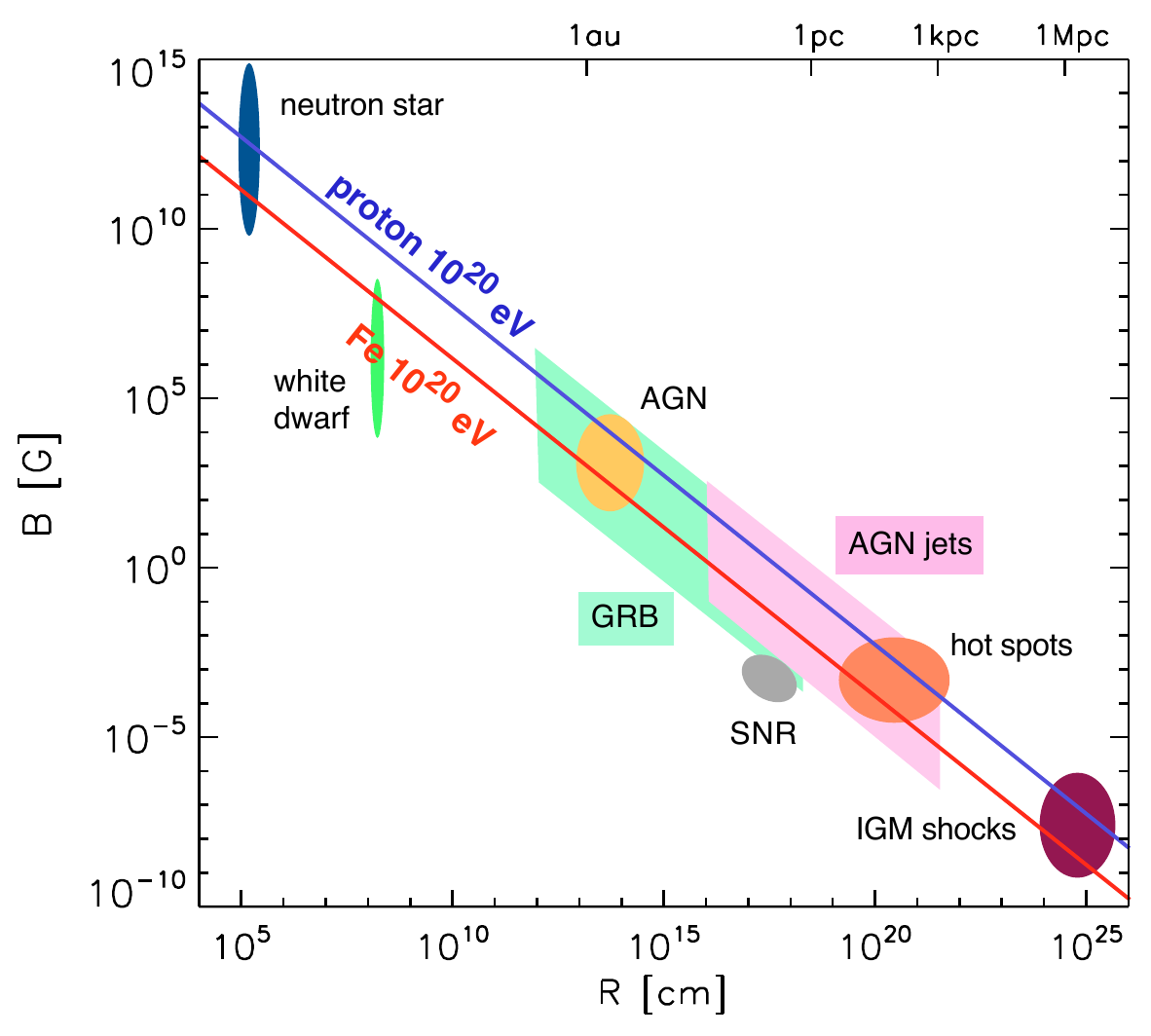}
\caption{Hillas plot \cite{Hillas:1985is} obtained by imposing the condition $r_L(E)<R$, dashed line corresponds to protons with energy $10^{20}~{\rm eV}=100$ EeV while the two solid lines correspond to protons with $E=10^{21}~{\rm eV}=1$ ZeV and iron nuclei with $E=10^{20}$ eV as labeled. Also shown are several astrophysical objects that meet the requirements of size/magnetic field needed for the acceleration process. The figure is taken from \cite{Kotera:2011cp}.}
\label{fig:HillasPlot}   
\end{figure}

We do not know what the sources of UHECRs are. Hence, it is important to define general requirements that a hypothetical accelerator should fulfil in order to reach the extreme energies observed. If the size of the accelerator is $R$, a necessary condition to reach the energy $E$ is that particles with this energy would remain trapped inside the source, hence the Larmor radius of the particle should be: $r_L(E)<R$. This condition fixes a relation, at the base of the so-called Hillas plot, between the size of the accelerator $R$ and the magnetic field $B$ in the acceleration region. In figure \ref{fig:HillasPlot} we show the Hillas plot \cite{Hillas:1985is} with the curves relative to the condition $r_L<R$ (see Eq. (\ref{eq:rL})) in the case of protons and iron nuclei, the corresponding energy of accelerated particles, and several astrophysical objects that match this condition. The acceleration of nuclei, thanks to larger electric charge, is less demanding and can be easier achieved with respect to the case of protons. 

Following \cite{Waxman:2005id,Blasi:2012yx}, the general idea at the base of the Hillas plot can be further refined by assuming that the accelerator moves (as for shocks) with either relativistic or non-relativistic velocity. Let us consider first the non-relativistic case. The condition $r_L(E)<R$ on the magnetic field can be easily transformed in a condition on the magnetic energy density $\epsilon_B=B^2/4\pi$. This quantity should be lower that the total ram pressure $\epsilon_B<\rho V^2$ and this fixes a lower limit on the luminosity of the source: 
\begin{equation} 
L=4\pi R^2 V \frac{\rho V^2}{2} >2\pi R^2 V \epsilon_B \simeq 1.6\times 10^{45} Z^{-2} \left (\frac{E}{10^{20}{\rm eV}}\right )^2 \beta ~{\rm erg/s}
\label{eq:L_limit}
\end{equation}
being $Z$ the electric charge of the particle and $\beta=V/c$ the velocity of the accelerator. As discussed in \cite{Blasi:2012yx}, the condition (\ref{eq:L_limit}) is somewhat uncertain in the scaling with $\beta$, as it depends on the details of particles transport in the acceleration region. What is more relevant in Eq. (\ref{eq:L_limit}) is the strong dependence on the electric charge of the particle $Z^{-2}$ that softens the limit in the case of nuclei.

In the case of a relativistic motion of the acceleration region, i.e. with a Lorentz factor $\Gamma \gg 1$, it is useful to introduce the co-moving reference frame, all quantities in this frame will be indicated with a prime. The condition discussed above for acceleration in the co-moving frame becomes: $r_L(E')<R'=R/\Gamma$, using the Lorentz contraction of length. Moreover, since the energy density transforms as $\Gamma^2$, we can rewrite the condition on luminosity in the case of a relativistic motion of the accelerator as: 
\begin{equation}
L>4\pi R^2c\Gamma^2\epsilon_{B'}\simeq 10^{47}\Gamma^2 Z^{-2} \left (\frac{E}{10^{20} eV}\right )^2 ~{\rm erg/s}~.
\label{eq:L_lim_rel}
\end{equation}

The two conditions, Eqs. (\ref{eq:L_limit}) and (\ref{eq:L_lim_rel}), are very stringent in the case of protons, while in the case of nuclei with high electric charge, as for iron $Z=26$, the two conditions become less constraining.

It is also important to keep in mind that the constraints obtained above only apply to stationary sources, or sources that can be considered as stationary on time scales relevant for the propagation of UHECRs. The constraints do not apply in a straightforward manner to bursting sources.

Another quantity that can be used to constrain sources of UHECRs is the energy input per unit volume and time, i.e. the sources emissivity. The flux of UHECRs can be roughly estimated as: 
\begin{equation}
J(E)\simeq \frac{c}{4\pi} \dot{n}(E)\tau_{loss}(E)
\label{eq:flux}
\end{equation}
where $\dot{n}$ is the injection rate per unit volume and $\tau_{loss}$ the time scale of energy losses. Comparing Eq. (\ref{eq:flux}) with the observed flux we can immediately deduce, at fixed energy, the expected emissivity of the sources: at the level of $3\times 10^{45}$ erg/Mpc$^3$/yr at $10^{19}$ eV. This quantity can be compared with the known emissivity of well defined sources. For instance, AGN typically show bolometric luminosities in gamma rays in the range between $L_{bol}\simeq 10^{43}$ erg/s, for Seyfert galaxies and radio-quite quasars, and $L_{bol}\simeq 10^{47}$ erg/s, for radio-loud quasars, with typical number density in the range $10^{-5}\div 10^{-4}$ Mpc$^{-3}$ \cite{Woo:2002un,Jiang:2006vv}. Hence AGN would meet the energy requirements if they emit a fraction in the range $10^{-4}\div 10^{-3}$ of their bolometric luminosity in the form of UHECRs. 

In the following we will address three general categories of possible astrophysical sources distinguishing among the acceleration mechanisms at work: relativistic and non-relativistic shocks and strong electric fields, as those produced by rotating magnetised stars.

\subsection{Non relativistic shocks and large scale structures} 
\label{sec:non_rel_acc} 

There are not many examples of non-relativistic shocks able to accelerate particles at the extreme energies of UHECRs. The most noteworthy case is certainly represented by shocks produced during the formation of clusters of galaxies \cite{Kang:1995xw,Kang:2012nz,Hong:2014yda}. Generally speaking, the formation of large scale structures in the universe naturally leads to supersonic motion of plasma that, fueled by gravitational forces, gives rise to shock waves. These shocks can be formed either during cluster mergers or during accretion of cold gas on an isolated cluster.

In filaments, that develop between clusters during large scale structure formation, typical flow velocities are around $10^3\div 10^4$ km/s with a background temperature at the level of $10^{5}$ K. The corresponding shock waves are usually relatively strong, with typical Mach numbers of order $M_s\simeq 10-100$. On the other hand, shocks produced during cluster mergers are weaker ($M_s\ll10$) because formed in the intra-cluster medium with a typical temperature at the level of $10^8$ K, being clusters already virialised structures \cite{Gabici:2002ie}. In both cases of accretion and merging, the available luminosity in the form of accelerated particles can reach $L\simeq 10^{45}-10^{46}$ erg/s, compatible with the estimates discussed above.

These kinds of accelerators operate for very long time, of the order of the age of the universe, and the accelerated particles are typically confined inside the cluster volume \cite{Berezinsky:1996wx}. The maximum energy attainable with these mechanisms is constrained by the magnetic field at the shock, which fixes the acceleration time, and by the energy losses of the accelerated particles. Particularly relevant are losses due to photo-pion production and photo-disintegration that limit the maximum energy at the level of $\sim 5\times 10^{19}$ eV \cite{Kang:1995xw,Kang:2012nz,Hong:2014yda}. Hence, shocks in clusters of galaxies, also called cosmological shocks, may represent an option for UHECR acceleration only in the case of no substantial flux observed at energies higher than few$\times 10^{19}$ eV. It is unlikely that a substantial fraction of heavy nuclei may be accelerated at these pristine shocks, because the intergalactic medium in the outskirts of clusters is probably not reach enough in such heavy elements. Finally, the spectrum of the accelerated particles produced by cosmological shocks can be determined applying the general theory of particle acceleration at non-relativistic shocks which is relatively well understood and discussed in section \ref{sec:accelera}. 

\subsection{Relativistic shocks} 
\label{sec:rel_acc} 

As discussed in section \ref{sec:accelera}, particle acceleration at shock fronts requires multiple shock crossings, which in turn depend upon the return probability from downstream being sizeable. While this does not, in general, represent a problem in the case of newtonian shocks, it becomes critical in the case of relativistic shocks. In fact, for such shocks both the particles and the front move at speeds very close to the speed of light and this results in a large anisotropy of the particle distribution at the shock. The first point to notice is that a relativistic shock is superluminal for all orientations of the background magnetic field that form an angle $>1/\Gamma$ with the shock normal. For large values of the shock Lorentz factor $\Gamma$, it becomes hard to avoid this condition. The consequences are quite important for shock acceleration: for $\Gamma\gg 1$, the shock velocity in the frame comoving with the downstream plasma is $\sim c/3$. On average a particle takes a time $\tau = 2\pi r_{L}/c$ to cover one Larmor rotation. In this time the shock moves by $\tau c/3=(2\pi/3)r_{L}>r_{L}$, namely the particle is trapped downstream and its probability of returning upstream is greatly reduced. This fact leads to expect steeper spectra for acceleration at relativistic shocks, as discussed in \cite{Lemoine:2006gg}.

The first time that a particle traverses the shock from upstream to downstream and back its energy can increase by a large factor of the order of $\simeq 4\Gamma^2$. For large values of $\Gamma$, as in GRBs that can achieve $\Gamma\sim 300$, the energy gain can be remarkable with particles that acquire energies at the level of $10^{5}\div 10^{6}$ GeV, note that this energy scale will be a low energy cutoff in the spectrum of accelerated particles. After the first shock crossing particles distribution will be beamed within an angle of the order of $1/\Gamma$ around the shock normal. The highly anisotropic distribution of particles implies a much lower energy gain at any subsequent shock crossing, if any, of the order of $\Delta E/E\simeq 2$. Note that the simple picture discussed here applies to planar shocks, it might become somewhat different (and more complicated) for non-planar relativistic shocks as, for instance, in the case of relativistically moving plasmoids as those observed in AGN jets. 

Based on the simple argument illustrated above, one can infer that reaching ultra high energies at a relativistic shock requires efficient scattering downstream, on scales smaller than or comparable with the Larmor radius of particles. The presence of such turbulence is in fact implicitly present in all models of particle acceleration at relativistic shocks \cite{Schneider:1987aa,Ellison:1990aa,Kirk:2000yh,Vietri:2002kq,Vietri:2003te,Blasi:2005qd}, which show a general consensus on the accelerated spectrum being $N(E)\propto E^{-2.3}$ following from the assumption of small pitch angle scattering downstream. However, the nature of the magnetic turbulence in the shock region determines whether the maximum energy achieved by accelerated particles is of interest for UHECRs or not. If the power spectrum of the turbulence upstream of the shock is peaked at scales larger than the Larmor radius of the particles, then the compressed field downstream of the shock front is again quasi-perpendicular and the same argument above holds, namely the return probability is small and the spectrum of accelerated particles is steep, thereby making the process of poor interest for the application to UHECRs. On the other hand the turbulence upstream may be on scales much smaller than the Larmor radius, as would be the case for the field generated through Bell-like instabilities. Such phenomena appear to be needed to explain typical strengths of the field of $10\div 100$ $\mu$G \cite{Li:2006ft,Li:2010zx} as inferred from X-ray observations. In this case, the return probability from downstream may not be small, but the scattering is weak, thereby making the acceleration time longer and the maximum energy correspondingly lower. These arguments, that here we described in a rather qualitative manner, have been formalized and investigated in detail in the context of PIC simulations in \cite{,Sironi:2015oza,Sironi:2013ri,Marcowith:2016vzl}.

Recently it has also been proposed that UHECRs may in fact result from re-acceleration of lower energy CRs (around $10^{17}$ eV): CRs accelerated to such energies in the host galaxy of an AGN may penetrate the jet of the AGN sideways, thereby receiving a one-shot boost in energy by a factor $\Gamma^2$ that allows the particles to reach ultra high energies \cite{Caprioli:2015zka}.

\subsection{Acceleration inspired by unipolar induction} 
\label{sec:unip_acc} 

The rotation of a magnetised star implies potentially large induced electric fields that, in turn, could accelerate particles to ultra high energies \cite{Ginzburg:1990sk}. Several astrophysical objects show strong magnetisation and may be suitable to accelerate UHECRs: noticeable examples are black hole magnetospheres and fastly spinning Neutron Stars (NSs). NSs were proposed as sources of UHECRs some time ago \cite{Venkatesan:1996jw,Blasi:2000xm,Arons:2002yj} while such models were recently updated mainly to accommodate the new Auger findings \cite{Fang:2012rx,Fang:2013cba,Kotera:2015pya}, with special attention for the mixed composition.

Magnetised, fast spinning NSs present important advantages as sources of UHECRs with respect to more sources more traditionally associated to UHECRs, such as AGN and gamma ray bursts: first, the energy budget is favourable with a NS population density $\dot{n}_s=3\times 10^{-3}$ Mpc$^{-3}$yr$^{-1}$ \cite{Lorimer:2008se} and a very large reservoir of rotational energy, at the level of 
\begin{equation}
E_{rot}\simeq 2\times 10^{52} \left (\frac{I}{10^{45} g cm^2}\right) \left (\frac{P}{10^{-3} s} \right)^{-2}~erg
\label{eq:Erot}
\end{equation}
with $I$ the moment of inertia and $P$ the rotation period of the star \cite{Lorimer:2008se}. Comparing these numbers with the emissivity expected from UHECR sources (see Eq. (\ref{eq:flux})), one finds that only a tiny fraction of NSs, $\leq 10^{-4}$, is expected to contribute to the observed flux of UHECRs.

Another important point in favour of the NS hypothesis is represented by the NS surface which is naturally rich of heavy elements that may potentially be accelerated thereby providing a possible explanation of the composition observed by Auger \cite{Kotera:2015pya}, while in the case of other kind of sources, such as GRB, the acceleration of heavy nuclei seems challenging \cite{Lemoine:2002vg,Pruet:2002hi,Horiuchi:2012by}.

The crust of a NS is thought to be made of condensed matter tightly bound in long molecular chains oriented along the magnetic field lines \cite{Ruderman:1972aj,Ruderman:1974aa}. These chains are thought to be made of iron nuclei ordered in a one dimensional lattice with an outer sheath of electrons. The binding energy of iron nuclei can be estimated as $\sim 14$ keV and the lattice spacing $d\simeq 10^{-9}$ cm \cite{Ruderman:1975ju}. Hence, the electric field needed to extract an iron nucleus is ${\mathcal E}_0=14 keV/(Zea)\simeq 1.4\times 10^{13} /(Z d_{-9})$ V/cm, being $Z$ the electric charge of the extracted nucleus ($Z=26$ for iron) and $d_{-9}$ the lattice spacing in units of $10^{-9}$ cm. 

Extraction of nuclei can be achieved by the electric field generated at the NS surface by the star rotation,  estimated to be \cite{Arons:2002yj}
\begin{equation}
{\mathcal E}=\frac{2\pi B R_s}{P c} \simeq 6.3\times 10^{14}\left (\frac{B}{10^{13} G}\right) \left (\frac{R_s}{10^{6} cm}\right) \left (\frac{P}{10^{-3} s}\right)^{-1} ~ \frac{V}{cm},
\label{eq:E_NS}
\end{equation}
where $B$ is the strength of the magnetic field of the star at its surface, $R_s$ is the radius of the NS and $P$ the star rotation period. 

The main effect of the electric field in Eq. (\ref{eq:E_NS}) is to extract electrons from the NS crust electrons, since they are much less bounded than nuclei. Such electrons suffer curvature losses in the strong dipolar magnetic field of the star. Photons produced as curvature radiation by electrons can in turn give rise to pairs by scattering on the virtual photons of the magnetic field, pairs will in turn generate other curvature photons, giving rise to an electromagnetic cascade. This chain of events leads to a multiplication of the number of electron-positron pairs that eventually fill the magnetosphere of the star. The number of pairs generated by a single extracted electron is in the range $10\div 10^{4}$ depending on local conditions. Pairs in the magnetosphere have a screening effect, which reduces the effective electric field available for particle acceleration. 

At least in principle, the total potential drop available for particle acceleration in the magnetosphere is \cite{Arons:2002yj}:
\begin{equation}
\Phi=\frac{2\pi^2 B R_s^3}{P^2 c^2}\simeq 7\times 10^{19} \left (\frac{B}{10^{13} G}\right) \left (\frac{R_s}{10^{6} cm}\right)^3 \left (\frac{P}{10^{-3} s}\right)^{-2} ~ V,
\label{eq:phi_NS}
\end{equation}
that would correspond to a maximum particles' Lorentz factor $\gamma_{\Phi}=Ze\Phi/(A m_p c^2)$, namely an energy exceeding $10^{20}$ eV for iron nuclei. In fact, the maximum energy of nuclei accelerated in the NS magnetosphere is limited by curvature losses. Assuming that the total potential drop $\Phi$ is available over a gap of length $\xi R_L$, being $R_L$ the radius of the light cylinder\footnote{Distance at which the peripheral velocity of the star reaches the speed of light $2\pi R_L/P=c$ and the magnetic field lines spiral outwards along the azimuth.} of the star $R_L=cP/2\pi$, and equating curvature energy losses with energy gain one gets an upper bound to the acceleration Lorentz factor as \cite{Kotera:2015pya}
\begin{equation}
\gamma_{curv}=\left (\frac{3\pi B R_s^3}{2ZeP\xi c} \right)^{1/4} \simeq 10^8 \left (\frac{\xi Z}{26}\right)^{-1/4} \left (\frac{B}{10^{13} G}\right)^{1/4} \left (\frac{P}{10^{-3} s}\right)^{-1/4} \left (\frac{R_s}{10^6 cm}\right)^{3/4} ~,
\label{NS_gmax}
\end{equation}
hence the actual maximum energy that particles can attain within the co-rotating magnetosphere will be set by $\gamma_{max}=min(\gamma_{cur},\gamma_{\Phi})$. The parameter $\xi$ takes into account the screening effect of pair creation and it can be estimated at the level of $O(1)$ \cite{Kotera:2015pya}, signalling that the gap cannot be too far from the star surface. 

If the NS wind has a Lorentz factor larger than $\gamma_{max}$, particles that end up in the wind will be advected with it at the Lorentz factor of the wind irrespective of the energy reached in the magnetosphere. This mechanism discussed in \cite{Kotera:2015pya} provides a way of increasing UHECRs energies independently of the curvature losses.

The discussion above focuses on the most classical scenario of particles acceleration nearby the stellar surface. There are in literature other scenarios in which acceleration happens at the light cylinder or further out \cite{Cheng:1986qt} that are not discussed here. 

The spectrum of UHECRs accelerated by a NS is determined by the evolution of the rotational frequency: as the star spins down the energy of the accelerated particles decreases (see Eq. (\ref{eq:phi_NS})). In general a NS is powered by the rotational kinetic energy and loses energy by accelerating particle winds and by emitting electromagnetic radiation. Because of this, the rotation frequency of the star decreases with time following the relation $\dot{\nu}=-K\nu^n$ where $n$ is the braking index ($n=3$ for a pure magnetic dipole) \cite{Ruderman:1972aj} and $K$ is a positive constant, which depends on the moment of inertia and on the magnetic dipole moment of the star \cite{Ruderman:1972aj}. The spectrum of UHECRs accelerated by the NS integrated over the history of the star, is found to be $N(E)\propto E^{\frac{1-n}{2}}$ \cite{Blasi:2000xm}, which can be as hard as $N(E)\propto E^{-1}$ in the reference case $n=3$ and even harder for braking index $n<3$, which are actually the norm.

Once nuclei are extracted from the stellar crust and accelerated by the potential gap $\Phi$ they are advected with the NS wind and interact with the environment of the star, suffering mainly photo-hadronic interactions \cite{Fang:2012rx,Fang:2013cba,Kotera:2015pya}. As a result, both the energy of accelerated particles and their mass composition change in time.

The nature of the nuclei that manage to escape the wind region at later times is fully determined by how effective photo-disintegration of nuclei on the thermal photons coming from the star's surface is in breaking heavy nuclei into lighter ones. As discussed in \cite{Protheroe:1997er,Kotera:2015pya}, for reasonable values of the NS surface temperature ($T<10^{7}$ $^\circ$K), nuclei are not completely destroyed, hence the mass composition of UHECR nuclei from NSs is naturally expected to be mixed \cite{Fang:2012rx,Fang:2013cba,Kotera:2015pya}.

\section{Exotic models} 
\label{sec:td}

The extreme energies of UHECRs, as high as $10^{11}$ GeV, eleven orders of magnitude above the proton mass and "only" eight below the Planck mass, are a unique workbench to probe new ideas, models and theories which show their effects at energies much larger than those ever obtained, or obtainable in the future, in accelerator experiments.

There are two general classes of exotic theories that can be tested using UHECRs: top-down models for the production of these extremely energetic particles and models that imply extensions and/or violations of Lorentz invariance, as in certain theories of quantum gravity.

\subsection{Top-down models and super heavy dark matter}

In top-down models, UHECRs are not accelerated particles but rather result from either the decay or the annihilation of particles with very large masses, produced as relics of early universe phenomena. The idea of generating UHECRs in this way arose in the 90's in the aftermath of the AGASA claim of absence of the GZK feature in the data \cite{Takeda:1998ps}. 

The two main classes of top-down models are associated with topological defects and super-heavy relics (see \cite{Berezinsky:1998ft} for a discussion of these two classes). The former are usually associated with symmetry breaking of some type; known instances of topological defects are monopoles, cosmic strings and necklaces. The latter, super-heavy relic particles, may form as a consequence of quantum processes during inflation and have been widely discussed as candidates for dark matter, the so-called super heavy dark matter (SHDM) to distinguish it from the more familiar WIMP candidates. In fact, the connection of super heavy relics to dark matter is, nowadays, the main scientific motivation for pursuing this type of investigation. 

The existence of SHDM has been postulated based on the possibility of particle production due to a non-adiabatic expansion of the background space-time acting on vacuum quantum fluctuations. In quantum theories the possibility of producing particle pairs through the effect of a strong (classical) external field is well known; consider, for instance, the case of $e^{\pm}$ pair creation by strong electromagnetic fields. The idea to apply such a mechanism to the cosmological context using external strong gravitational fields dates back to E. Schr\"odinger in 1939 \cite{Schro:1939}.

The theory of particle creation during the expansion of the Universe has been developed through the last 40 years, starting with the pioneering work discussed in Refs. \cite{Chernikov:1968zm,Parker:1968mv,Grib:1970mv,Zeldovich:1971mw,Grishchuk:1974ny}. More recently, in the framework of inflationary cosmologies, it has been shown that particle creation is bound to be a common phenomenon, independent of the specific cosmological scenario. Moreover the resulting SHDM particles (labeled as $X$) have been shown to potentially account for most dark matter, $\Omega_X(t_0)\simeq 1$ \cite{Chung:1998zb,Kuzmin:1998uv,Chung:1999ve,Chung:2001cb,Aloisio:2006yi,Kolb:2007vd,Fedderke:2014ura,Aloisio:2015lva}. This conclusion can be drawn under three general hypotheses: (i) SHDM in the early Universe never reaches local thermal equilibrium (LTE); (ii) SHDM particles have mass $M_X$ of the order of the inflaton mass $M_{\phi}$; and (iii) SHDM particles are long-living with a lifetime exceeding the age of the Universe, $\tau_X\gg t_0$.

Precision measurements of CMB polarisation and observations of UHECRs up to energies $\simeq 10^{20}$ eV enable a direct experimental test of the three hypotheses listed above. As discussed in \cite{Aloisio:2015lva}, the first two hypotheses can be probed through the measurement of CMB polarisation. The third hypothesis, particle life-time longer than the age of the Universe, is a general requirement of any DM model based on the existence of new particles. As in the case of WIMPs, discrete gauge symmetries protecting particles from fast decays need to be introduced (see \cite{Aloisio:2003xj,Aloisio:2006yi} and references therein).

\begin{figure}[!h]
\begin{center}
\includegraphics[scale=0.45]{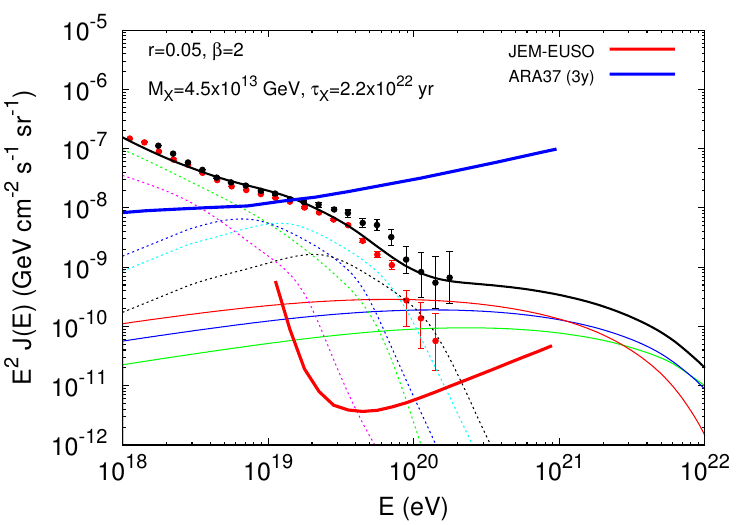}
\includegraphics[scale=0.45]{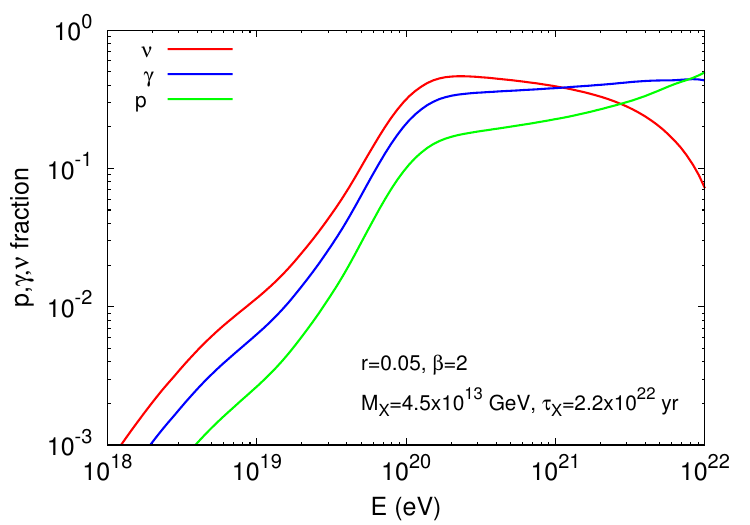}
\end{center}
\caption{[Left Panel] Flux of UHECR from the decay of SHDM (thin solid lines) with parameters as labelled together with the flux expected in the framework of the mixed composition model of \cite{Aloisio:2015ega}. Also shown is the sensitivity to SHDM decay products: of the proposed JEM-EUSO space mission (thick red solid line) and, for UHE neutrinos, the upcoming ARA observatory (thick blue solid line). Experimental data are those of Auger (red points) \cite{Aab:2015bza} and TA (black points) \cite{Ivanov:2015pqx}. [Right Panel] Fraction over the total flux of protons, photons and neutrinos by SHDM decay as follows from fluxes in right panel. Both figures are taken from \cite{Aloisio:2015lva}.}
\label{fig9}
\end{figure}

The best way to test the existence of SHDM is through the indirect detection of its annihilation and/or decay products (direct detection is unattainable). In general, since the annihilation cross section of a (point) particle is bound by unitarity, $\sigma_{ann}\propto 1/M_X^2\sim 1/M_\phi^2$, the annihilation process results in a small annihilation rate \cite{Aloisio:2006yi}, although alternative theoretical models can be constructed (see for instance \cite{Blasi:2001hr}) in which this limitation is overcome. Below we will specialize our discussion to decay channels.

If SHDM particles decay, under general assumptions on the underlying theory (see \cite{Aloisio:2003xj,Aloisio:2006yi} and references therein), we can determine the composition and spectra of the standard model particles produced. Typical decay products are neutrinos, gamma rays and nucleons with a relatively hard spectrum, that at the relevant energies can be approximated as $dN/dE \propto E^{-1.9}$, independently of the particle type, with a photon/nucleon ratio of about $\gamma/N\simeq 2\div 3$ and a neutrino nucleon ratio $\nu/N\simeq 3\div 4$, rather independent of the energy range \cite{Aloisio:2003xj}. The most constraining limits on SHDM lifetime are those coming from the (non) detection of UHE photons and neutrinos. 

Auger observations provide very stringent limits on the flux of photons at energies above $10^{19}$ eV: at the level of $2\%$ \cite{Aglietta:2007yx}. This fact already constrains the SHDM lifetime to values $\tau_X\ge 10^{21} \div 10^{22}$ yr, depending on the underlying inflationary potential, and implies that SHDM models can be effectively probed only by the next generation of UHECR experiments, those designed to maximise statistics at the highest energies \cite{Ebisuzaki:2014wka,Cline:1999ez,Petrolini:2009cg}, together with new and more refined observations of the CMB polarisation pattern.  

In the left panel of figure \ref{fig9}, as discussed in \cite{Aloisio:2015lva}, we plot the flux of UHECRs coming from the decay of SHDM in a specific model of inflation with $M_{X}=4.5\times 10^{13}$ GeV and $\tau_X=2.2\times 10^{22}$ yr (solid lines); we also show the expected sensitivities of the proposed JEM-EUSO space mission (thick red solid line) \cite{Ebisuzaki:2014wka} and, for UHE neutrinos, of the upcoming ARA observatory (thick blue solid line) \cite{Allison:2011wk,Allison:2014kha}. In the right panel we show the corresponding fractions of fluxes of protons, photons and neutrinos produced by the decay of SHDM. 

\subsection{Tests of Lorentz invariance} 

Lately a general consensus has emerged that UHECR observations can be used as a powerful tool to put under experimental scrutiny the validity of Lorentz Invariance (LI). The possibility of testing LI at scales not probed so far is interesting {\it per se}, as any new experimental scrutiny of theoretical foundations. Moreover, the need for extensions and/or violations of LI can be connected to the long-standing problem of the construction of a quantum theory of gravity (QG). 

Our universe is very well described by quantum mechanics at small scales and general relativity at large scales, but a unified theory of QG is still out of reach. While all other fundamental interactions propagate through space-time, gravity turns out to be a property of space-time itself. This simple statement, at the base of General Relativity, has important implications for the construction of a quantum theory of gravity, as it implies that the structure of space-time itself has quantum fluctuations. In other words, at the scales where quantum effects of gravity arise, space-time could have a granular (or foamy) structure where the size of space-time cells fluctuates stochastically causing an intrinsic uncertainty in the measurements of space-time lengths and, indirectly, of the energy and momentum of particles. The typical scale at which quantum gravitational effects are supposed to become relevant is the so called Planck length, i.e. the length scale given by $l_p=\sqrt{\hbar G/c^3}\simeq 1.6\times 10^{-33} $ cm. It is generally argued that measurements of distances (times) smaller than the Planck length (time) are conceptually unfeasible, since the process of measurement collects in a Planck size cell an energy exceeding the Planck mass ($M_P=\sqrt{\hbar c/G}\simeq 1.2 \times 10^{19}$ GeV) hence forming a black hole in which information is lost. 

As was immediately realised after the proposal of the GZK suppression \cite{Kirzhnits:1972sg}, in the reference frame in which astrophysical photon backgrounds are isotropic, an UHE nucleon only needs a fractional gain in energy at the level of $10^{-22}\div 10^{-21}$ to perform the transition to its final state (photo-pion production or photo-disintegration). LI guarantees that this is exactly the same to what happens in the reference frame in which the nucleon is at rest and the photon has $10\div 100$ MeV energy. But this also shows that even very tiny violations of LI are bound to give, in some selected reactions at least, observable effects. The kind of reactions typically very sensitive to LI violations are those characterised by a particle production threshold \cite{Aloisio:2000cm,Aloisio:2002ed,Aloisio:2005rc,Aloisio:2006nd}. 

In recent times LI violating models have been investigated in depth and their implications compared with available experimental data \cite{Liberati:2013xla,Kostelecky:2008ts}. Particularly interesting is the approach of Effective Field Theories (EFT) in which LI or CPT symmetry (and renormalizability) is no longer a guide; in this kind of theories the number of possible terms violating LI is very high. Those that can be tested experimentally (several hundreds) are described in \cite{Kostelecky:2008ts} and can be generally modelled through modifications of the dispersion relation of particles \cite{Carmona:2012un} (which in the EFT approach corresponds to modifications of kinetic terms in the Lagrangian density), such as: 
\begin{equation}
E^2-p^2=\mu^2(E,p,M_P)\simeq m^2 + f E^{2+n}/M_P^n,
\label{eq:mashell}
\end{equation}
where $\mu$ is a generalised "mass" that can be always approximated as the mass of the particle $m$ plus terms that violate LI at the strength fixed by $f$ ($f=0$ corresponds to the standard invariant relation). 

The firm experimental evidence of the suppression in the spectrum of UHECRs around few$\times 10^{19}$ eV implies very stringent limits on the possible violations of LI. Using the parameterisation introduced in Eq. (\ref{eq:mashell}), the case $f>0$ is strongly excluded by observations, because in this case thresholds for particles' production move to lower energies and new exotic processes are allowed, such as vacuum \v{C}erenkov $p\to p\gamma$ for which very strong bounds exists \cite{Klinkhamer:2008ss}. As soon as $f$ moves toward negative values, thresholds for particle's production slightly increase up to the point where the process becomes kinematically forbidden. In this case, limits of LI violations obtained from the observed spectral steepening are reported in literature \cite{Jacobson:2005bg,Saveliev:2011vw}. These limits, however, depend crucially on the assumption that the steepening in the flux is originated by the propagation of UHECRs. As discussed in section \S \ref{sec:prop}, Auger data can be very well accommodated in models in which the flux suppression is connected with low values (see section \S \ref{sec:mix}) of the maximum acceleration energy at the source. In this case no relevant limit on LI violations can be placed using the observed flux of UHECRs \cite{Aloisio:2014dua}. 

Violations of LI can also produce important effects in the development of showers produced by the interaction of UHECRs with nuclei of the Earth's atmosphere. These effects typically reduce the kinematical phase space for certain processes modifying the particles content of the cascading shower. The most important process in the physics of cascades is the neutral pion decay $\pi^0\to \gamma\gamma$, which has a reduced kinematic phase space in the case of LI violations ($f<0$) with stable neutral pions at energies larger than $E>(M_P^n m_\pi^2/|f|)^{\frac{1}{2+n}}$ \cite{Aloisio:2014dua}. This modification of particles' cascade has the net effect to move the shower maximum to higher altitudes as the electromagnetic part of the shower consumes faster. Moreover, it produces an increased number of high energy muons in the shower due to the interaction of "non-decaying" neutral pions. As of today, observations of the shower development in the atmosphere are not able to exclude LI violations effects, that are however much weaker and difficult to tag than in the case of UHECRs propagation. 

\section{Transition between galactic and extragalactic cosmic rays} 
\label{sec:trans}

In \S \ref{sec:transport}, \S \ref{sec:accelera} and \S \ref{sec:trans} we discussed the transport of galactic CRs, their acceleration, mainly in SNR shocks, and the transport of extragalactic CRs respectively. We left on purpose the definition of galactic and extragalactic CRs somewhat vague. In this section we discuss the transition between the two in the different scenarios that we introduced earlier in this review paper. 

Historically, the transition from galactic CRs to extragalactic CRs has been assumed to take place at the ankle: CR iron nuclei were assumed to be accelerated up to energies in excess of $\sim 10^{19}$ eV , where they would leave room to extragalactic CR protons, injected with spectrum $\sim E^{-2}$. This picture is typically invoked in models of GRBs as sources of UHECRs \cite{1995PhRvL..75..386W,1995ApJ...453..883V}.

This picture remained virtually untouched until it was realized that an ankle-like feature would naturally appear in the spectrum of extragalactic CRs due to the combination of the expansion of the universe and pair production of protons propagating in the CMB photons, the so-called dip scenario  \cite{Berezinsky:2002nc,Aloisio:2006wv}, discussed in \S \ref{sec:dip}. In this model the galactic CR spectrum is required to end with a heavy composition at much lower energies, $\sim 10^{17}$ eV, two orders of magnitude below the energy of the ankle. This picture is roughly consistent with the idea that galactic SNRs may accelerate protons up to the knee. 

As discussed in \S \ref{sec:pevatrons}, acceleration of CRs to PV rigidity at SNR shocks is also all but trivial: only in core collapse supernovae exploding in the wind of their red giant companions there seem to be the conditions to reach maximum energies around $E_{max}\sim 10^{15}$ eV for protons \cite{Schure:2013kya,Bell:2013kq,Cardillo:2015zda}, reached at the beginning of the Sedov-Taylor phase, that occurs about $\sim 30$ years after the explosion for these supernovae. In fact it was  pointed out that particles could be accelerated to somewhat higher energies at earlier times, but the number of particles processed is smaller, so that the transition between the ejecta dominated and the Sedov phase leads to a broken power law rather than an exponential cutoff at $E_{max}$. 

This finding was used in Ref. \cite{Cardillo:2015zda} to calculate the shape of CRs between the end of the galactic component and the beginning of the extragalactic one, assuming that in the energy region between $10^{17}$ and $10^{18}$ eV the flux of light CRs as measured by KASCADE-Grande \cite{Apel:2011mi,Apel:2013dga} is of extragalactic origin. 

A careful investigation of the transition region in the context of the dip and ankle models was carried out in Refs. \cite{Aloisio:2006wv,Aloisio:2007rc}, assuming an exponential cutoff in the galactic CR component. The authors concluded that the ankle model is basically ruled out by measurements of the depth of shower maximum, $X_\mathrm{max}(E)$, and its rms fluctuations. 

As mentioned above, a turning point in the investigation of the transition region is represented by the recent measurement of the spectra of the light and heavy components of CRs in the energy region $10^{17}-10^{18}$ eV by KASCADE-Grande \cite{Apel:2011mi,Apel:2013dga}. These measurements found evidence for an ankle-like feature in the light component at $\sim 10^{17}$ eV and a knee-like feature in the heavy component at roughly the same energy. The former was interpreted by many as a signature of the transition to a light extragalactic CR composition, while the latter was interpreted as the end of the galactic heavy CR component. 

The main reason to believe that the light KASCADE-Grande component is of extragalactic origin is the low level of anisotropy observed by Auger at $10^{18}$ eV, that seems to be at odds with a galactic origin of protons at those energies \cite{Giacinti:2011ww}. This view has been recently questioned in Ref. \cite{Eichler:2016mut}, where the authors proposed that CRs might all be generated in occasional galactic GRBs, at all energies, and advocate that the low observed anisotropy can still be compatible with this picture \cite{Kumar:2013jaa} if the transport of CRs in the galactic magnetic field is properly taken into account. Further investigation is probably needed to identify the reasons for the different conclusions reached by different numerical simulations of CR propagation. 

It is worth stressing that if indeed the light CR component measured by KASCADE-Grande is of extragalactic origin, the transition region becomes weakly dependent upon whether extragalactic CRs are all light (dip model) or have mixed mass composition, since in the energy region $10^{17}-10^{18}$ eV the expected composition is light in both scenarios (see discussion in \S \ref{sec:mix}). What is less clear is whether this light component reflects a class of sources which are different from those that produce nuclei (with a hard injection spectrum) or rather the light component is due to interaction processes inside the sources, as advocated in Ref. \cite{Globus:2015xga}.

\section{Conclusions} 
\label{sec:conclu}

While writing this review article on the origin of the cosmic radiation, we often found ourselves in need of editing the manuscript to update it with the most recent ideas and/or measurements appearing in the literature. This is a clear evidence that one century after the discovery of cosmic rays, this field of research remains extremely lively, mainly driven by new exciting experimental findings.

This wealth of data pouring in the field has the effect of continuously challenging models, even those that in time appear to be well established. In this context, the main role has been played by measurements of spectra and mass composition. Although almost all spectra in CR science have a power-law-like shape, it is the deviations from such power laws that typically provide clues to the underlying physics. For instance, the knee, probably the most prominent and best studied feature in the all-particle CR spectrum, is by most considered as emerging from the combination of two factors: 1) the maximum energy of accelerated protons, and 2) the rigidity dependence of the acceleration process, so that a knee results from the overlap of the spectra of elements with different atomic number $Z$. This implies that the mass composition at the knee should be predominantly light (H and He), a picture that is supported by the KASCADE data. Recent data by ARGO-YBJ and by CREAM cast a doubt on this picture, showing some evidence of a flux reduction in the light component (H+He) at energies well below the knee in the all-particle spectrum, which would then correspond to intermediate mass nuclei. 

At first sight it may seem surprising that decades after the experimental discovery of the knee there are still uncertainties on its interpretation, but the truth is that this is the first time that we are actually probing this region with direct measurements on one side, and the first time that we studied EAS very close to the shower maximum (high altitude), and its core, with full coverage arrays. 
This is the crucial point: understanding the physics of the knee requires a credible measurement of the mass composition in this energy region. The fact that 
available data return somewhat conflicting outcomes provides us with an estimate of the systematic uncertainties intrinsic in experimental approaches, mainly 
coming from indirect measurements. As long as these systematic uncertainties are not understood the issue of the physical origin of the knee will remain open.
In this aspect, the extension to larger energies of direct (space based) detections is a key issue.

On the other hand, whenever the mass composition has been measured accurately the improvement of our knowledge has been astonishing: a clear instance is the measurement of the spectra of protons and helium nuclei at energies $\leq 1$ TeV by PAMELA and AMS-02, that showed the existence of a spectral break at rigidity $\sim 300$ GV and systematically harder spectrum of helium nuclei. This finding is leading to the development of alternative models of CR transport and the investigation of the implications for secondary nuclei is only starting now with the most recent measurements of the B/C ratio by AMS-02. 

A similarly problematic situation exists in the ultra high energy region: while the all-particle spectrum has been reliably measured and the existence of features such as the ankle and the so-called GZK suppression has been confirmed, their interpretation is still subject of an active debate, mainly triggered by an experimental assessment of the mass composition that is far from being completely understood. The Auger Observatory measured $X_\mathrm{max}$ and its fluctuations and concluded that the mass composition is bound to be mixed, being predominantly light around $10^{18}$ eV and gradually heavier at higher energies, though iron seems to be basically absent. In this picture the GZK feature is not associated with energy losses but mainly to an intrinsically low value of the maximum energy at the sources. The Telescope Array, with a smaller surface and different systematics, also measures the GZK suppression but such feature seems to be compatible with a light composition.
Again, as for the case of the knee, the physical picture remains unclear and can be possibly outlined only when the observational situation will be better assessed. 
The common effort of the Auger and TA collaborations made setting up joint working groups to compare data and analysis tools will certainly
help to assess the experimental situation.
%% Again, as for the case of the knee, since the observational situation appears to be unclear the corresponding physical picture is equally unclear. 
This situation also affects the related problem of transition from galactic to extragalactic CRs.

Therefore the reliable measurement of the mass composition of CRs appears to be central to the investigation of the origin of CRs throughout the spectrum. It is certainly positive that the issue is mainly experimental in nature: one should aim at building detectors with large exposure and accurate mass discrimination in the knee region in order to make sure that this region is well understood. ISS-CREAM and HERD are appropriate responses to this requirement. In the transition region the infill and low energy enhancements of Auger and TA are definitely a move in the right direction. 

At the highest energies the problem is somewhat more complex, this being mainly a consequence of the limited exposure of fluorescence data, that at present constitute the bulk of composition measurements. The  AugerPrime  upgrade  of  the  Pierre Auger Observatory  has  been  specifically  designed  to extend the energy range of these measurements. Yet the accurate measurement of the mass does not only rely on an increased statistics of events, but is also strongly dependent upon the understanding of shower development. Although much progress has been taking place in this field,  especially in the aftermath of LHC data collection, there are still aspects of the shower development that need improvement to be reliably applied to mass discrimination. 
Finally, new experimental approaches to the extreme energies, i.e. UHECR induced fluorescence light observation from space, 
are currently in R\&D phase and, if the sufficient mass resolution will be reached, would help understanding the mass composition
providing an important jump in exposure.

\begin{acknowledgement}
This review paper sprang out of the workshop 'Multiple Messengers and Challenges in Astroparticle Physics', held at GSSI from October 6 to 17, 2014. The authors acknowledge all participants for making many discussions possible during and after the workshop. The authors are also grateful to their colleagues at the GSSI, at the Arcetri Astrophysical Observatory, at the LNGS, at the University of Salento and the University of L'Aquila for stimulating discussions on the topics illustrated in the review.
\end{acknowledgement}

\bibliographystyle{spphys}
\bibliography{CosmicRays-GSSI}
\end{document}